# Stochastic resonance for exploration geophysics

Mensur Omerbashich
Eötvös Loránd Geophysics Institute of Hungary, Columbus u.17-23, Budapest
Ph. +36-1-252-4999 x151, Fax +36-1-363-7256
Correspondence: *omerbashich@gmail.com*, cc: *omerbashich@elgi.hu*

Stochastic resonance (SR) is a phenomenon in which signal-to-noise (S/N) ratio gets improved by noise addition rather than removal as envisaged classically. SR was first claimed in climatology a few decades ago and then in other disciplines as well. The same as it is observed in natural systems, SR is used also for allowable S/N enhancements at will. Here I report a proof-of-principle that SR can be useful in exploration geophysics. For this I perform high-frequency Gauss-Vaníček variance-spectral analyses (GVSA) of model traces characterized by varying levels of complexity, completeness and pollution. This demonstration justifies all further research on SR in applied geophysics, as energy demands and depletion of reachable supplies potentially make SR vital in a near future.

## Introduction

*Stochastic resonance* (SR) (Benzi *et al.*, 1981, 1983) is a rather elusive improvement in the signal-noise ratio (S/N), as it occurs by way of adding rather than removing the classical stochastic noise; for SR reviews see, e.g., McNamara and Wiesenfeld (1988), Gammaitoni *et al*. (1998), and for SR newer developments see, e.g., Wu *et al*. (2006).

I here demonstrate potential value of SR for exploration geophysics as typified by virtually complete, short-span (on the order of seconds-to-minutes) records of relatively large and densely packed data. Here by ‘densely packed’ I mean a complete record in which the ratio of time-span v. sampling size exceeds the time span by two orders of magnitude or more. The reasoning for using SR in exploration geophysics is to obtain quality by adding Gaussian noise *via* data pollution as well as data removal, where by new quality I mean improved S/R, which is of importance say in the vibroseis approach where signals are of generally small power (Seriff and Kim, 1970). I here tacitly assume that both random purification as well as stochastic pollution could add harmonic noise, thus expending on a common belief that the random purification can remove stochastic (harmonic-degenerating) noise. Therefore, purification in this context plays the role of enhancer of classical harmonic signal, in at least two unrelated ways.

I apply the Gauss-Vaníček variance-spectral analysis (GVSA) by Vaníček (1969; 1971), a method fundamentally different from the classically favored Fourier Spectral Analysis (FSA). Certain advantages of the GVSA v. FSA are well established; see, e.g., Taylor and Hamilton (1972), Press et al. (2003), Omerbashich (2006). This holds true particularly in low-frequency analyses of highly (up to 90%) discontinuous records of natural data of long spans (on the order of days-to-Myr+), such as in climatology, astronomy, or any other fields that deal with inherently unstable systems.

What gives leverage to the GVSA the most is its large insensitivity to discontinuities in a dataset. Besides, the GVSA is able to model periods, unlike the FSA that simply recovers periods which must be (made) fully resolvable in advance. The GVSA can be an alternative to its famed counterpart at least for some cases of high-frequency analyses of records typical of exploration geophysics as well. This because such records technically "suffer" from *information overload* i.e. they contain redundant contents in the sense of robustness as lent by any least-squares modeling (but not necessarily in the FSA sense). This becomes helpful as I use synthetic seismic traces of increasingly systematic complexity, which I then also expose to varying levels of random pollution by white noise as well as to varying levels and types of data purification *via* systematic and random removal. I thus hope to observe SR in typical exploration geophysics settings, as well as to evaluate potential benefits from SR. For general or blind performance tests of the GVSA, see, e.g., Abbasi (1999), or Omerbashich (2004), respectively.

**Why use GVSA for this testing: information exist outside Fourier application domain and jargon too**

By definition, any data analysis not concerned with gaps in data also does not depend on *dt*. Then, strictly speaking, it makes no sense to talk about the Nyquist frequency in the GVSA where the continuity (requirement) of classical mathematical analyses ceases to be a limiting factor. As it is well known, in the FSA and Fourier Transform realm of continuous data and continuous functions alike there is a frequency called the Nyquist limit: the "absolutely highest" frequency that may be taken into account at a given sampling rate in order to reconstruct the signal. But, reconstruction is a concept from the continuous and hence imagined world of mathematical analysis. As such, reconstruction by definition is not a goal of the GVSA, for in the GVSA a fitting occurs which abides by a perplexing nature of the Law of Large Numbers. Thus the GVSA is a fundamentally different method of spectral analysis than the FSA, and thereby a suitable a gauging tool.

Then one can duly object to the all-situations application of mathematics of continuous functions (here to exploration geophysics), and replace for testing purposes the usual mathematical analyses with the mathematical (here least-squares) *norming* i.e. norm-fitting, alone. It is of little help that sometimes continuous functions are being replaced with their crude discrete representations, as this is a mere processing necessity. Same goes for other trickery normally done out of necessity too, such as a mathematical "transform" of *real something* into *imagined nothing* that therein becomes *imagined something* and gets "sent" back into *real something*. But the latter is unavoidably dependent on some prime theoretical foundations, which in case of the FSA are based on continuity requirements (of data; of functions).

What exactly are we looking at then, for by using the GVSA, spectra of any size can be drawn from data of any size? In other words, is there a firm ground to stand on when interpreting GVSA results? For example, if there are less data values than spectral points, such a fitted spectrum could look jiggled, with slopes resembling anything but a smooth curve. But even so baleful looking a creature is far from being meaningless. Rigorously considering (the outskirts of) the FSA application domain, the smoothly curving spectra and their peaks – be them poorly or well resolved – look as such mainly due to inescapable limitations of the FSA method. This was demonstrated by Omerbashich (2007) who showed on a ~2000 pt sample representing some 3800 data values including placeholders for missing values, that by simply increasing the spectral resolution from 2000 to 50 000 pts one becomes able to obtain a natural (therein: lunar synodic month) period with the maximum accuracy attainable by the instrumentation used (therein: a superconducting gravimeter (SG)), from a dataset, which, in classical view, was relatively undersampled 12-25 times. But how is this apparent "black box" possible, given that merely a low pass Gaussian filter that properly weighs data for gaps, was applied in that case? It appears that there is much more than meets the eye in the periodicity game concerning natural data. Or, at least, more than what a FSA fundamental requirement (for continuous records) would have us believe. Thus it turns out that, contrary to the common idea, not only that the FSA is not a universal i.e. all-situations method, but also the Nyquist frequency is not a general parameter at all.

In the least-squares norming, which usually is more general than simple (linear, but the so-called nonlinear hardly score any better either) mathematical analyses, the classical *undersampling* means just that information was not loaded into processing to the fullest. By extension then, the classical *oversampling* can mean also an information overload in the GVSA sense as norm is being fitted and not, simplistically enough, met. Obviously then, the downsizing of such an overload in the GVSA can be expected to extend no significant consequences on the spectra, to a degree. It is indeed valid, if only by a sheer abundance of such categories in the FSA lingo, to assume that nature had actually intended an important role for what we *a priori* denounce as classical noises, aliases, natural interference, (harmonic) ghosts, (non)linear trends, vibroseis harmonic distortion, and so on; a role beyond reach of hereinafter primitive logics of the FSA application domain as normally based on *equidescription of equisampled* nature. This point of view deserves at least a benefit of doubt, since the reader surely can appreciate the fact that nature is anything but uniformly boxed. Of course this thought is trivial, but with advancements in measuring technology the questions remaining to be answered are in fact far from being trivial. This because the more isolated and better defined those last questions normally become, a clearer picture on fundamental shortcomings of the methodology used to answer most of the allegedly already answered questions emerges too, if only as a byproduct of scientific progress.

So in the Fourier application domain i.e. way of thinking, periods get literally reconstructed. For this operation to succeed however, the periods must be (made) *a priori* and fully resolvable. This is so because we do know what the simplistic norm therein is, what it looks like, and why it is that it works – the latter requiring rather elementary-continuous mathematics for anyone to reproducibly remonstrate. However, in the Gauss-Vaníček realm (of least-squares, variance-norming), we do as well know what it’s (entirely different from the Fourier) norm looks like, but we have not even a slightest idea as to why it is that *it* works (*via* fitting the data so as to satisfy the norm); because, again, the latter norm type is based on the never understood Law of Large Numbers. Then, the GVSA might also be able to shed new light on the Law of Large Numbers itself.

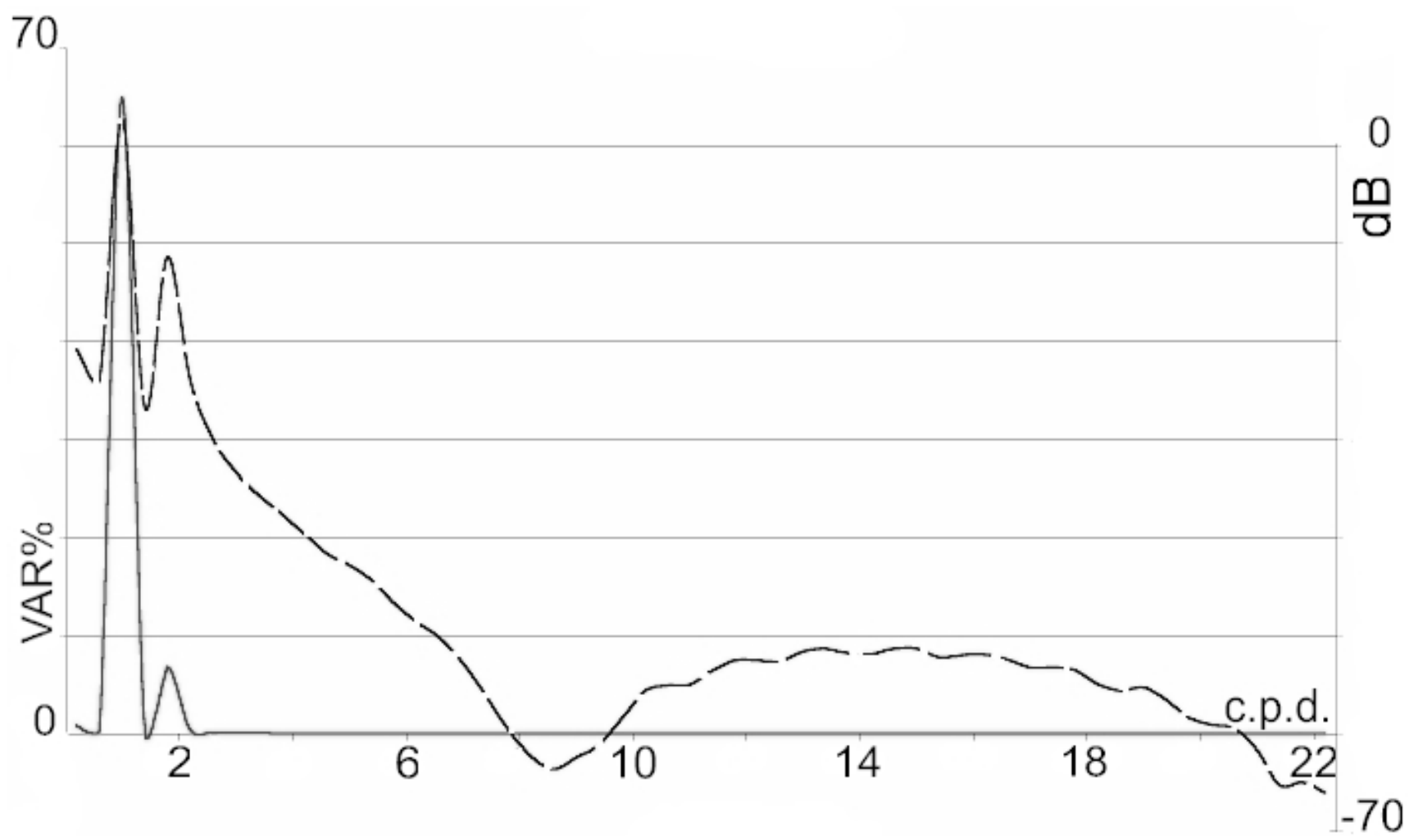

FIG.1. Variance-spectrum (solid) of one-week of SG reduced gravity, in var%, v. power-spectrum (dashed line) of the same data, in dB. Spectral peaks are 99% significant. Band: 0.2-22 cycles per day (c.p.d.).

The FSA, its derivatives as well as supporting techniques are by far the most used tools for spectral analysis in all sciences. As mentioned earlier, this is undeserved for some notable types of records. On the other hand, and in addition to its most remarkable (non-uniform-sampling-) advantage over the FSA, it is easily seen that comparatively the GVSA is also less sensitive to the presence of system noise. In fact, being the most natural descriptor of system noise (simply, variance feeds on noise!), GVSA spectra (GVS) are blind to the existence of gaps, to a significantly useful degree. This in turn means that original data can also be freely edited, to a degree and in such a manner that conventionally less reliable or noisier observations get purposely left out, and thus the record purified. This could be done in order to increase the reliability of harmonic components' estimates without significant losses in precision on frequency estimates.

Note that the GVSA requires neither preprocessing (to artificially "enhance" the time series) say *via* record padding by inserting invented data values, nor post-processing (to enhance the respective spectrum) say *via* stacking or other augmentation. Furthermore, the output variance-spectrum generally features linear background, i.e., the spectrum is rather zero everywhere except for the periodicities; see Fig.1. This gives unique definition and full meaning to the spectral magnitudes across the entire band of interest. Compared to other spectral analysis methods' the GVSA-preferred application domain is a most general one, as it encompasses virtually any set of numerical or quasi-numerical (originally non-numeric) data continuous or not, thus enabling for unbiased production of spectra. This makes the GVSA a preferred technique in practically all sciences that deal with inherently discontinuous records, or with generally noisy records (regardless of the completeness), the latter case being precisely that of exploration geophysics!

## GVSA time-domain filtering: data purification

In the following, by *data purification* I shall mean the removal to the greatest (non-significant) degree possible of the conventionally least reliable measurements from the dataset of interest in order to improve periodicity estimates from that dataset. Fig.2 demonstrates validity of the time-domain filtering by the data removal, for significant improvements in periodicity estimates (therein: of the lowermost end of the long-periodic part of the Earth's normal band of free oscillation).

In order to estimate the grave mode of the total Earth's mass oscillation, broadband recordings with SG at Cantley, Que., of gravity past three greatest shallow (focal depth at or below est.10 km) earthquakes from the 1990-ies, were used to compute the results of Fig.2. Thanks to the GVSA-unique ability to process gapped records, the differences were sought between the spectra of gravity records without gaps v. spectra of the same data but with gaps artificially introduced. Thus new, 5, 21, and 53 filter-step-long (8 sec steps) gaps were made in the three records respectively, where the order of earthquakes was selected at random. By observing the differences between the GVSA spectra of complete v. GVSA spectra of incomplete records, the first instance when this difference reaches zero was sought for. Since both the complete and incomplete records described the same instances and the same field (therein: the total Earth's gravity field) sampled during the three strongest distinct kinetic energy emissions, it is precisely this value which marked the beginning of the Earth's natural band of oscillation as well. To prove this setup correct, it sufficed to show that, the more gaps a test-record had, the more pronounced the impact of the non-natural information onto the spectra was. Indeed, more gaps resulted in a clearer distinction between the natural and non-natural bands, Fig.2. Thus the grave mode (i.e., the most natural period) of the total Earth's mass oscillation was measured as $T_o$' = 3445 s ±0.35%, where the uncertainty was based on 1000 pt spectral resolution (Omerbashich, 2004; 2007). This is in agreement with the seminal paper by Benioff (1958), and it represents an alternative way to the Rayleigh's Method for determining the fundamental frequency; see, e.g., Den Hartog (1985).

Since only raw data are required for data preparation, and subsequently for the GVSA as well, the procedure is justified by the most natural criterion of all: that of using raw (unedited) data rather than artificially edited datasets as normally created by augmenting the data in order to make input records artificially equispaced (padded) and thereby fit for feeding the data into a pre-selected processing algorithm, such as the FSA and its derivatives or patching procedures (padding, detrending, boosting, tapering, windowing, etc.).

Here I test the GVSA in a new filtering concept for the time-domain, or, smoothing in the frequency domain, by using a synthetic trace in a stationary setup (from one station). I dub this concept the *data purification procedure* (Omerbashich, 2007). Other terms used herein for the same approach are *successive decimation* and *even/odd decimation*.

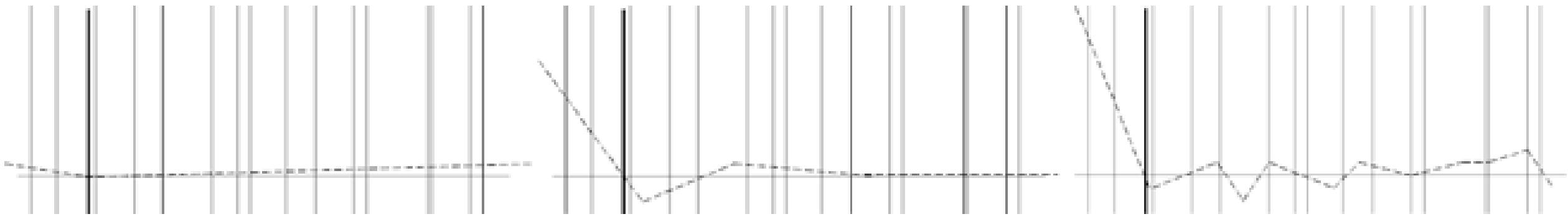

FIG.2. Validity of time-domain filtering *via* data purification, for achieving significant improvement in least-squares estimates of periodicities. Shown is the blind plot of a linearized effect (dashed line) of regarding a series as equidistant. Here for: 5 (left), 21 (middle) and 53 (right panel), 8-second gaps in the 1Hz diurnal gravity data from the Canadian SG, at one week past three strongest global shallow earthquakes of 1990-ies. The effect is computed as the difference between the GVS of complete v. the GVS of gapped SG gravity data. The effect shows an increase in gravity field's oscillation magnitude (disturbance) with decrease in data completeness (mimicking an increase in the energy supplied to an otherwise undisturbed system). See text for the statement on the effect's physical meaning. Vertical grid marks long normal periods of the Earth's free oscillation. Thick single line marks the grave-mode period of the Earth, ca. 57 minutes. Ordinates range: -0.01 to 0.05 %var (cf., Omerbashich, 2006).

**The test**

The test, composed of a series of computations, was devised in the following manner:

a. First: internal checks of GVSA usefulness for geophysical applications (Results given in Supplement). Create a single-station, synthetic sweep seismogram (the main test-series) S, complete with synthetic data, harmonic noise inclusive;
b. spectrally analyze S in the $f \in 10–200$Hz band, using GVSA (to produce variance- and power-spectra) and/or using Short-time Fourier Transform – SFT (to produce magnitude- and power-spectra);
c. examine the differences between, as well as the ratios of the so-obtained G-V variance-spectra (GV VS) v. G-V power-spectra (GV PS) (since the S is complete, here it does not matter whether the PS are computed in GVSA or in SFT, so for this first part of the test the GVSA can be used alone; besides, the first part of the test aims at establishing the GV VS and GV PS as valid spectral analysis tools in applied geophysics). Do it separately for the zero-padded and for the non-padded synthetic seismogram. Previously perform general internal checks to demonstrate the validity of the GVSA for applied geophysics, as using a stationary (single station) synthetic seismogram without modeling (single harmonic); the seismogram is to be expanded to more realistic models for subsequent use.
d. Second: purification by successive decimation of systematically-noised synthetic data. Create fifty test-series $A_i$, by successively decimating every other value of the synthetic data S, starting with 0.5% removal, then 1%...50%, from the first half of S (for a total of 25% of the entire dataset), where $i$ = 1…50, resolution is 0.5%, and there is correlation $A_i \rightarrow A_{i+1}$; this creates the worst-case scenario (of total one-sided biasness) even for the one quarter-removal;

e. compute the high-frequency GV VS response for such a simplistic (correlated) purification-by-decimation, by GV-spectrally analyzing each of the $A_i$ (and if needed for resolution purposes, create also an $A_j$' series consisted of $j$ sub-samples, say $j$ = 3.1, 3.2…3.9; or, expand the $A_i$ series to $i$ = 51…100 while keeping the original resolution);
f. compute the high-frequency SFT response as in the above, and, if better resolution is needed, do it also around individual elements of the $A_i$ for which the computed GV VS start underperforming (meaning: largest GV VS peaks drop below the 99% significance level – for this, select say a few marker peaks with highest fidelity, and use those markers also to observe the GVSA frequency response to purification; if selecting marker peaks turn out unpractical or/and such peaks indistinguishable, observe the response of fidelity instead);
g. for analysis purposes, create a diagram by stacking the $s_i^{GVSA}$ v. $s_i^{SFT}$ spectra all depicted in some simplified, say linearized representation, and compare whether the $s_i^{SFT}$ starts underperforming sooner than the $s_i^{GVSA}$ does, or vice versa.
h. <u>Third: purification by random removal of systematically-noised synthetic data</u>. Create fifty test-series $B_i$, by randomly removing synthetic data, first 1%, then 2%...50%, from S, where $i$ = 1…50, and $B_i \nrightarrow B_{i+1}$;
i. examine the high-frequency GV VS response to such an uncorrelated (more realistic) purification, by GV-spectrally analyzing each of the $B_i$ (if needed for resolution purposes, create also a $B_j$' series of $j$ sub-samples, say $j$ = 2.1, 2.2…2.9; or, expand the $B_i$ series to $i$ = 51…100 while keeping the original resolution);
j. do the same for SFT as in step f;
k. create a diagram in the same manner as in step g.
l. <u>Fourth: purification by random removal of randomly-noised synthetic data</u>. Create a *polluted synthetic seismogram*, S*, by adding random noise to S;
m. create fifty test-series $C_i$, by first randomly removing 1%, then 2%...50% from the polluted synthetic data S*, where $i$ = 1…50, and $C_i \nrightarrow C_{i+1}$;
n. examine the high-frequency GV VS response to such a highly-uncorrelated (most realistic) purification, by GV-spectrally analyzing each of the $C_i$ (and if needed for resolution purposes, create also a $C_j$' series of $j$ sub-samples, say $j$ = 2.1, 2.2…2.9; or, expand the $B_i$ series to $i$ = 51…100 while keeping the original resolution);
o. do the same for SFT as in step j;
p. create a diagram in the same manner as in step k.
q. <u>Fifth: test assessment</u>, in order to quantify the experiment, compare diagrams created in steps g, k and p; in order to quantify the experiment frequency-wise, compare period estimates, both for consistency (within the GV VS for varying degrees of purification; within the SFTS for varying degrees of purification) and method verification (the GV VS v. SFTS). If necessary, do this at mid-steps too.

Note that in the above, and in what follows, the complexity of the models is irrelevant in absolute terms, and what only matters is a relative gradual increase in model complexity.

**Purification by successive decimation of progressive-systematically noised model M1**

As outlined above, I create fifty test-series $A_i$, by successively decimating every other (evenly ordered) value of the M1 synthetic data, first 0.5%, then 1%...50%, from the first half of S, where $i = 1…50$, the resolution is 0.5%, and there is correlation $A_i \rightarrow A_{i+1}$. This makes the worst-case scenario (of total one-sided biasness) even for a semi-removal, i.e., removal of up to one quarter of total data; in other words, 50% of that one half of the data. It turned out that the GVSA responded to the successive removal in the above manner better than expected, so that by the test's end, i.e., before the total purification of one side of the dataset ended it did not become clear at which point during the successive purification the method actually started underperforming – here measured by the signal falling under the GVSA 99%-significance level. Therefore, the successive purification was continued even beyond one half of the data, and carried on throughout the other half of the dataset too. Accordingly, additional fifty series were created in the above manner, bringing the purification to the 50%-level of the entire dataset.

As seen on Fig.5, the high-frequency GV VS responded to the semi-removal by a drop in the signal magnitude to below the 99% significance-level. The drop occurred sharply at about 44% of the synthetic data removal, as done in an even-decimation fashion. Because the examined scenario is the least natural one of all, this experiment puts an upper limit on the method's insensitivity to the discontinuity of a geophysical trace. The sharp ending (frequency-wise) of the response change, as seen on Fig.5, is due to the type of a synthetic, and is not a bad news given that this response change also grows in magnitude as the signal band shrinks, maintaining some of the quality information contents, and thus performing as a "snapshot-carrier wave"; note that, while performing excellently frequency-wise, the secondary signal (the purification-insensitive part of the signal, so in the context of this testing it can be regarded as noise) never reaches the 99%-significance level while staying safely above the 67%-significance level. This means that the secondary signal's significance level is subjective, and could be improved to the 99%-significance level probably for the most part as well, just by enforcing i.e. removing the primary 99%-significant signal (the signal part sensitive to the purification). Fig.5 is also a demonstration as to why FSA of natural-data records which often are 90%-incomplete cannot be trusted.

It appears that purification in the above manner mostly affects i.e. subdues the longer-periodic signal but has little adverse effect as well as a more enhancing effect in higher frequencies, Fig.3. As it can be seen from Fig.3 the total, 50%-evenly-decimation of the synthetic record, results in a GV VS that is identical to that of the complete synthetic dataset, Fig.4. This a complete agreement (magnitude-wise, but for all practical purposes frequency-wise as well) with the theoretical should-be case, makes a perfect test-score while demonstrating the main advantage of a least-squares approach to estimating periodicities in relatively large complete dense datasets: up to ~44% of such data may be safely thrown out if pollution or/and low reliability of data are suspected, in the perfect case scenario here represented by the M1 model. Note that the 44% should be split onto a number of windows covering the record – equally to all windows if they are of a same step, or proportionately per different windows' step sizes as relative to the total data size or to some other pre-adopted criteria. Given such an absolute accuracy of the least-squares fit for a long and entirely sieved synthetic dataset, Fig.4, gains could perhaps be made *via* purification of real (polluted or not) geophysical data as well, by expelling perhaps up to a few tens of percent of raw measurements per window.

Note that Figs.3 and 5 serve also as a modeling mould for minor-to-moderate pollution (in this testing, by noise I mean non-fundamental harmonics). Fig.6 shows that no synthetic signal remained beneath the 67%-significance level, so that all the information contents that were meant to be picked up were picked up, and simple enforcement or removal of the primary component would boost the secondary component to above the 99% significance. One should keep in mind that, while enabling the signal-noise separation by time-domain filtering, the maximal (50%) evenly decimation also ends up doubling the 99%-significance level; but this is spurious in the context of physical signal (i.e., not due to physics but rather to a data manipulation alone), hence the dashing out the significance-level lines starting at the 45% purification point and on, Fig.6. Obviously, using the GV VS beyond the ~44% purification makes no sense then.

The fact that the GV VS of the long complete geophysical dataset and the GV VS of the same dataset after it was 50%-evenly purified look identical magnitude-wise (and for the practical purpose of noise detection, frequency-wise as well) represents a potential benefit from the GVSA. Namely, this could be made useful in the following manner:

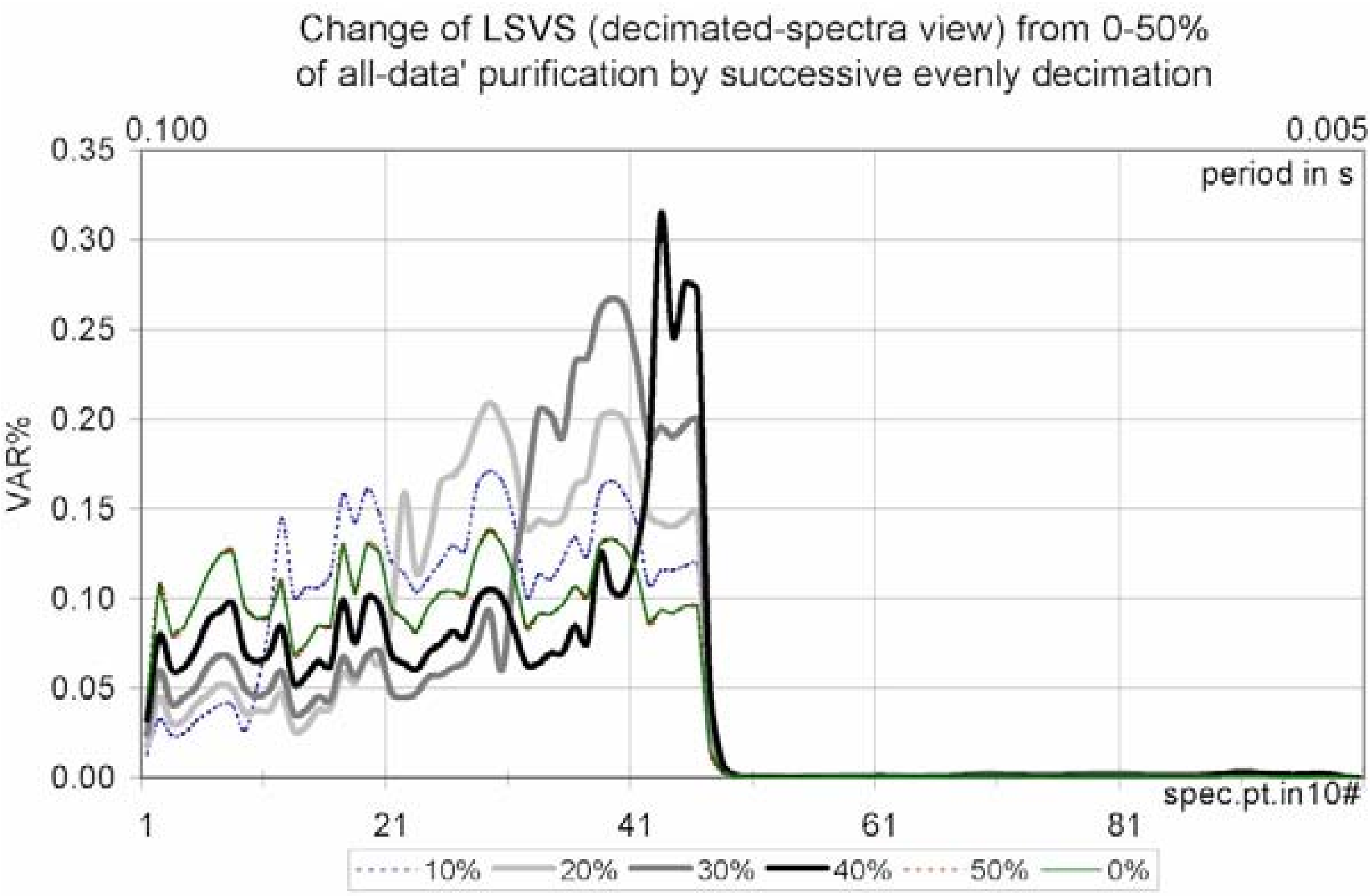

Figure 3. Note the GV VS for 0% (raw data) and 50% (record halved; span preserved) are identical. (Here the halved-record's spectrum was left intentionally unsmoothed so that a non-existent "distinction" between green and red lines could be discerned.)

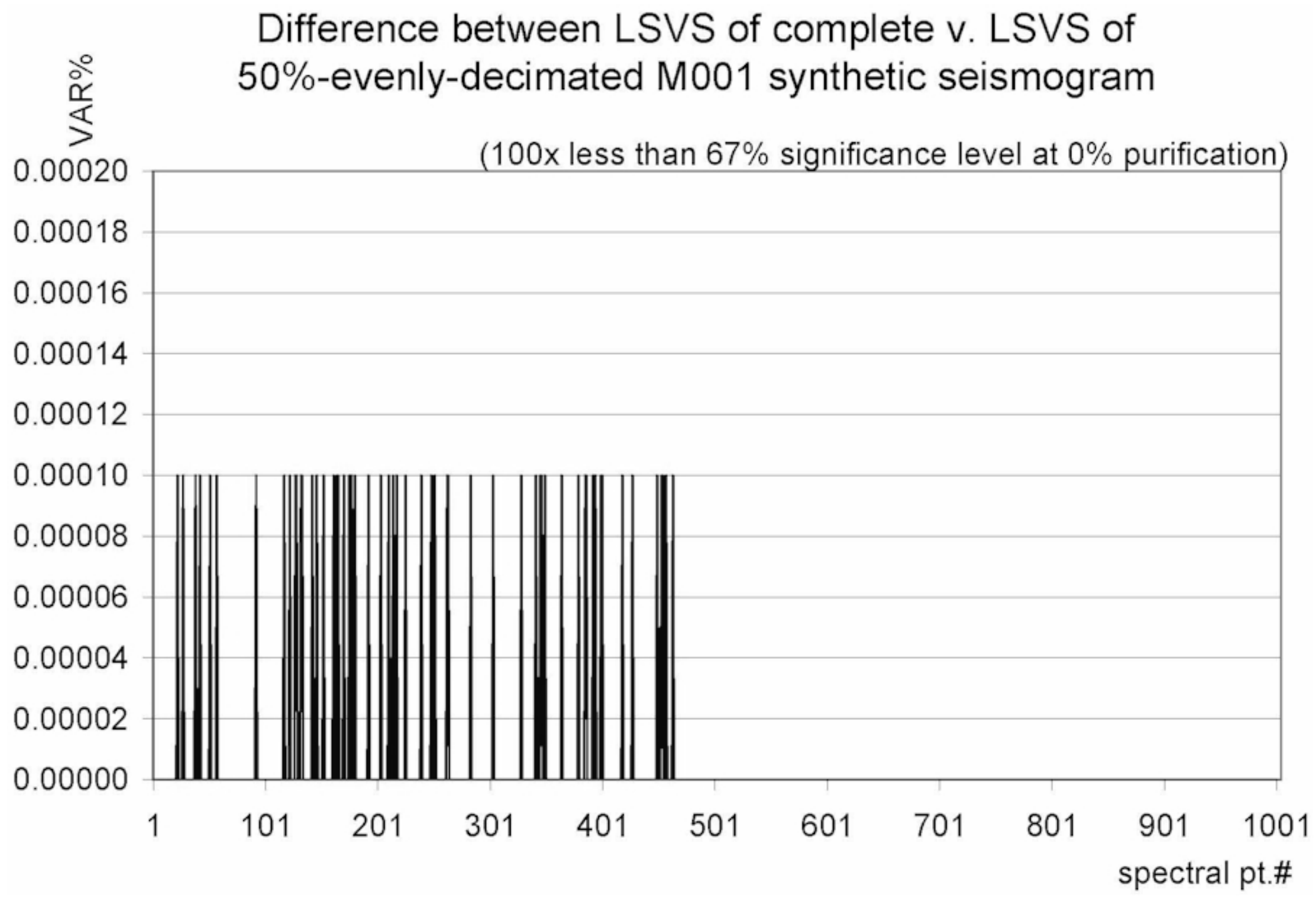

Figure 4.

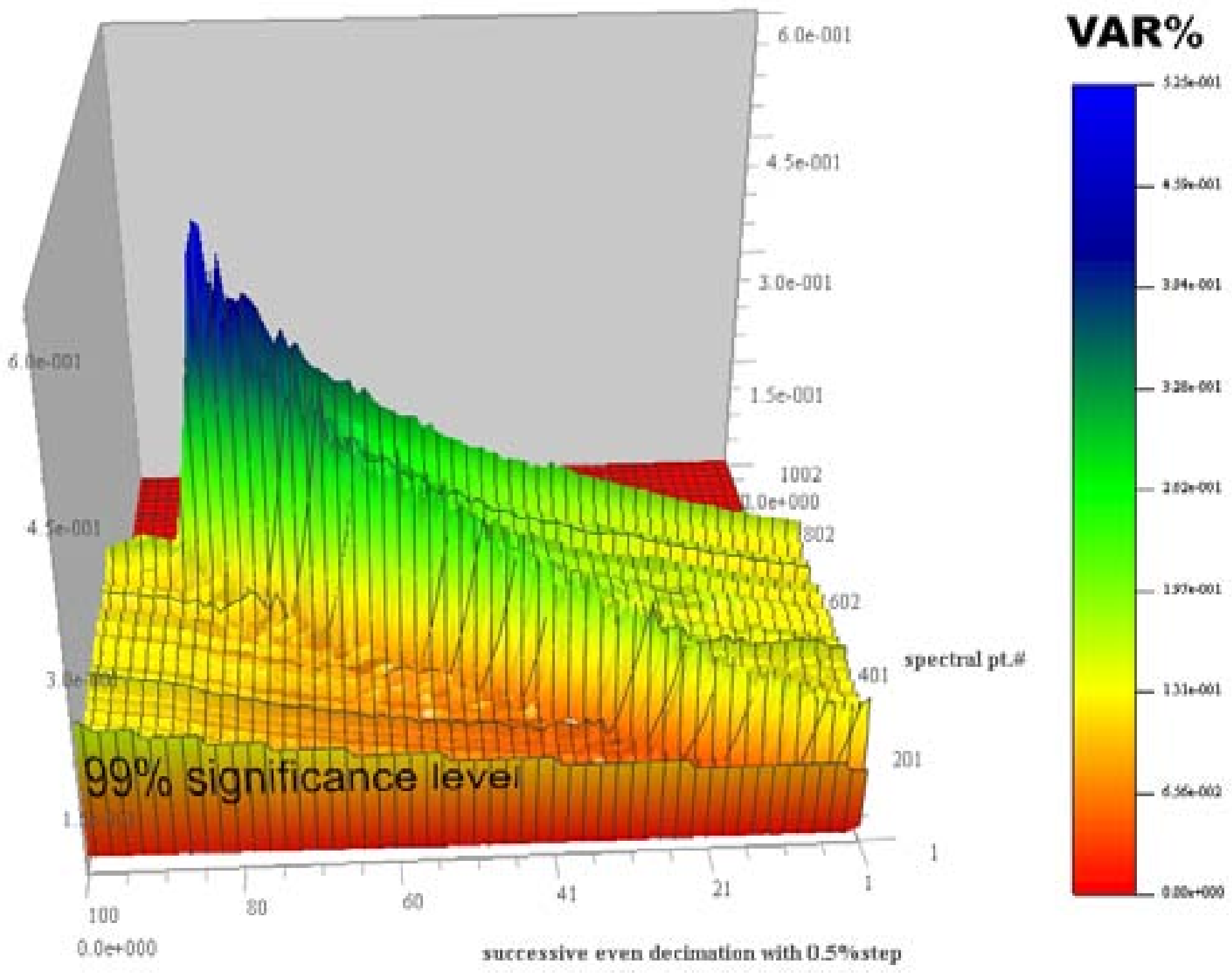

Figure 5. A demonstration as to why FSA of natural-data records which often are 90%-incomplete cannot be trusted. Side view; see Supplement for additional views.

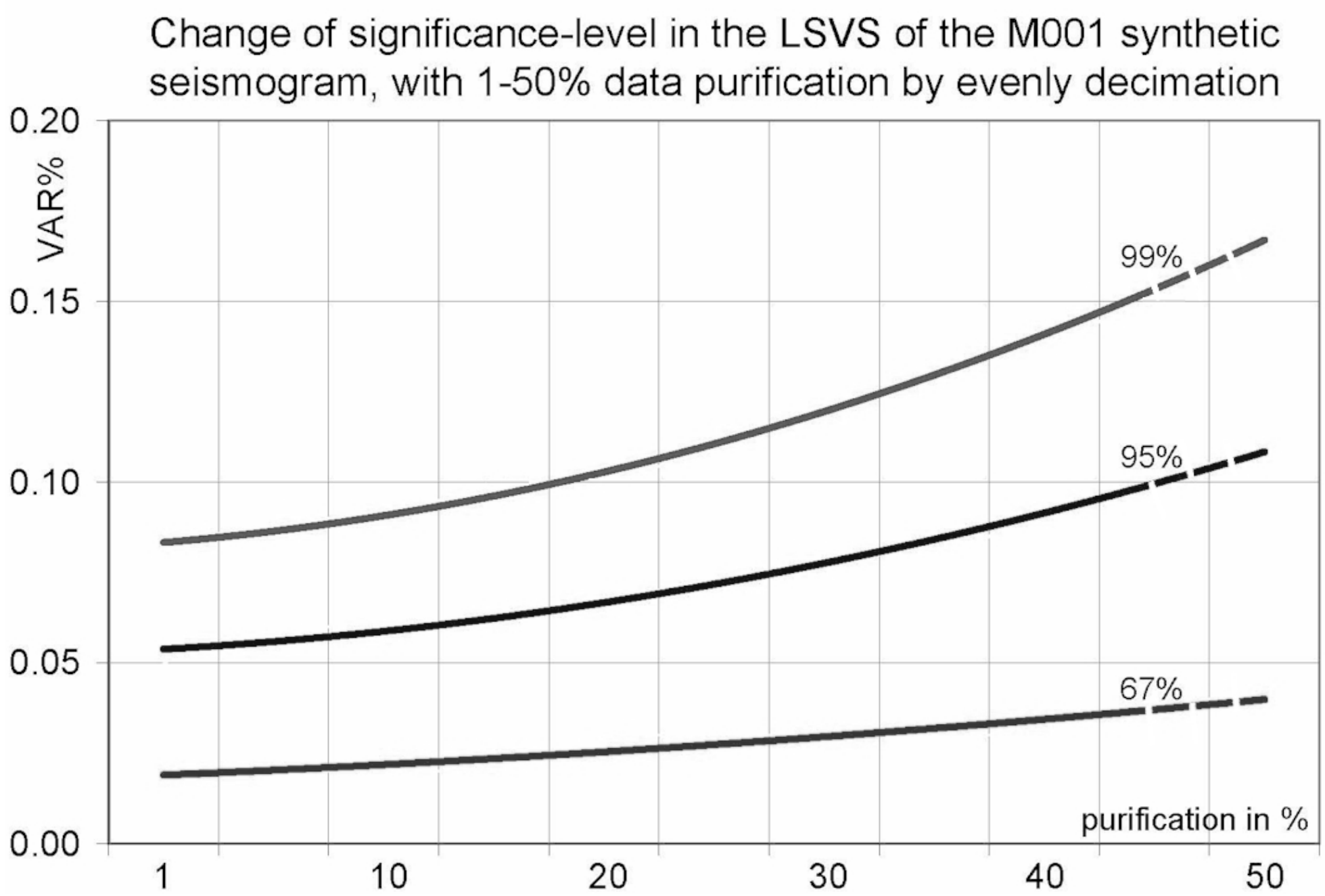

Figure 6. Change smoothed.

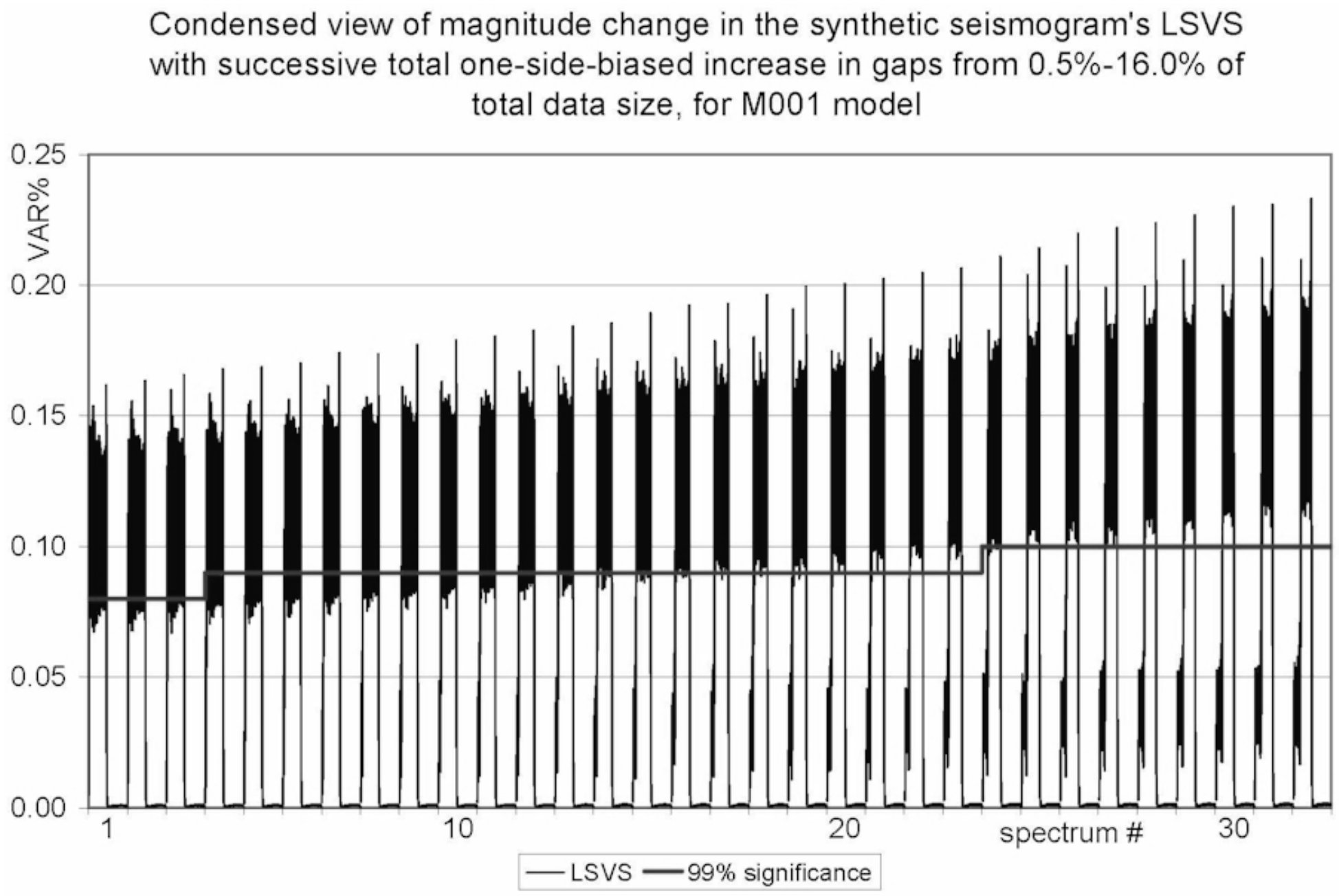

Figure 7.

- make note of the *evenly purification* and the *oddly purification* of real data;
- when noise is suspected, compute in the vicinity of the suspect the respective *evenly spectrum* (E GV VS) and the *oddly spectrum* (O GV VS) of that data subset, say a few % of the original data, or say the whole window or part(s) of it, that contains the suspect(s) of approximately known location;
- subtract the two spectra to obtain the *spectral differential*, $\Delta \boldsymbol{s}^{\mathrm{GV}}$, i.e., a *purified window*;
- in theory (so far limited to the synthetic M1 but more complex synthetics should produce not too different results either), the spectral differential should equal to zero-spectrum;
- in case of real data however, the variance-spectral differencing has a potential for producing a variance-spectrum of pure noise, i.e., such a spectrum in which the entire signal is actually geophysical noise in the classical sense;
- this *noise spectrum* enables in turn the identification of the systematic noise in real data by zooming-in (iterating/repeating the GVSA until all noise suspects are localized to the best of the GVSA ability – meaning the analytical "modelability");
- the effect of systematic such noise can be annulled in the GVSA itself, in the procedure called *enforcing*, which is to be performed in a repeated GVSA of the raw dataset, where one enforces some or all significant periodicities found in any one of the two parity spectra E GV VS or O GV VS, but not in both (where this applies to periodicity-matching only, but not necessarily to the magnitude-matching);
- to identify systematic-noise sources:
  i. Let us adopt the following notation
     1. GV VS → $\boldsymbol{s}^{\mathrm{GV}}$
     2. E GV VS → ${}^{e}\boldsymbol{s}^{\mathrm{GV}}$
     3. O GV VS → ${}^{o}\boldsymbol{s}^{\mathrm{GV}}$,

     and note that (for now: of limited success due to M1 model):

     $\boldsymbol{s}^{\mathrm{GV}} = {}^{e}\boldsymbol{s}^{\mathrm{GV}} = {}^{o}\boldsymbol{s}^{\mathrm{GV}}$; $\boldsymbol{s}^{\mathrm{GV}\prime} = \boldsymbol{s}^{\mathrm{GV}} + \boldsymbol{\varepsilon} : \ \boldsymbol{\varepsilon} \approx \boldsymbol{0} \ \wedge \ \varepsilon \in \Re$,

     $\Delta^{e}\boldsymbol{s}^{\mathrm{GV}} = \boldsymbol{s}^{\mathrm{GV}\prime} - {}^{e}\boldsymbol{s}^{\mathrm{GV}\prime}$,

     $\Delta^{o}\boldsymbol{s}^{\mathrm{GV}} = \boldsymbol{s}^{\mathrm{GV}\prime} - {}^{o}\boldsymbol{s}^{\mathrm{GV}\prime}$,

     so finally, the spectral differential is

     $\Delta \boldsymbol{s}^{\mathrm{GVS}} = \Delta^{e}\boldsymbol{s}^{\mathrm{GV}} - \Delta^{o}\boldsymbol{s}^{\mathrm{GV}} = {}^{o}\boldsymbol{s}^{\mathrm{GV}\prime} - {}^{e}\boldsymbol{s}^{\mathrm{GV}\prime}$;
  ii. Left- and right-handedness are interchangeable in the above, because "negative variance-spectrum" is nonsense, so, if, say

     ${}^{e}\boldsymbol{s}^{\mathrm{GV}} - {}^{o}\boldsymbol{s}^{\mathrm{GV}} > 0$,

  most of noise sources are located in the evenly spectrum, or, inversely, if the difference is negative, most of noise sources are to be found in the oddly spectrum;
- to identify largest singled polluters (blunders, outliers, etc.): usually a single run is sufficient – namely, apply the evenly or oddly procedure for 50% of the entire dataset without need for previous knowledge on the suspects' location. Say, take the E GV VS from the above inequality, and if not finding any significant differences amongst E GV VS and GV VS, consider the E GV VS continuous. If necessary, re-partition the E GV VS in the second run so as to create the E E

GV VS and the O E GV VS, and keep on zooming always the total-parity-wise (50% of the full subset), till major singled polluters are identified;
- other (arithmetic, logical, etc.) operations over the E GV VS and O GV VS could also be physically justifiable, which is worth examining further also.

But in order for the above concept to actually work on real data ("be real"), one first has to make sure that the above-detected feature (of identity between the original and evenly variance-spectra) was not perhaps a special case as due to the synthetic function(s) choice, in which case the whole scenario should be abandoned altogether, and the identity amongst the GV VS of complete v. GV VS of evenly (or oddly) data could not be regarded as true any more, i.e. as trivial and thereby useful by definition. To that effect I computed the O GV VS of the M1 model, and checked whether it too would have restored the original variance-spectrum, just as the E GV VS does.

As it turned out, the spectral differentials for the M1 model, amongst the oddly, evenly and raw spectra, all reproduce similar spectra of negligible magnitude, Fig.8. However, an interesting observation appears frequency-wise: the (OVS-EVS) and the (VS-EVS) seem to coincide better, meaning that the OVS and the VS were in this case apparently more similar in-between than the EVS and the VS were. Also, over the total of 1000 spectral points (the spectral resolution used throughout this first test), there were quite more (OVS-EVS) differentials than those of the other two possible kinds: 93 of the (OVS-EVS) v. 56 of the (VS-EVS) v. 37 of the (OVS-VS), which ads realism (note again that magnitude-matching and frequency-matching should be observed separately) to the claim that the GVSA is sensitive to the apparent "data parity", constantly for data manipulations totaling up to ~44% of the total dataset size, as applicable to the synthetic model M1 at least. Hence, while being able to reproduce the original dataset exactly i.e. trivially magnitude-wise, the GV VS at the same time do manage to reflect a disparity in noise distribution in a dataset, frequency-wise. This means a potential for the GVSA to show the same type of quality in high frequencies as it has in low frequencies where it was shown to provide a simultaneous descriptor of field's relative dynamics and eigenfrequencies (Omerbashich, 2006b).

It is for the above reasons that the claim of *spectral parity of GV VS* for up to 44% of data removal, and the related discussion above, could be founded for real high-frequency data within a window, as well. Therefore it is justified to continue on to the next step – of checking whether the above claim is real for a randomly polluted and possibly blundered synthetic dataset as well. Here, blunders could be implanted as say, one blunder in an odd placeholder, then one blunder in both odd and even placeholders, and, finally, ten blunders all of which ought to be placed in either odd or even placeholders but not in both types of placeholders. Besides being out of the scope of this paper, such a testing for blunders (outliers) is somewhat trivial too, and it has also been already solved in practice in better ways, so I leave it untested.

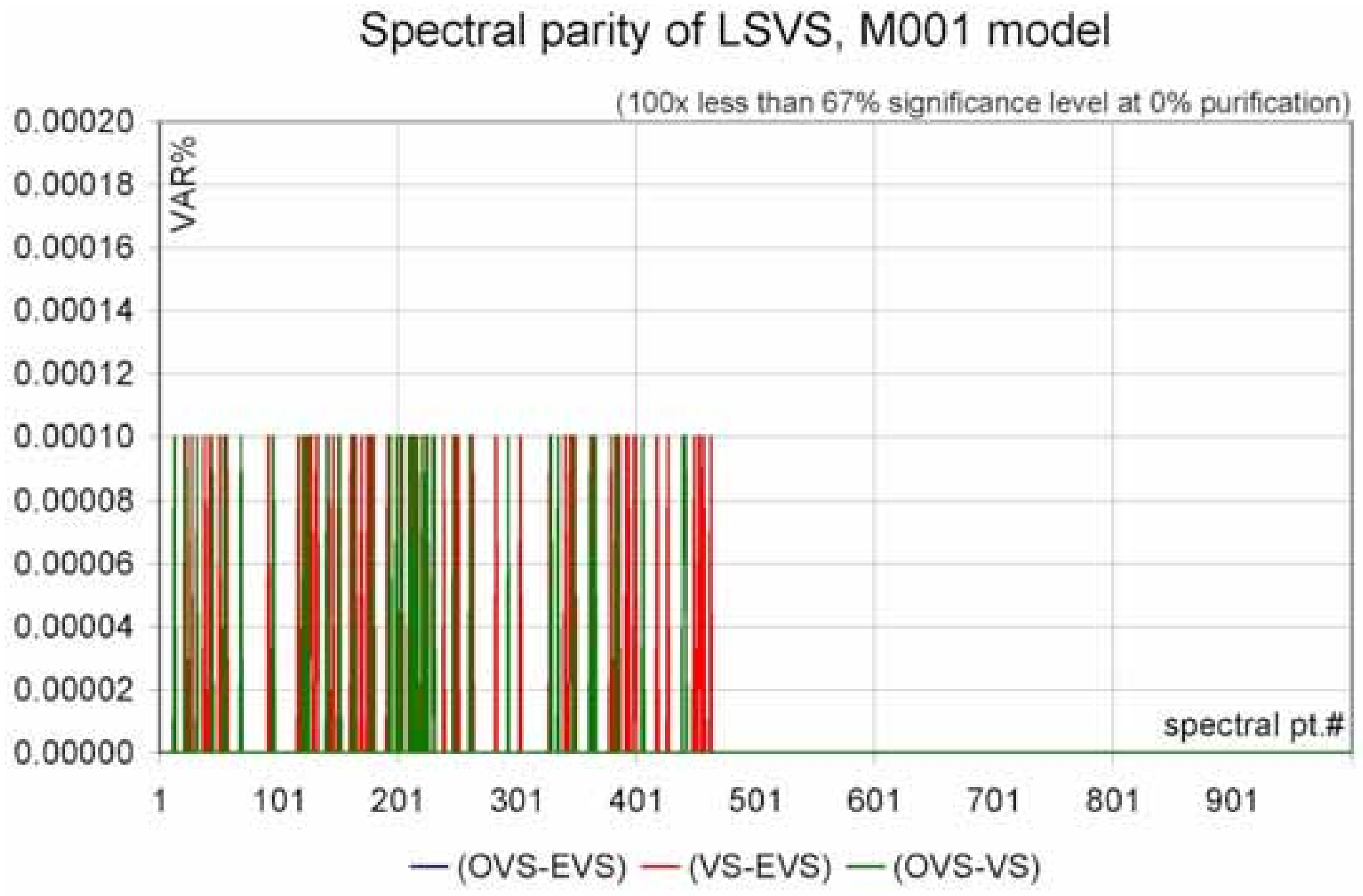

Figure 8. Note that (OVS-EVS), blue line, coincides at times with (VS-EVS), red line, and at times with (OVS-VS), green line. This indicates that the parity phenomenon, as used herein, is not a result of the choice of the synthetic function(s), and that it could be real, i.e., provided considerably larger presence of random data contents as well.

But why should the spectral parity be real in the physical world, and for subsets of size less than 50% of the original data size? To answer that, consider the fact that what works as a detriment for the Fourier and other transforms of long complete datasets (the case of exploration geophysics), works only to the benefit of the GVSA. More specifically: while short-span, relatively large complete records make the cumulative effect of FSA drawbacks only magnify with increase in length, density or discontinuity/purification of a raw dataset, all of those three noteworthy increases actually make a least-squares estimation of such a raw dataset fit even tighter. That's the main reason why a GV VS of a subset from such data should, compared to the corresponding O GV VS or/and E GV VS, enable identification of noise. (To that effect, note also a confirmation of the method's high sensitivity; see Supplement, Fig.7-10 and Figs.17-20.) For example, least-squares adjustments of astronomical and geodetic measurements proved to be a most sensitive tool for blunder detection, and this actually was one of the main reasons why C.F. Gauss was so interested in the least squares norm – to reduce astrolabe and tachymeter observations along great circles as well as in classical national triangulation networks. Not to mention that, in all physical sciences without exception, only raw data should be taken and used at their face value (Wrinch and Jeffreys, 1921, Omerbashich, 2006a), while everything else means manipulations that inevitably end up tempering with data (and, unavoidably, their spectra too), as well as involve intermediary procedures tasked with patching up numerous drawbacks of the FSA and its derivative-methods.

For the above mentioned three reasons as well as for others, when computing spectra the window enlargement (to lower the project workload) should pan out better in the GVSA than in the SFT. But the window shortening (to increase the project resolution) should also end up at least of the same ability in the GVSA after indiscriminate removal of data within the window, resulting in the individual window's number of data points either remaining the same or lowering, while keeping only the time-span apparently increase or decrease. Creating the GV VS (or spectrograms for visualization purposes) would then consist of finding a tradeoff between these two end-approaches, where at least one, here the former (the enlarging) must be firmly anchored i.e. entirely method-justified, and the other, here the latter (the shrinking), can also rely on some "art" considerations as usual for spectral analyses. At the same time, one should be careful not to allow for such "art" considerations to get out of hands, something which could be described as:

$$\Delta t \rightarrow \Delta t_{\text{enlarged}} \Rightarrow n_{\text{enlarged}} > n$$
$$\wedge$$
$$\Delta t \rightarrow \Delta t_{\text{shortened}} \Rightarrow n_{\text{shortened}} \leq n,$$

where $n$ is the number of data points within a window that are processed, and $\Delta t$ is the individual window's time span.

To round up the study of the M1 model, I also looked more closely into the frequency-behavior of the response. But even taking a quick look at how the GVSA-computed statistical fidelity (admittedly a somewhat obscure concept) behaves for strongest peaks detected, it seems that in reality one would have to give up some 10% or so in the method's ability to sustain reliably the time-domain decimation-per-window as done in this first test. Even more reality provided – that ability would probably shrink some more – for another 10% or so in the decimation degree, so as to stabilize safely at some 20-25% of the total-data decimation removal. This can be tested on a randomly-polluted synthetic record with both systematic and random removal applied; later on, more complexity can be added to the synthetic, or one can move on to real data as well – even as an intermediary step. But the 44% for spectral magnitude responsiveness, plus the spectral parity (claimed frequency-wise only) for up to ~44% data removal, seem like promising results as they give plenty of room for further investigations. This is also the best outcome one could hope for at this stage of GVSA test for exploration geophysics.

**Strong purification, models M1, M1A, M1B**

Another way to apply the purification as described for per-window usage stems from the above-mentioned fact that the GVSA norming works better the more data one has at disposal. Then one could also resort to a *strong purification*, by performing the spectral zooming in the following manner.

After say the evenly, $^{e}l$ (or oddly, $^{o}l$), decimated raw record has been created, regard that new record as a whole, and create its purified record, $^{ee}l$ (or $^{oo}l$). Now, repeat the procedure so as to create the $^{eee}l$ (or $^{ooo}l$). Do it yet again, so as to obtain the $^{eeee}l$ (or $^{oooo}l$). And, once again, and obtain $^{eeeee}l$ (or $^{ooooo}l$). Now, the latter three subsets represent the 87.500%-, the 93.750%-, and the 96.875%-purified raw complete dataset, *l*, for the purpose of GVSA, respectively. Note that a general stipulation by Omerbashich (2006a), on preserving as many raw data as possible when computing GV VS, was in reference to long-periodic analyses of sparsely populated (highly gapped) records with up to 90% or more sparseness, and not in reference to high-frequency analyses of relatively large, dense and complete records such as the case here. Then a strong purification as described in the above should result in GV VS that preserve the strongest signal in *l* the most, while at the same time virtually eradicating (by ignoring) the rest of the periodic information. Note that the least-squares fitting in the GVSA (as a measure of variance contribution) first applied to the strongest signal in the information, and then to other significant contents too.

The strong purification in the above manner will be tested on randomly polluted synthetic trace as well. Optional tests within this testing of the strong purification would be to examine possible benefits from the permutated-indices (*e*; *o*) purification, as well as from applying the strong purification on individual windows, though clearly the drastic reduction in data size in that case (as windows possess considerably less information than the whole record) would probably blight the result at least somewhat or even severely.

But before testing the above claims (about the strong purification) on a randomly polluted synthetic trace, one has to test those claims first on the systematically-polluted synthetic trace M1 and variations. This was done and the results of that testing are shown on Fig.9: it can be seen that not only the strong purification preserves successfully (where success means 3-of-4 or 4-of-4 estimate matches) period estimates for the strongest information (my signal for synthetic-testing purposes), but it also successfully (at the 8-of-9 level) resolves any estimate ambiguities seen here as spectral slopes' inflections in the left-handed half of the spectrum (10-100Hz), depicted by dashed lines on Fig.9. In addition, the highest-frequency spectral contents, here from 100-200 Hz, seen in the right-handed half of Fig.9 (mostly gray and blue grid), get resolved consistently at least at the 2-of-4 level in 19 out of 22 short periods, of which a total of 5 estimates were at the 3-of-4 level. This overall success rate (out of 40 periods, 34 or 85% were picked up at least twice, and 19, or 48% of those, thrice) is entirely due to the removal of most of the data and therefore model-independent; it is an alternative to the enforcing (detrending) of the strongest signal alone and then possibly re-running of the spectral analyses. In the same sense, one should not confuse this 100-200 Hz periods' excellent resolve (there were no high-frequency contents in the original spectrum at all!) as a spurious "boost" but as an operation that forces the least-squares fit to smooth out the information overload, and is another hint that the strong purification concept is real. The virtually uniform consistency in longer-periods' v. shorter-periods' working ability of the strong purification is due to different strength of the signal in those two parts of the band. This strength-proportionate treatment of data by the GVSA that got preserved before and after strong purification is yet another indication that the purification is real, and that it works for the entire band.

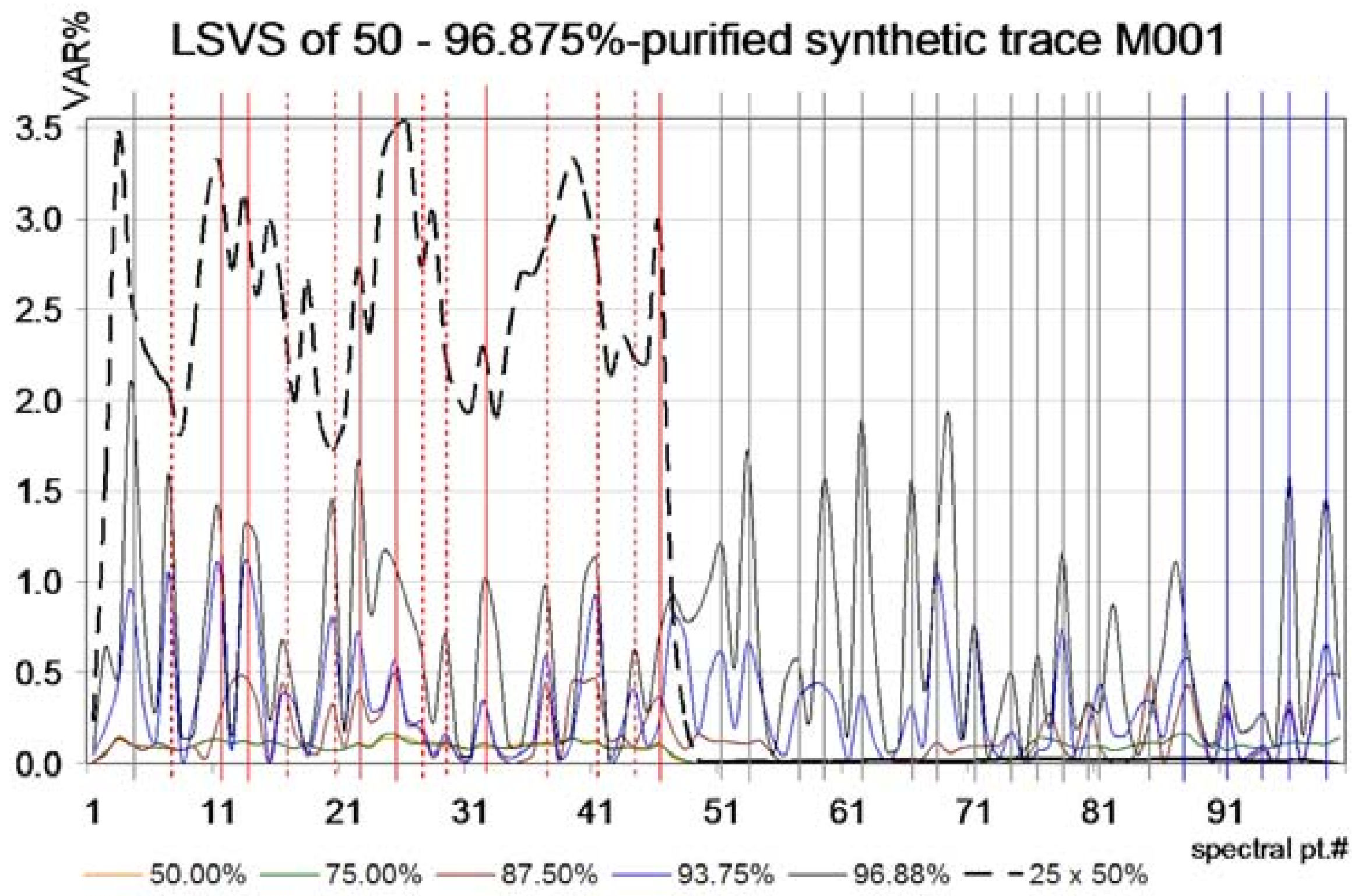

Figure 9. Comparison of the GV VS of 50-96.875%-purified synthetic trace M1. Purification was done in the evenly-only spectral-zooming fashion. Red vertical grid represents successful (where success means 3-of-4 or 4-of-4 estimate matches) period estimates when going from one zoom level to another for 10-100Hz (lower frequencies); blue grid is the same for 100-200Hz (higher frequencies); gray grid marks semi-success, i.e. 2-of-4 estimate matches. Dashed vertical grid represents instances when strong purification actually resolved a spectral estimate's ambiguity as seen in the reference (dashed) spectrum. The 50%-purified (reference-) spectrum (orange) was scaled up 25 times for clarity (dashed spectrum). Spectral resolution: 100 pt.

In order to utilize the strong purification as described above (if proved real on yet more complex synthetics), the one thing one has to bear in mind is how the significance-level changes with purification. For this, note the computed change of the 67%, 95% and 99% significance-levels with increase in the strong purification, Fig.10. As that figure shows, the significance level exactly doubles with the strong purification as defined in the above. This means that the change can be modeled exactly, here as the geometric progression, thus making the here stipulated strong purification, if real, uniformly credulous too. Random noise in the record should somewhat hamper the geometric progression model.

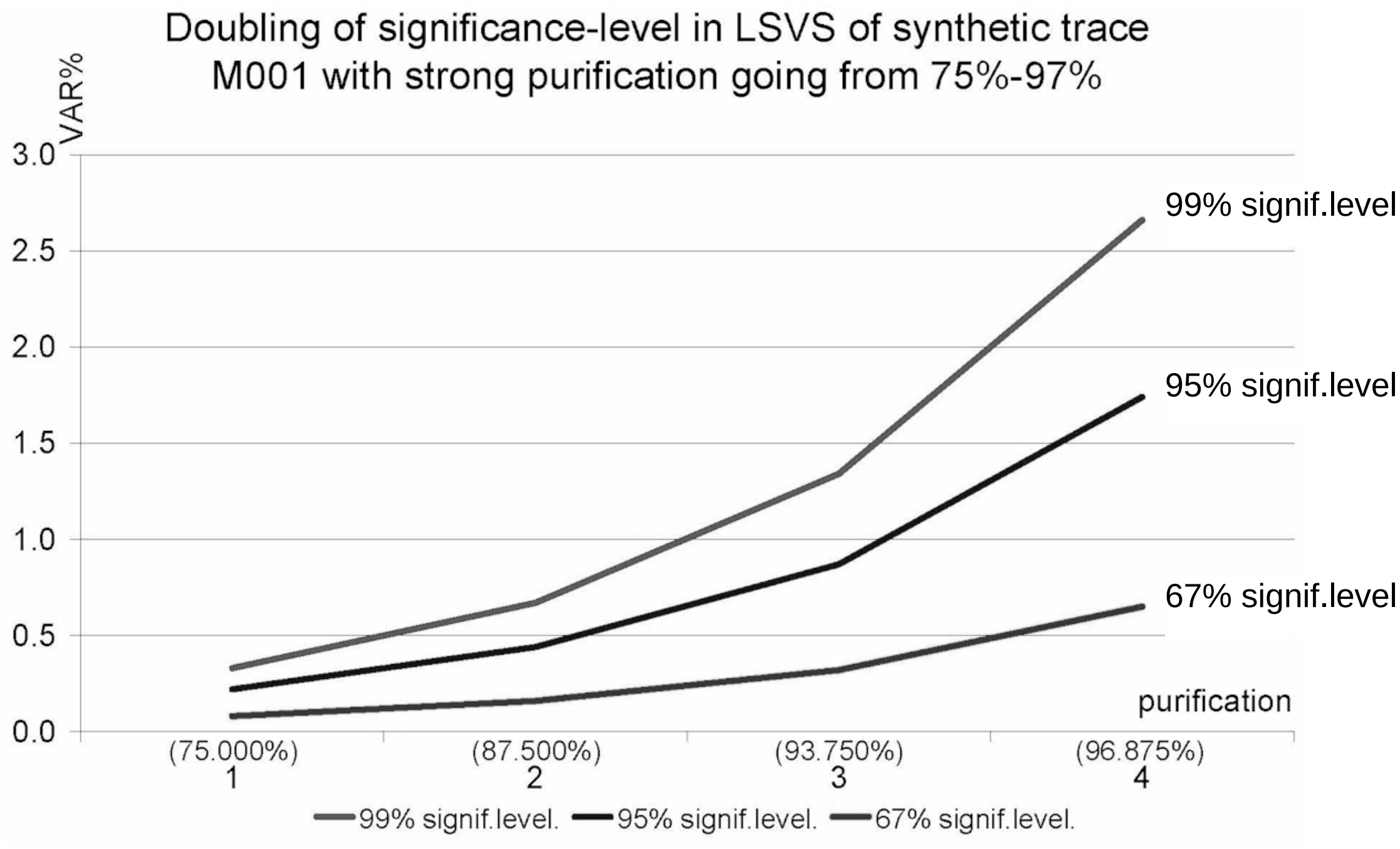

Figure 10.

Practically, the fact that the 10-200Hz spectra from the Fourier transformation return numerous components of the systematic information contents is regarded as *information overload* in the context of the GVSA strong purification. Overestimated solutions also mean smoothing in the sense of classical information theory; see Fig.11. In our case, the least-squares estimation is turned into a smoother *via* stationary time-domain filtering, i.e. purification of data by decimation that significantly regresses at the geometrical-progression rate.

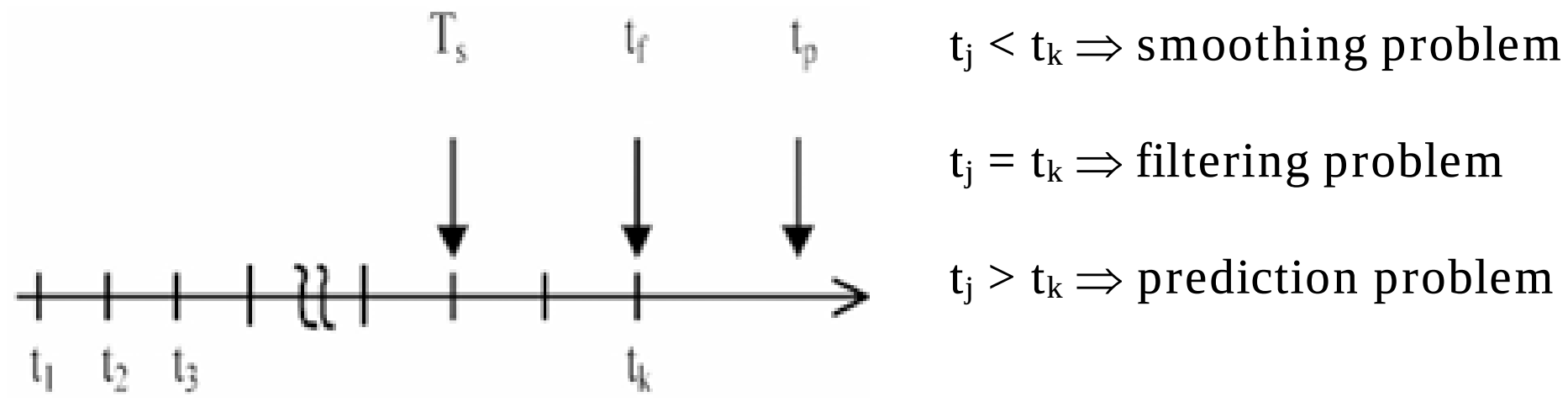

Figure 11. Three types of estimation problems (Omerbashich, 2002).

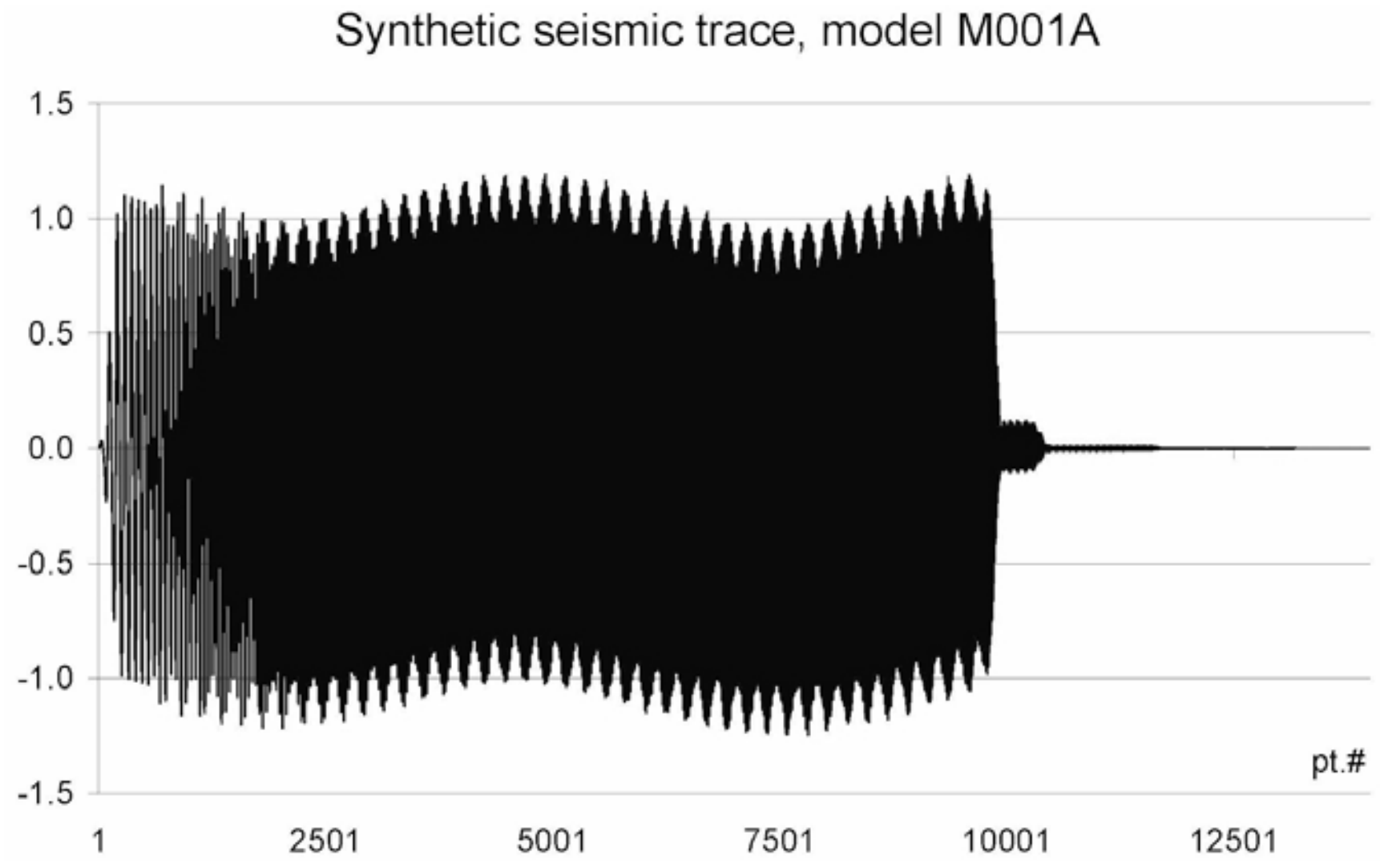

Figure 12. Matlab-created complete synthetic seismogram, no padding, model M1A (amplitudes: 1.0, 0.1, 0.01, 0.001, 0.0001; offsets: 1, 1.5, 2.8, 4.3, 5).

As already noted, the GVSA feeds on noise, so real data should actually enhance this blinding of the secondary contents at the benefit of the primary (strongest) contents. The presence of random noise in reality should enhance the GV VS, leading us to expect that the GVSA will act as an efficient stochastic resonance modulator. Thus the strong purification seems to be real, at least based on the above analyses of a synthetic trace polluted by systematic noise. Besides using a randomly polluted synthetic trace for testing the strong purification, models obtained by the synthetic trace being systematically polluted with 5, and then 20 harmonics, here designated as (models) M1A, Fig.12, and M1B, respectively, were used for testing, since these do model real-world situation closer while together they enable depiction of the GV VS response-change with change in (the degree of) complexity of a real-world situation. Finally, pending its passing of all the tests, the strong purification has potential to enable spectra computation from real data while responding mostly to the harmonic relatively strongest in terms of var%.

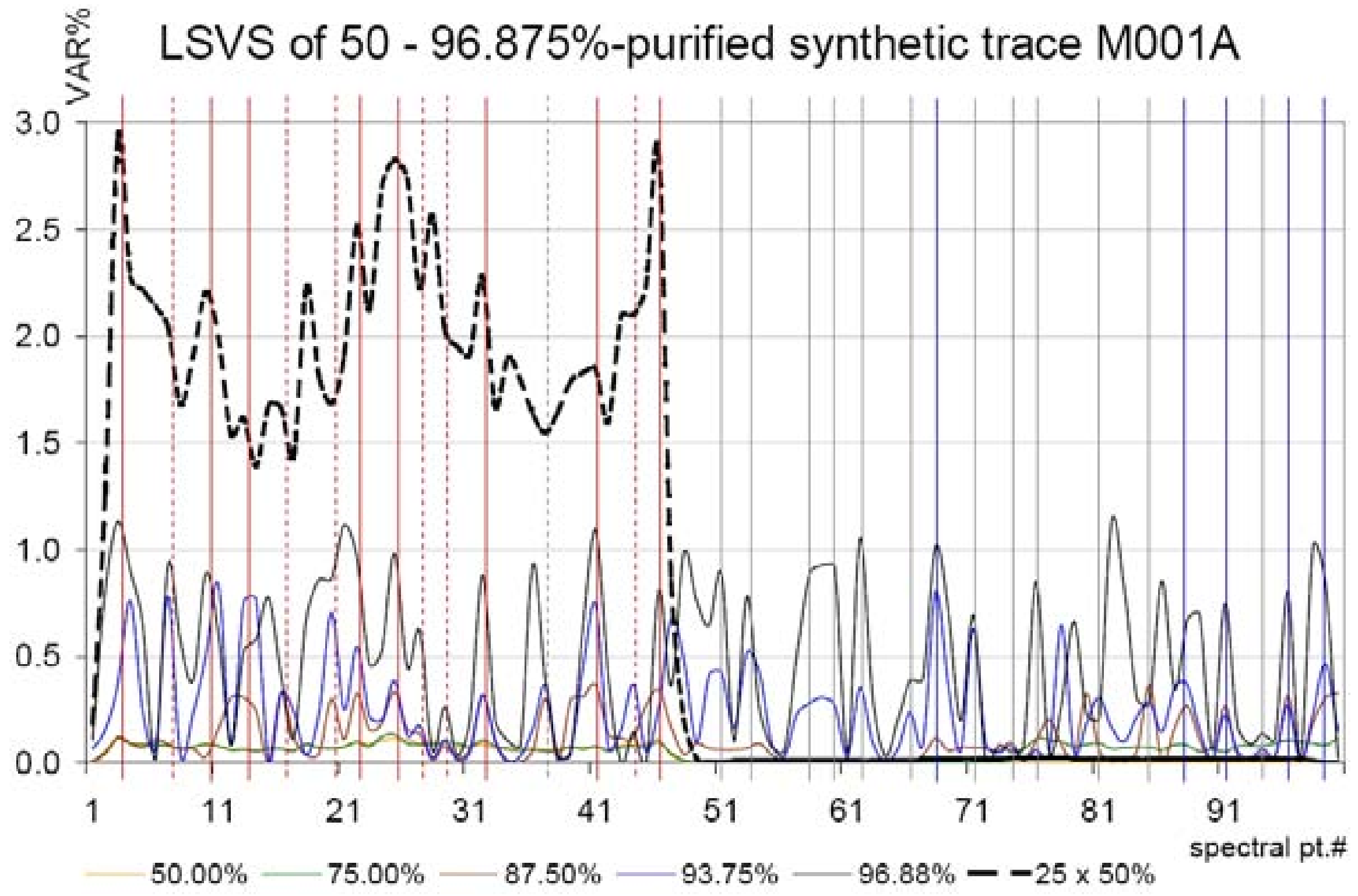

Figure 13. Comparison of the GV VS of 50.000-96.875%-purified synthetic trace M1A. Purification was done in the evenly-only spectral-zooming fashion. Red vertical grid represents successful (where success means the 3-of-4 to 4-of-4 estimate matching) period estimates when going from one zoom level to another for 10-100HZ (lower frequencies); blue grid is same for 100-200Hz (higher frequencies); gray grid marks half-success, i.e. 2-of-4 estimate matches. Dashed vertical grid represents instances when strong purification resolved a spectral estimate's ambiguity as seen in the reference (dashed) spectrum. The 50%-purified (reference-) spectrum (orange) was scaled up 25 times for clarity (dashed spectrum). Spectral resolution: 100 pt.

It can be seen from Fig.13 that not only does the strong purification preserve successfully (where success, again as on Fig.9, means either 3-of-4 or 4-of-4 matches per estimate) period estimates for the strongest information (my signal for synthetic-testing purposes), but it also, and as successfully as in the M1 model (7-of-10 here v. 8-of-9 for M1, Fig.9), resolves any estimate ambiguities (seen as spectral slopes' "inflections") in the left-handed half of the spectrum (frequencies up to 100Hz).

Note here that FSA spectra (FS) are non-discriminatory, i.e., all periods are treated equally in a FSA. In the GVSA however, periods are treated in *parental hierarchy*, i.e. the period which is strongest i.e. both physically (amplitude) and statistically (variance) most pronounced in a record (thus, "oldest"), occupies most of the spectral magnitude too – hence the straightforward significance-level regime in the GVSA. Of course, the M1B model is to be used as well, in order to make a final ruling on the applicability of the GVSA to the problems of exploration geophysics.

The highest-frequency spectral contents, here again from 100-200 Hz as depicted in the right-handed half of Fig.13 (mostly gray and blue grid), get resolved consistently in 17 out of 25 short periods at least at the 2-of-4 level, and of which periods 5 were at the 3-of-4 level. Again as in the M1 model, Fig.9, this represents a success rate hardly attainable by the original or the reference-spectrum, unless the enforcing of the strongest signal was performed first. The overall success rate now is 32 out of 48 periods (or 67%) picked up at least at the 2-of-4 level, of which 19 (or 40%) at the 3-of-4 level. The model M1B, with its 20 harmonic components (v. 2 harmonic components in model M1; v. 5 harmonic components in model M1A), should help determine the level of degradation of the data purification concept with increase in systematic noisiness of data.

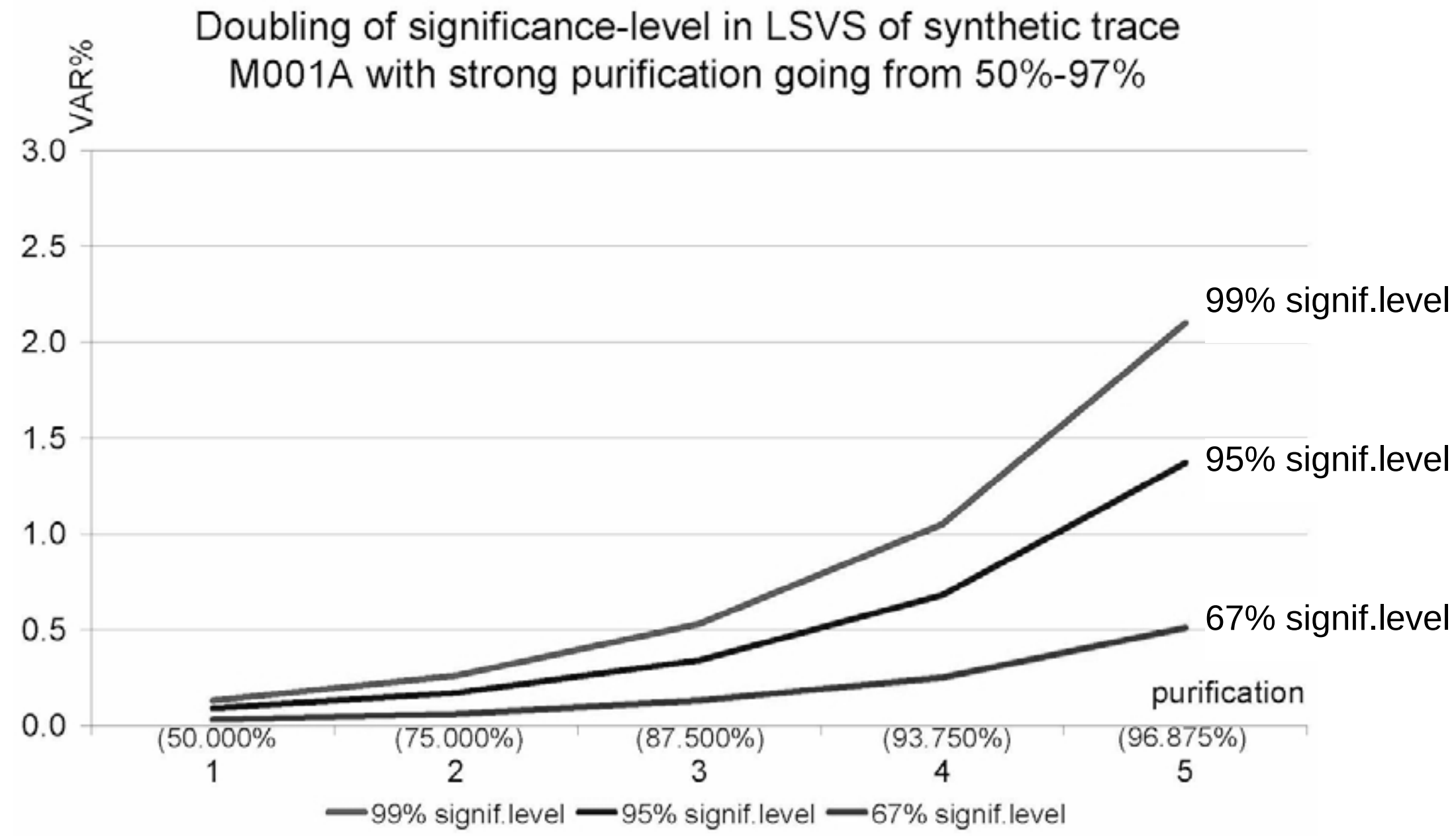

Figure 14.

As seen from Fig.14, the 99% significance level change can again be modeled exactly (also as the geometric progression), thus making the here claimed strong purification more uniformly credulous given that the same was found for the M1 model also; see Fig.10. Note that it can be expected for random noise in the record to somewhat obscure this geometric progression model.

The M1B model had 20 spikes with offsets: 1, 0.1, 0.01, 0.05, 0.03, 0.09, 0.016, 0.025, 0.045, 0.091, 0.06, 0.011, 0.017, 0.055, 0.07, 0.065, 0.089, 0.064, 0.04, 0.08. Those had respective amplitudes of: 1, 1.5, 2.8, 4.3, 5, 5.6, 6.2, 7.9, 8.5, 9.3, 10.4, 11.8, 12.7, 14.1, 15.6, 16.3, 17.8, 18.6, 19.8, 20. Since the time resolution was all the time kept at 1 ms, this increase in the number of peaks has resulted in an increase in the data density. This synthetic trace is depicted on Fig.15.

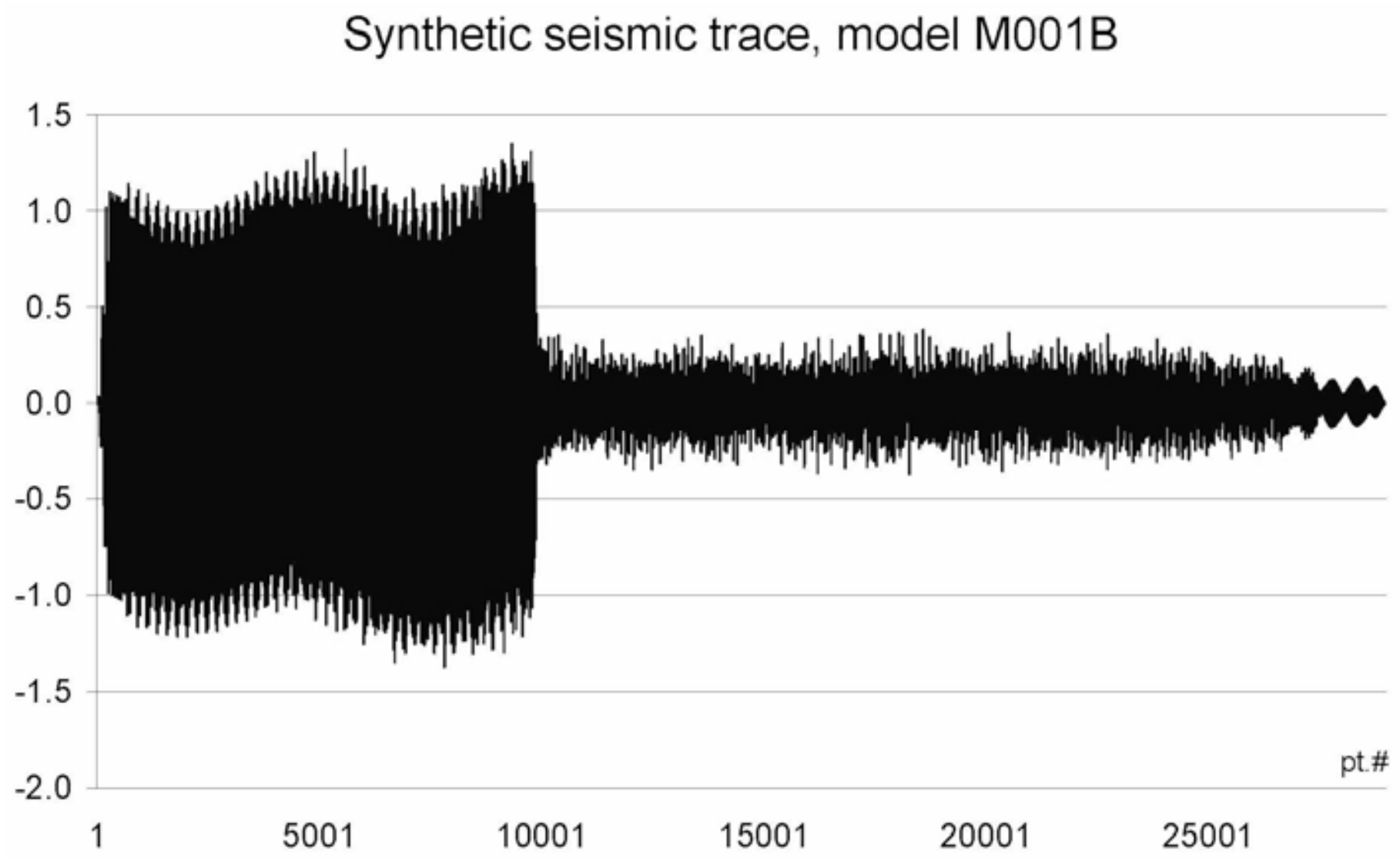

Figure 15. Matlab-created complete synthetic seismogram, no padding, model M1B.

The overall success rate now turns out to be: 36 out of 42 periods (or 86%) picked up at least at the 2-of-4 level, of which 21 (or 50%) at a rigorous 3-of-4 level, Fig.16. According to the less stringent criterion (i.e., looking at the 2-of-4-or-better success rate in period estimate matches as purification gets stronger), the M1B model performed best of the three (meaning: the more non-fundamental systematic noise, the better the solution), while according to the more stringent criterion (the 3-of-4-or-better success rate), its performance was in-between the M1 and M1A models.
Most importantly, in no case did the M1B model underperform when compared to either of the less realistic models. Absent some peculiarly coincidental relationships amongst the randomly selected amplitudes in the three models, this makes yet another hint that the strong purification is real, as well as that it improves with data density. Also, the M1B model, being the most complex of the three models used, did not affect the previously observed geometrical progression in the significance-level change as the strong purification intensified; see Fig.17 v. Fig.14 – note magnitude-wise scaling.

Note here also that there is no sense to analyze the GV VS-period-estimates' change with the change in strong purification, since the GVSA normally computes period estimates at always the same pre-specified frequencies, so situations like that shown on Fig.18, emerge.

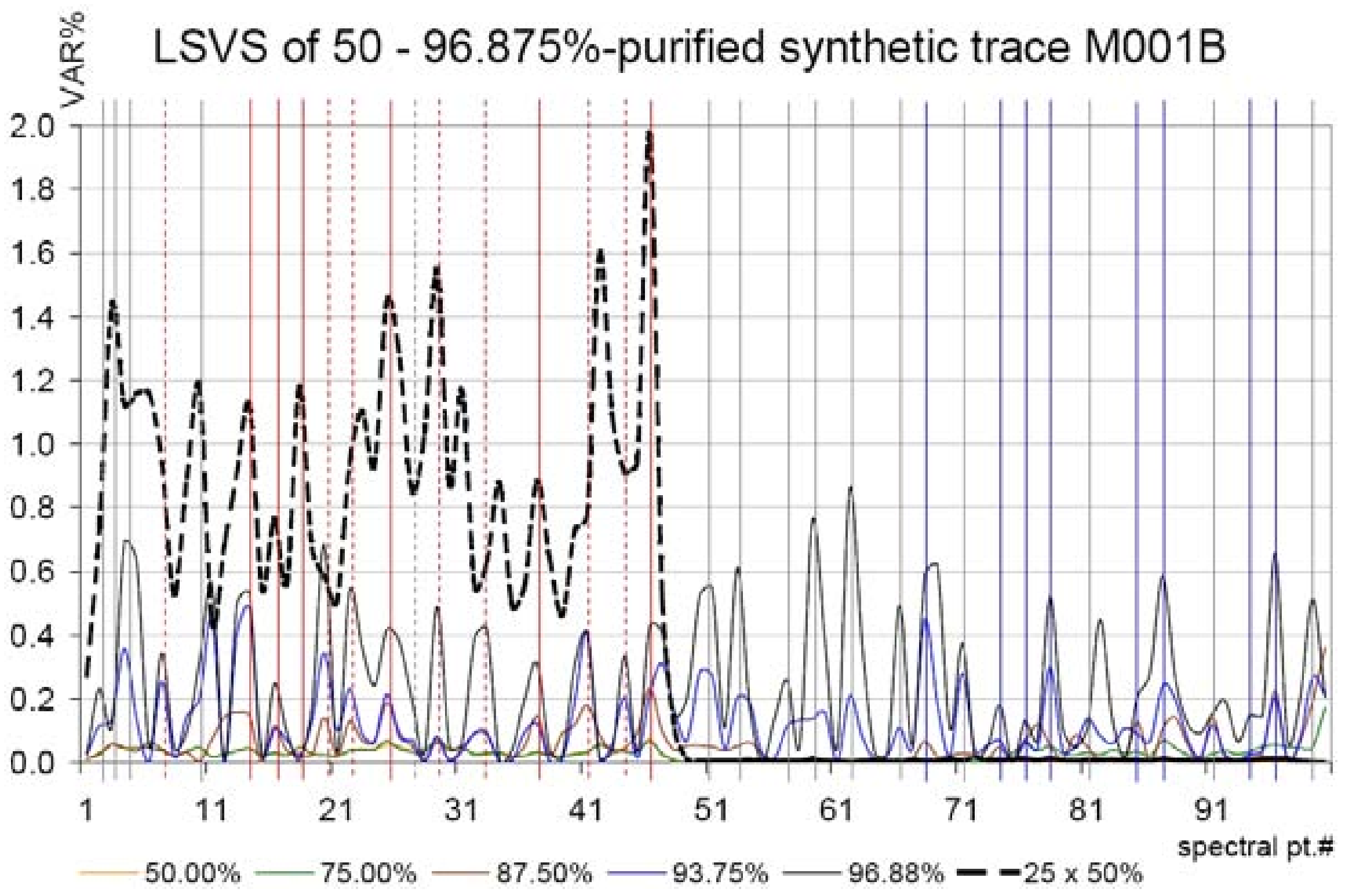

Figure 16.

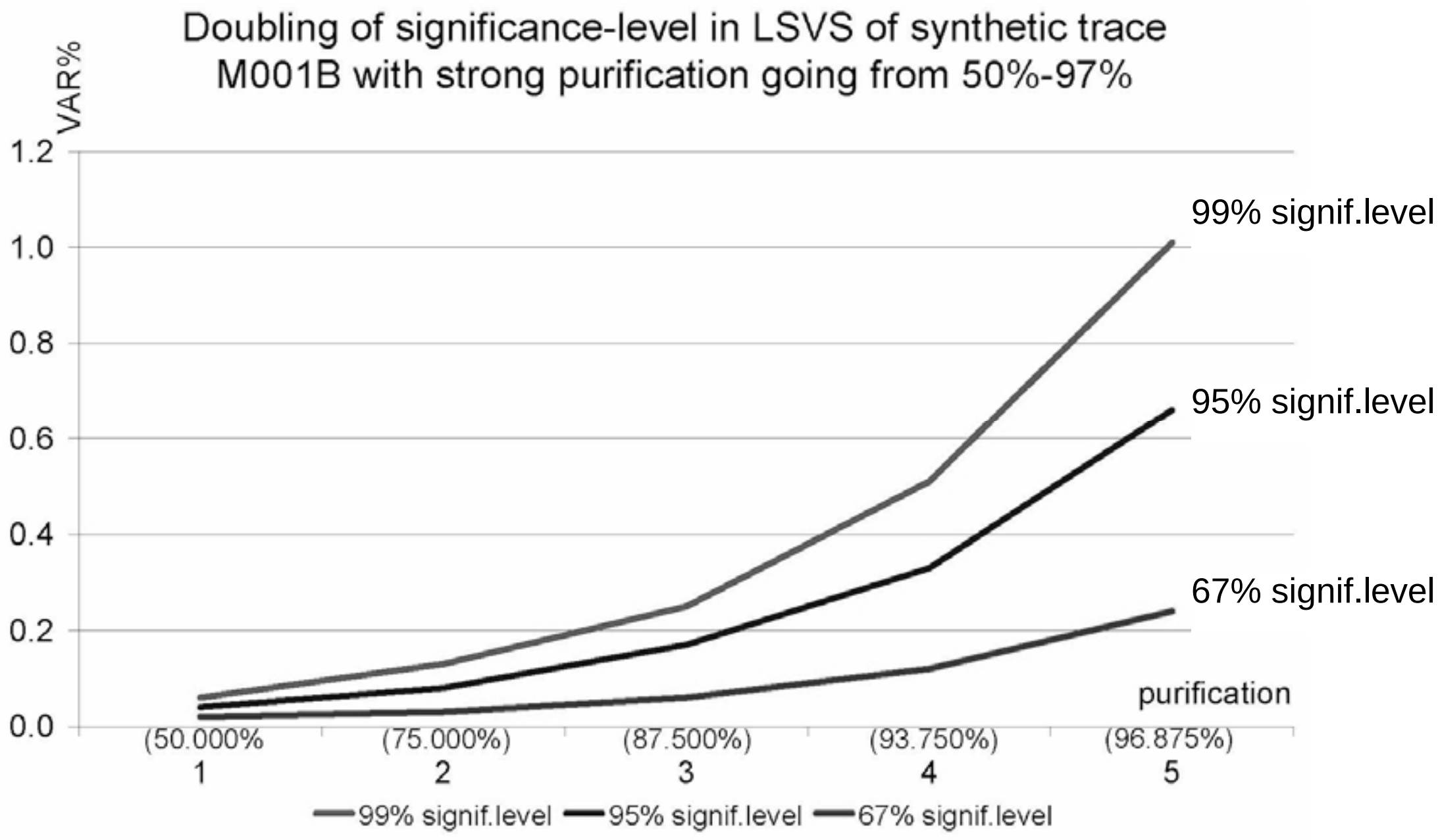

Figure 17.

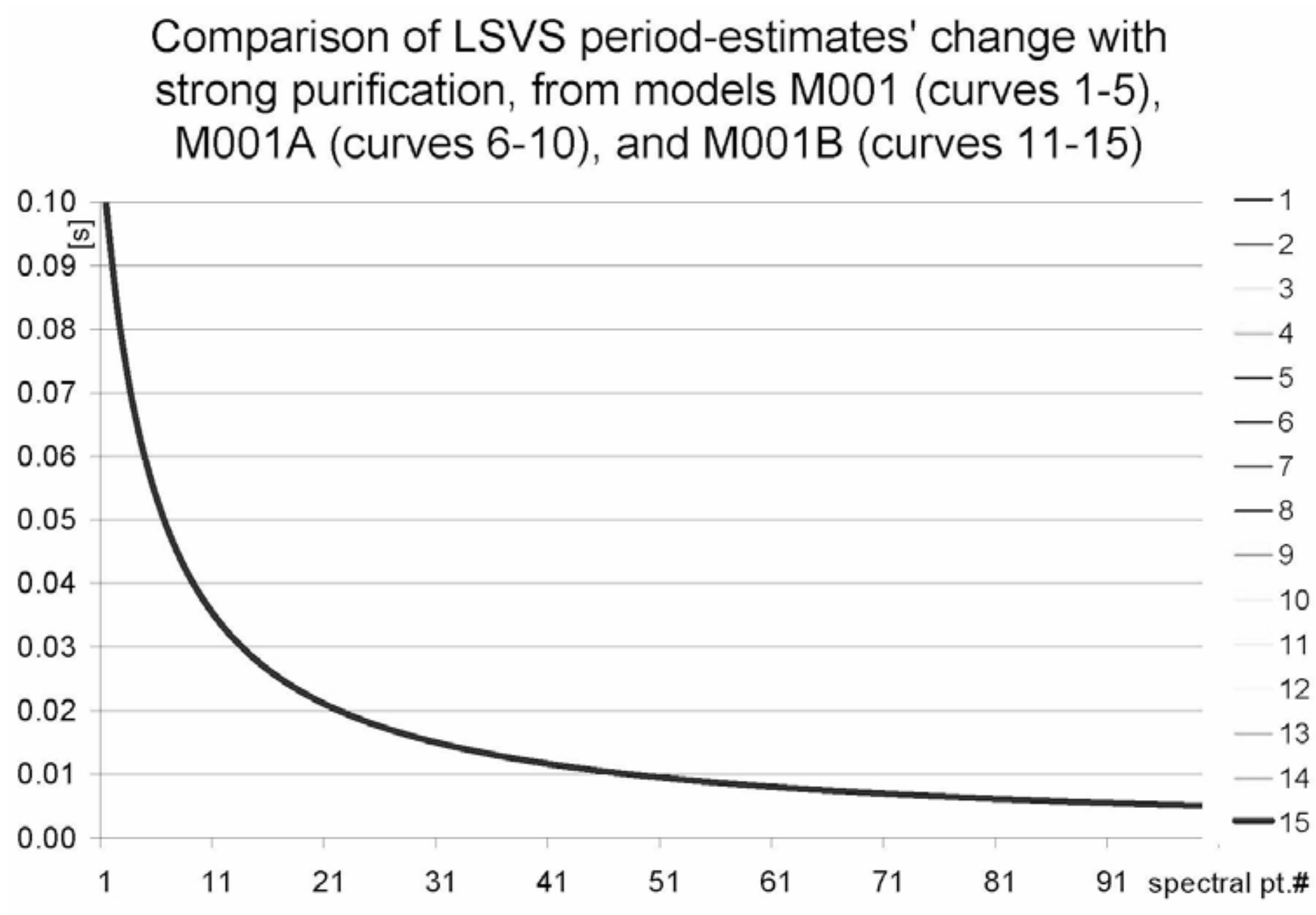

Figure 18. Change of GV VS period-estimates with increase in strong purification going from 50% (curves 1, 6, 11), to 75% (curves 2, 7, 12), to 87.5% (curves 3, 8, 13), to 93.75% (curves 4, 9, 14), to 96.875% (curves 5, 10, 15) of systematic synthetic trace. Note the coincidence in all period-estimate curves is due to the GVSA method.

**On fidelity in this testing**

Standard statistics textbooks describe the fidelity of a model as a parameter (or a set) describing how faithfully that model represents a real-world situation. Concepts most commonly related to fidelity are the accuracy, precision, resolution, and model validation. Fidelity comprises the methods, norms/metrics, and descriptions of models used to compare those models to their real-world referents. As applicable to this study, fidelity is the degree to which a model of the periodicity estimation under the least-squares norm reproduces the state and behavior of real-world cyclicity or the perception of real-world cyclicity, in a measurable manner. More generally, it is a measure of realism of a model, which is telling of that model's faithfulness.

Thus in cases opposite to our case, i.e., in largely incomplete physical datasets of very low data density, which, while being non-populous, still do reflect all sorts of data distributions or/and noises, the fidelity number qualifies a least-squares period estimate for a passing grade usually at 12.0 or above values (Omerbashich, 2006a). In my testing however, which is based on the M1 synthetic, regardless to distributions or random noise fidelity is rather small in absolute terms and is in the order of 1E-5. Importantly however, its relative consistency in describing the model's truthfulness successively from one degree of purification to another, meaning period-to-period as well as period-to-flat spectrum, remained preserved, Fig.19. So the GVSA passes this internal check, which is rather sufficient to hope that the GVSA is able to respond to the purification of real traces also.

This consistency is an indicator of the high accuracy of the GVSA too, and by extension (by virtue of scalability) also an indicator that the above-claimed spectral parity is real for subsets as well, meaning it could be useful to apply it on real trace- or sweep-data for accurate treatments of data, i.e. potential gains from the so-improved windowing process.

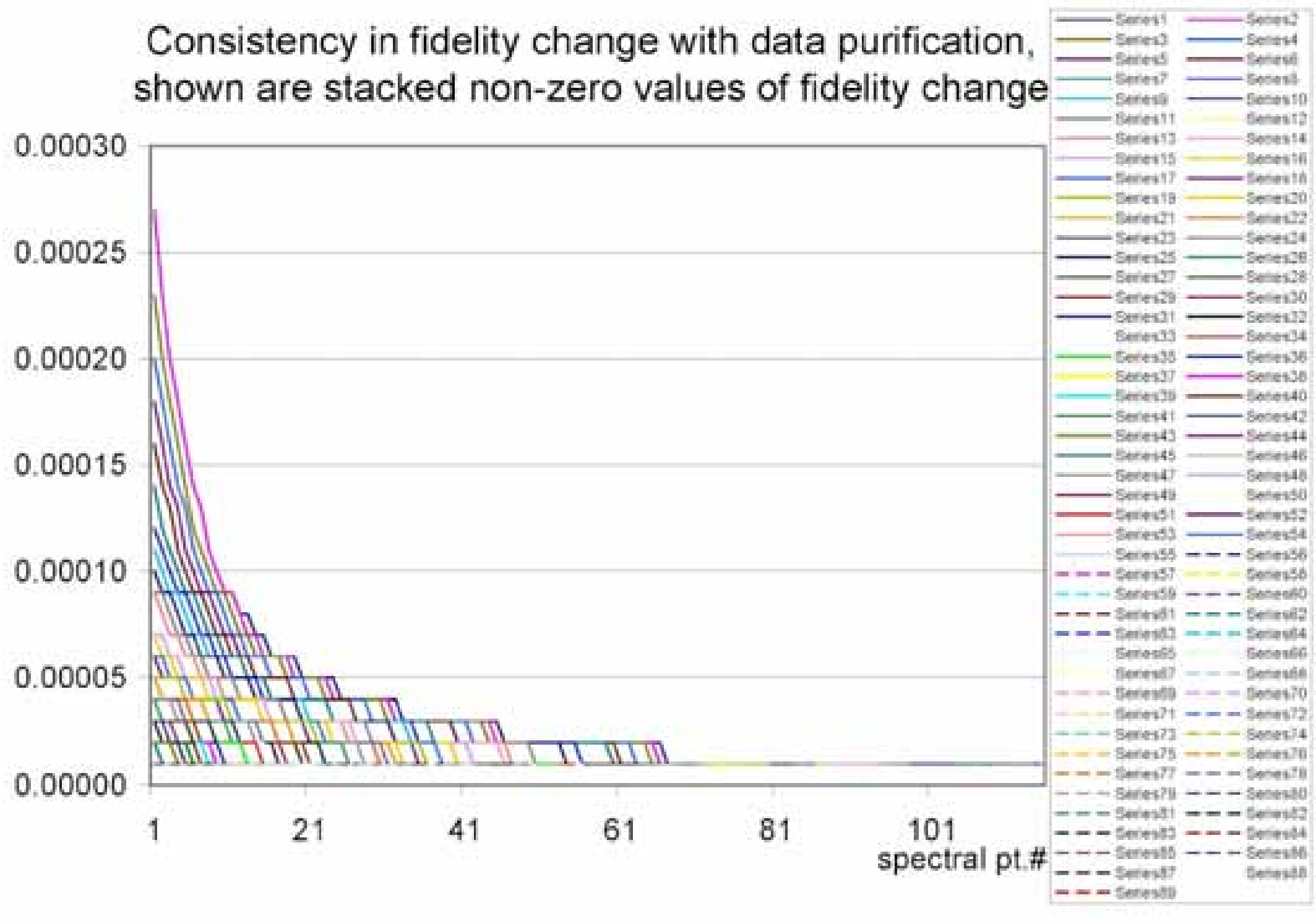

Figure 19.

The progressive systematic purification promises the following general advantages of the purification concept. Namely, the presence of classical noise and blunders can be significantly reduced in a geophysical record or even removed and thereby signal better resolved, *via*:

A.) <u>discriminate removal of less reliable/noisy data</u> for what no accurate knowledge of noise is necessary. Approximate locations of noise sources (i.e., the data of significantly lower, or of more suspicious quality or/and reliability) in the record would suffice. The rest is taken care of by systematic-noise enforcement, or by decimating the data that are in the vicinity of suspected noise-sources, say a few %of data before/after the suspect(s).

B.) <u>indiscriminate removal of substantial quantities of data of unknown characteristics</u> through evenly/oddly purification. The effect of classical systematic noise (most of secondary harmonics) could perhaps be alleviated by the strong purification of up to ~97% of the trace. Admittedly, while using the GVSA spectral parity, one has to account for the 99%-significance level's progressive increase with successive purification. However, this could be alleviated either by modeling (creating empirical expressions or tables) or by devising a calibration procedure each time data collection is started in the field or by a combination of the two approaches. For instance, as we saw above, the empirical model for spectral zooming in the strong purification of the systematically-only polluted synthetic trace is simple – it turned out to be the geometrical progression.

Note here that there was no need to create an $A_j$ series, as originally outlined under the point (e) of the test plan, because the selected 0.5% resolution proved to have been sufficient already after 25% of data were removed; see Fig.20.

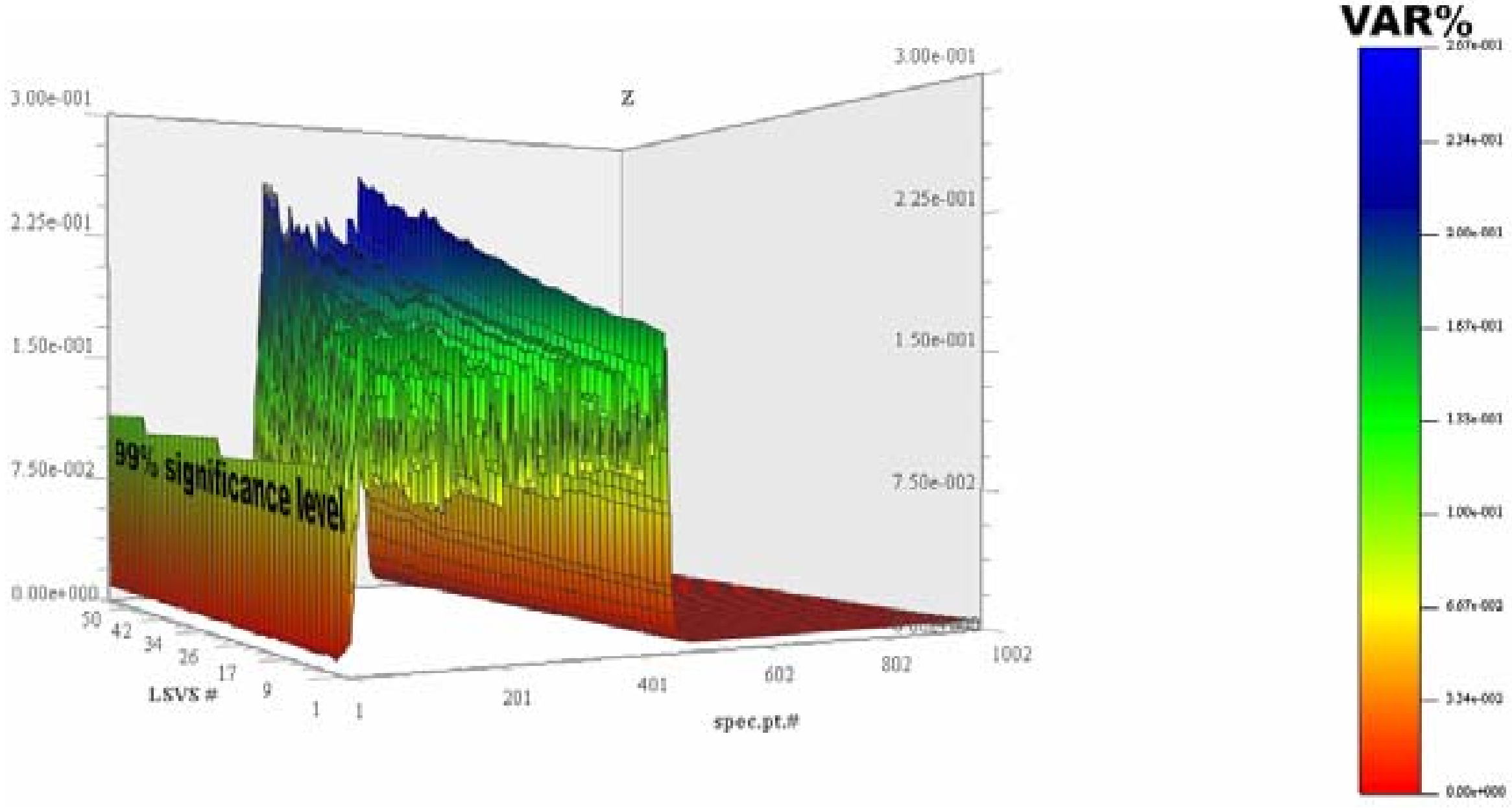

Figure 20.

Comparison between the SFT spectra of original synthetic seismogram as based on M1 model, v. the SFT spectra of 50%-purified M1, shows that the SFT underperforms when compared to the GVSA in spectral smoothing i.e. time-domain filtering by successive data purification. Namely, spectral parity, as established above for the GVSA of the same synthetic seismogram, although trivial in theory, is not observed as clearly for the SFT of the same synthetic seismogram, in the here examined stationary (single station) case; see Fig.21, so:

$$\boldsymbol{s}^{\mathrm{SFT}} \neq {}^{e}\boldsymbol{s}^{\mathrm{SFT}}$$

$$\boldsymbol{s}^{\mathrm{SFT}\prime} \approx 2\boldsymbol{s}^{\mathrm{SFT}} + \boldsymbol{\varepsilon}_1 : \ |\boldsymbol{\varepsilon}_1| >> \boldsymbol{0} \ \wedge \ \varepsilon \in \Re.$$

It can be further seen, Figs.25-26, that the SFT of the halved dataset also nearly halved in spectral magnitudes too, after a successive 50%-purification (in this case *via* total evenly decimation, but the outcome makes it uninteresting to check it also for the total oddly-decimation case). Note also that this doubling does not seem as predictable as the spectral parity from the GVSA, but rather appears to be erratic, unrevealing and likely with one or more harmonic noise signatures of its own. This however, unlike with the GVSA variance-spectra (see caption of Fig.8), can be expected to change from case to case, Fig.22. (Note that, as expected, the GVSA variance-spectra and SFT power-spectra perform about the same in absolute terms magnitude-wise, as the effect of the full i.e. 50%-decimation becomes measurable only at ~4 orders of magnitude below the original magnitudes in both types of spectra; see Fig.22 v. Fig.8.) Hence, the SFT power-spectra mean less utility in the purification setup than the GVSA variance-spectra do.

This stage of research suggests it would be useful to look into the inverse GV transformation, as well as to continue the testing with the random and the strong random purification of both raw and random-randomly polluted synthetic traces as well. (Here by random-randomly I mean that placeholders of the values polluted by white noise are selected at random as well.) The inverse is necessary in order to retrace the spectral contents back to the time domain where the most useful geophysical analyses normally take place. Alternatively, one may also want to re-formulate the problem of purification for the FSA methods as well, in which case the inverse Fourier transform would be readily available for time-recreating the data that were previously smoothed in the frequency domain using the GVSA. However, using the FSA on heavily (down to several tens of % of the raw record) purified data is likely to distort the fundamental harmonic too.

In what follows, randomly polluted synthetics as well as the random and the strong random purification are attempted. I note here that the 1-D inverse GVSA transform has been derived by Craymer (1998), who claimed that the same could also be done for multidimensional cases as well (*ibid.*). Note that only the 1-D inverse GV transform exists so far. However, the problem faced here with can actually be decomposed into 1D.

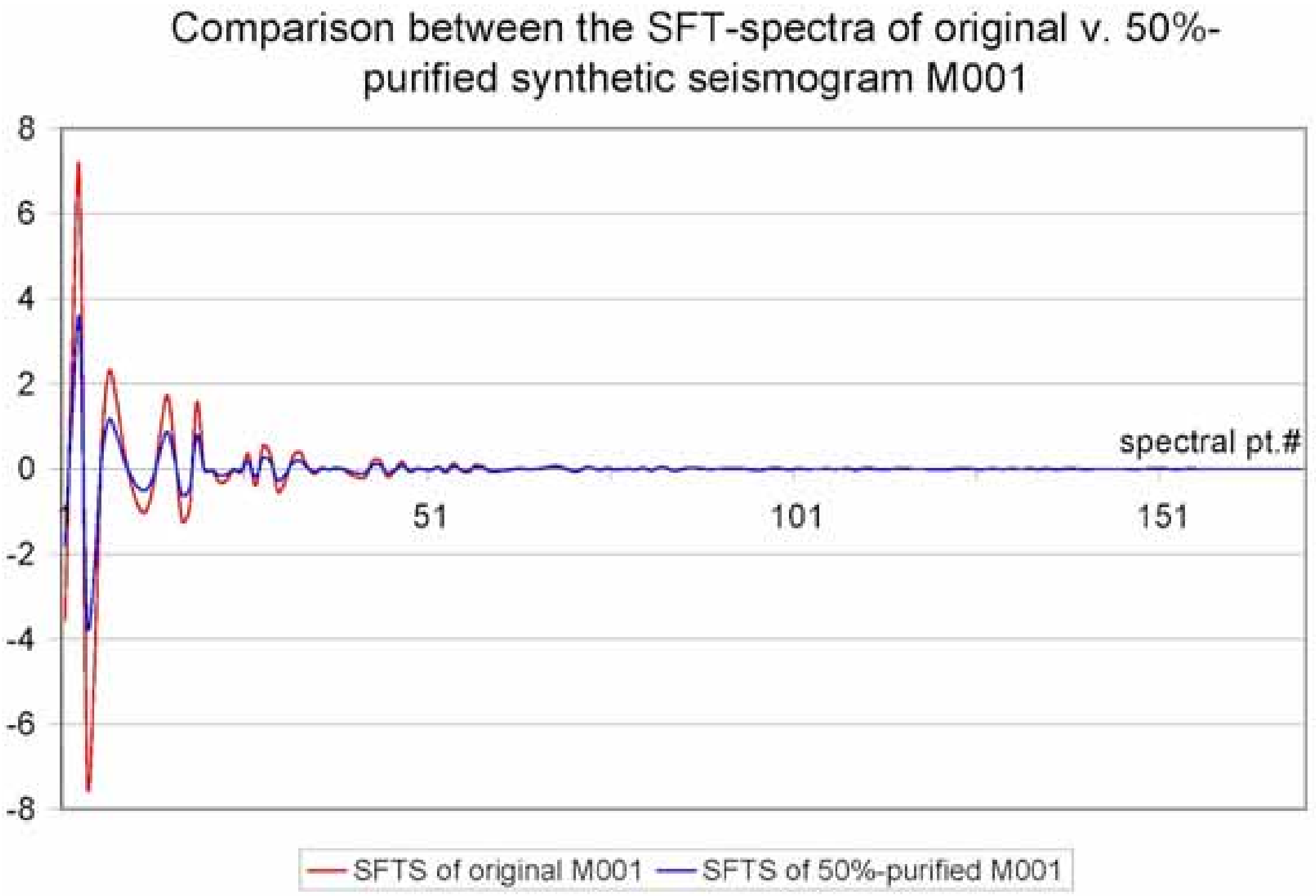

Figure 21.

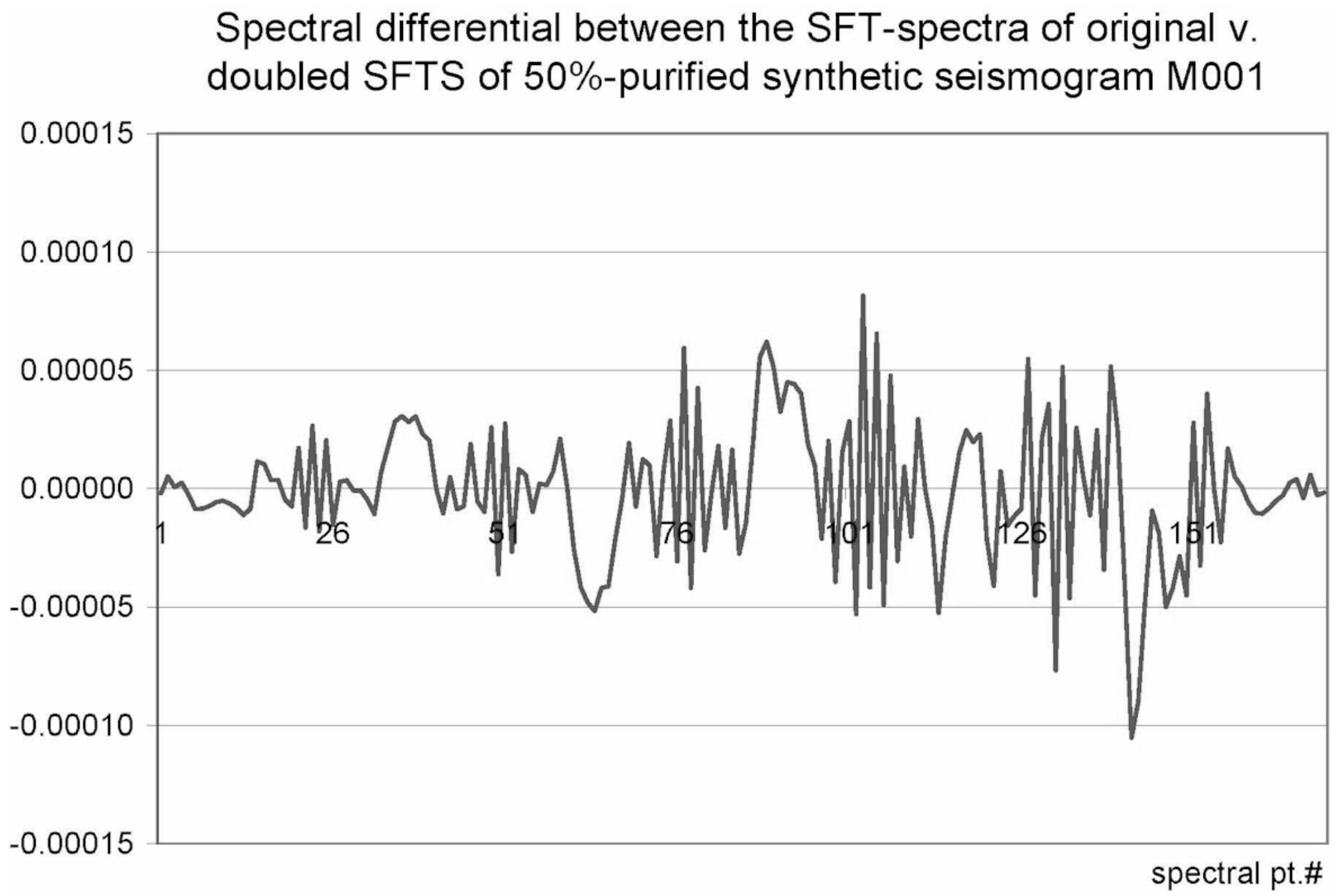

Figure 22.

## Random purification and random pollution

I next carry out the random purification of randomly polluted models M1, M1A and M1B, at various levels of pollution and for different degrees of random purification. All combinations of the complexity so imposed have resulted in 108 variance spectra computed and presented herein. The idea behind this final part of the test was a simple one: should any type of trending in quality with increase in complexity and/or drop in sheer information volume (and by extension: noise volume too) be observed amongst the computed GVSA variance-spectra, it would mean a demonstration that the purification of densely packed (overloaded in the sense of GVSA) records is a valid concept when used with the GVSA. This constitutes my starting physical hypothesis.

That this hypothesis works can be expected, in theory, because not only that such data are standard nowadays in exploration geophysics, but by the virtue of being such, they should also be significantly susceptible to the impact of *stochastic resonance* (improved signal resolvability by adding rather than subtracting noise), an elusive concept first discovered a few decades ago in climatology and subsequently in other disciplines as well. The commonsense validity of assuming such susceptibility raises hopes that even the most purified real data can, when fed into a gaps-insensitive spectral analysis, rely on stochastic resonance alone to help such data maintain their S/N at near-absolute-high levels; as shown above for a non-polluted case. Thus, as it is theoretically sound for the random case, and if a significant portion say over 67% of the information (signal + noise) contents could be recovered from significantly purified synthetic data using this (random) approach, then the approach would be automatically useful for raw data as well.

Here the polluting was done by normally distributed (Matlab generated) white noise, where randomness in selecting the placeholders of data values to be polluted was presumably uniformly distributed (again, trusting Matlab). Note that, unlike in the preceding, where by the purification I meant successive decimation, from now on and unless stated otherwise, by the purification I should mean data removal at random.

In summary, the test's final results show that the theoretical background of purification is sound and that it could work – thanks to a combined effect of the stochastic resonance and the GVSA modeling capacity. However, more work needs to be done in order to make the herein proposed purification practically useful as well. It can be seen from Figs.21-32/Supplement, that the sheer number of matches increased on average for about 50% despite the criteria not having been lowered though allowed to if reference to the successive purification was to be taken. The success rate, as established therein for the successive purification by decimation, now increased also – for about 50% in case of random purification, most likely due to stochastic resonance. Unfortunately, the impact of the stochastic resonance was insufficient to be practically useful (at 67%+ matches). At the same time however, the effect of stochastic resonance can indeed be seen in lower frequencies (10-100 Hz) – as a roughly-doubling of the number of perfect (4-of-4) matches, and in higher frequencies (100-200 Hz) – as a roughly-doubling of the number of at-least-3-of-4 v. 2-of-4 matches. This is same as previously (no stochastic processing or stochastic noise), but now also often with shifting to the neighboring higher frequency.

The observation of shifting-and-improving of period estimates with random purification gets entirely lost in the most (75 %-) polluted data, which is an additional observation in favor of the starting physical hypothesis. Obviously then, the here observed stochastic resonance effect is likely due to a combined effect of added random noise and data random removal. Remarkably and thanks to the use of the GVSA, this test demonstrates that the stochastic resonance indeed can arise classically in exploration geophysics as well, solely due to incrementally increasing presence or absence of noise in data, thereby depending upon data density only, Fig.25, rather than model complexity, and therefore in most real situations as well, Figs.23-24 and Table 1. I discuss this below.

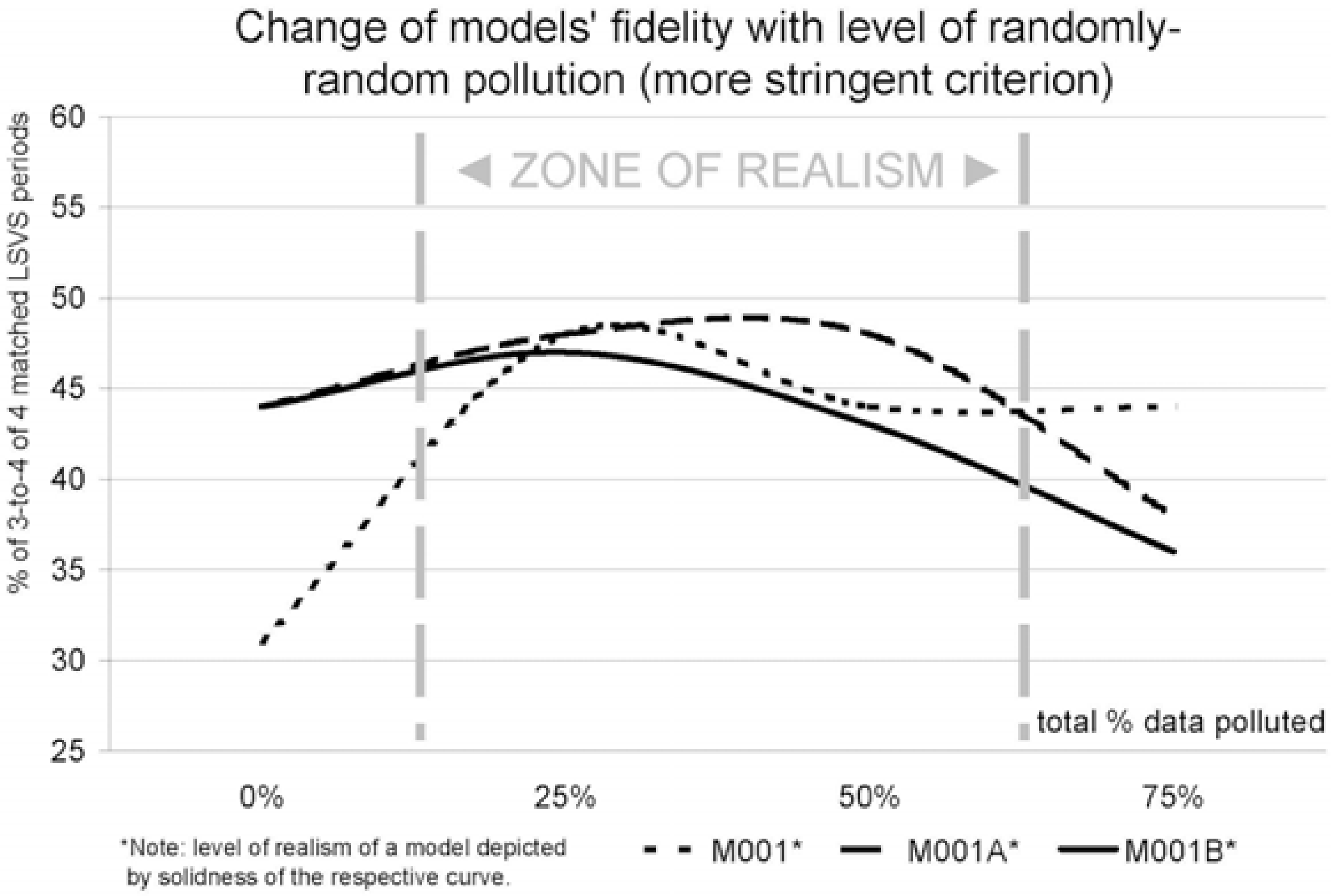

Figure 23.

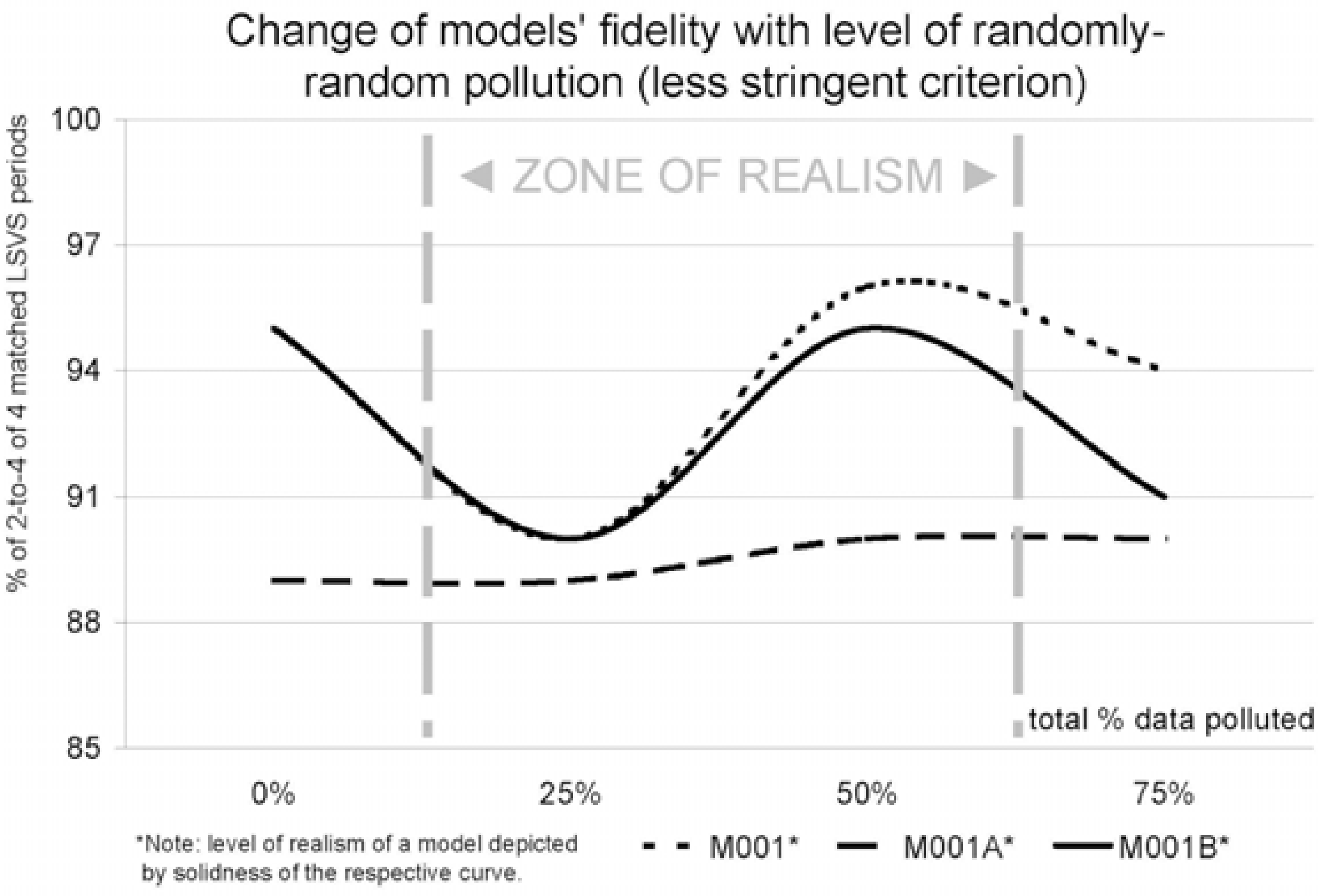

Figure 24.

| | (total # of periods) | @(2-or-better of 4) | in % | of which @(3-or-better of 4) | in % | | | 0% | 25% | 50% | 75% |
|---|---|---|---|---|---|---|---|---|---|---|---|
| 00M1 | 77 | 73 | 95% | 24 | 31% | | M1 | 31 | 48 | 44 | 44 |
| 00M1A | 79 | 70 | 89% | 35 | 44% | | M1A | 44 | 48 | 48 | 38 |
| 00M1B | 73 | 69 | 95% | 35 | 44% | | M1B | 44 | 47 | 43 | 36 |
| 25M1 | 73 | 66 | 90% | 35 | 48% | | | | | | |
| 25M1A | 80 | 71 | 89% | 38 | 48% | | | 0% | 25% | 50% | 75% |
| 25M1B | 72 | 65 | 90% | 34 | 47% | | M1* | 95 | 90 | 96 | 94 |
| 50M1 | 80 | 77 | 96% | 35 | 44% | | M1A | 89 | 89 | 90 | 90 |
| 50M1A | 79 | 71 | 90% | 38 | 48% | | M1B | 95 | 90 | 95 | 91 |
| 50M1B | 75 | 71 | 95% | 32 | 43% | | | | | | |
| 75M1 | 79 | 74 | 94% | 35 | 44% | | | | | | |
| 75M1A | 81 | 73 | 90% | 31 | 38% | | | | | | |
| 75M1B | 76 | 69 | 91% | 27 | 36% | | | | | | |

Table 1. Data values plotted on Figs.23-24. Left-hand section values: per-level-of-pollution; right-hand section values: per-model.

Note that, unlike with the preceding test of the purification by successive decimation, the GVSA variance-spectra of 0%-polluted (raw) data were used here as the reference spectra for testing the random-randomly purification. This because the spectral differential between the 0%- and the 50%-random-purified non-polluted (M1) data, though not exceeding the 99%-significance level, still remains above the 67% significance level for the most part, Fig.26. Note also that I use the term "random-randomly" also in order to distinguish the here presented test-case scenario from (also possible) successive-randomly purification – as a somewhat albeit not significantly less stringent approach of the two random purification approaches possible. Then in order to obtain a proof-in-principle (i.e., a demonstration of concept's validity on synthetic data) for the starting physical hypothesis, it sufficed to examine the random-randomly case only.

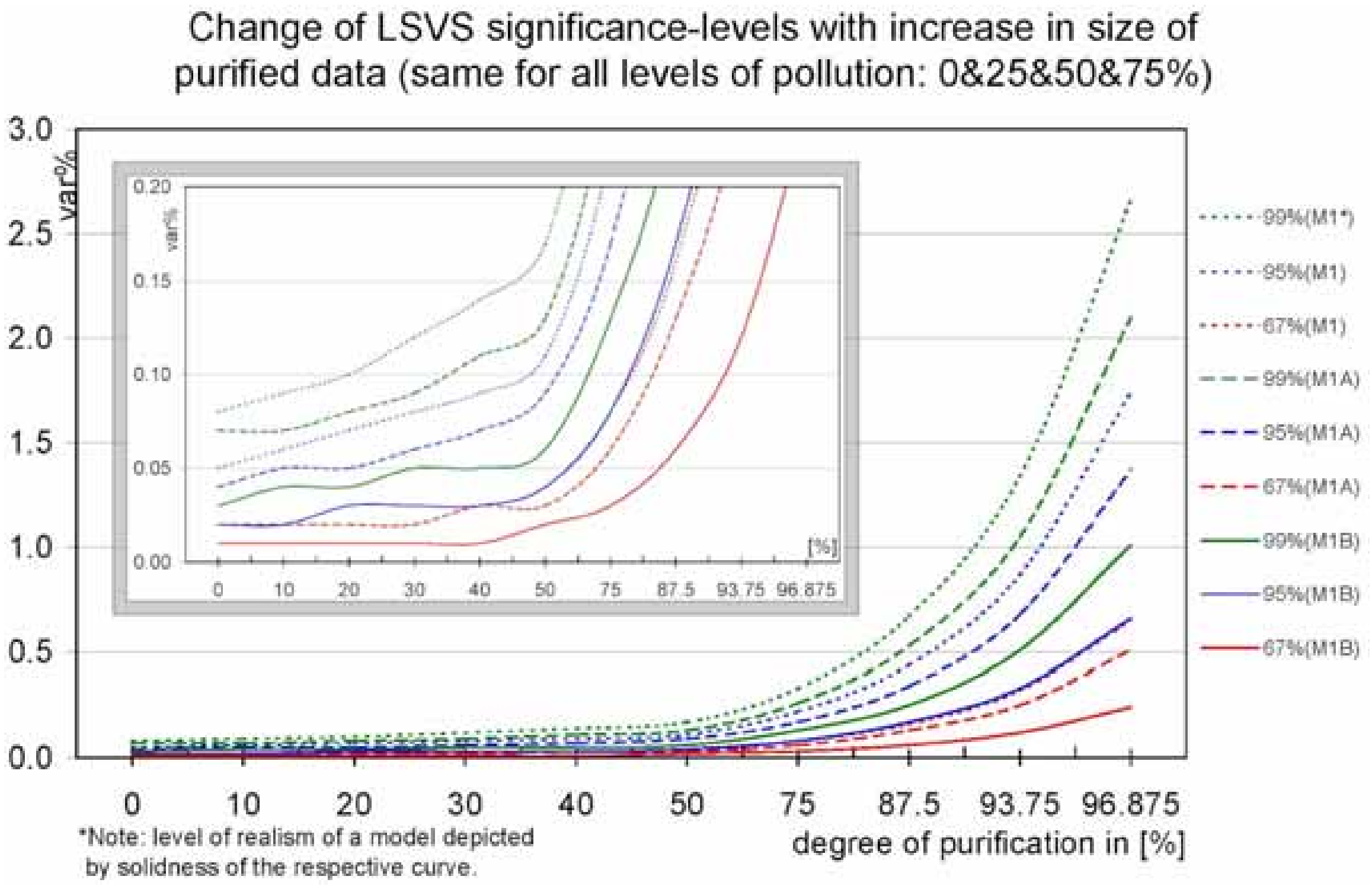

Figure 25. Change of significance level depends only on data density, whereas the change occurs irrespective of pollution levels. Expectedly then, significance levels in the M1B model (longest record) turned out to be least sensitive to the double drop in data density, while significance levels in the M1 model are most sensitive. Overall and clearly the 67% significance level is least sensitive amongst all significance levels as well, while the 99% significance level is most sensitive. Callout: ordinate-zoomed (x15) view shows that the same holds consistently for the successively intensified (linear-progressive) random purification (10-50%) as well. A general sturdiness of the 67%-significance level may have a physical meaning (i.e., be of a value in case of real data), since there the 95%- and 99%-significance levels are at the same time and generally not well separated in-between. The observed touching but not crossing of separate curves in the callout indicates real physical meaningfulness of the entire approach used for demonstrating SR.

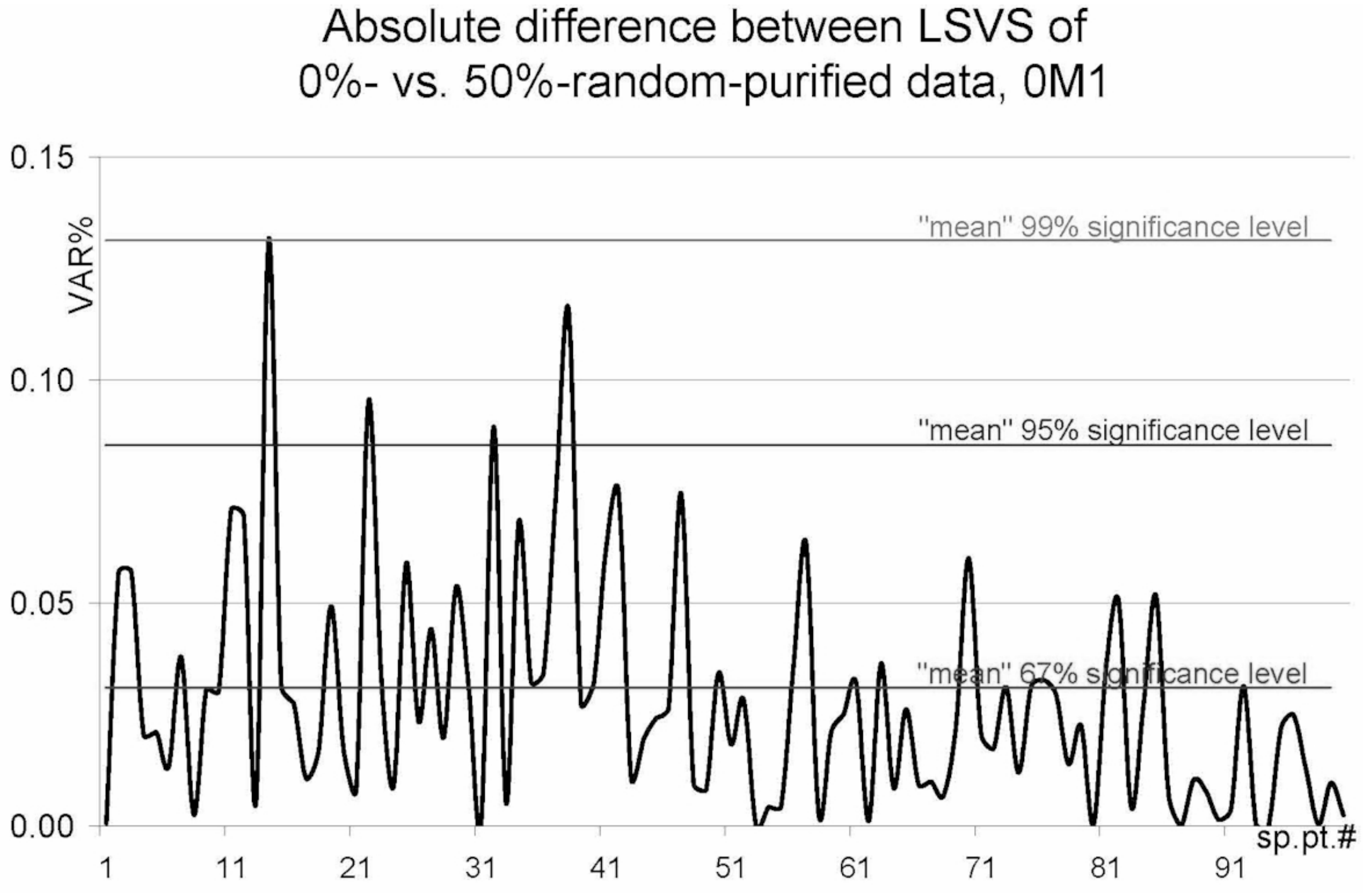

Figure 26. Arbitrarily smoothed spectral differential.

The effect of stochastic resonance is shown here using the GVSA of synthetic geophysical trace data of a varying (i) type (systematic v. random), (ii) complexity (2 v. 5 v. 20 harmonics), and (iii) normal-random pollution (0% v. 25% v. 50% v. 75%). Those results are summarized on Figs.31-33, and were based on another physical hypothesis: that a more complex model (i.e., the one which approaches the reality closer) should also reflect the stochastic resonance better than any simpler model(s) can be expected to. Then if real, the effect should be best discerned from the raw v. polluted data; this is the case here, as can be seen from Fig.29: when the raw data show little or no gradual response to increase in complexity, the adding of random noise introduces some stochastic resonance which enhances the data response to model complexity. Also, the same effect is seen from Fig.28 albeit with 0-50% and 25-75% parities, as the linear-progressive random purification affects the stochastic resonance by extinguishing it, which is expected given that linearity in removing of data means introducing systematic rather than stochastic noise. On the other hand, the exponential progression in data removal intensity only further enhances the benefits from the stochastic resonance, Fig.29. Thus the 0-50% parity, observed in linear-progressively so purified systematic models, is also seen in case of at least one randomly polluted such model, here the M1A, Fig.28. This constitutes an observation of (the effect of) the stochastic resonance in a set of exploration geophysics-like data.

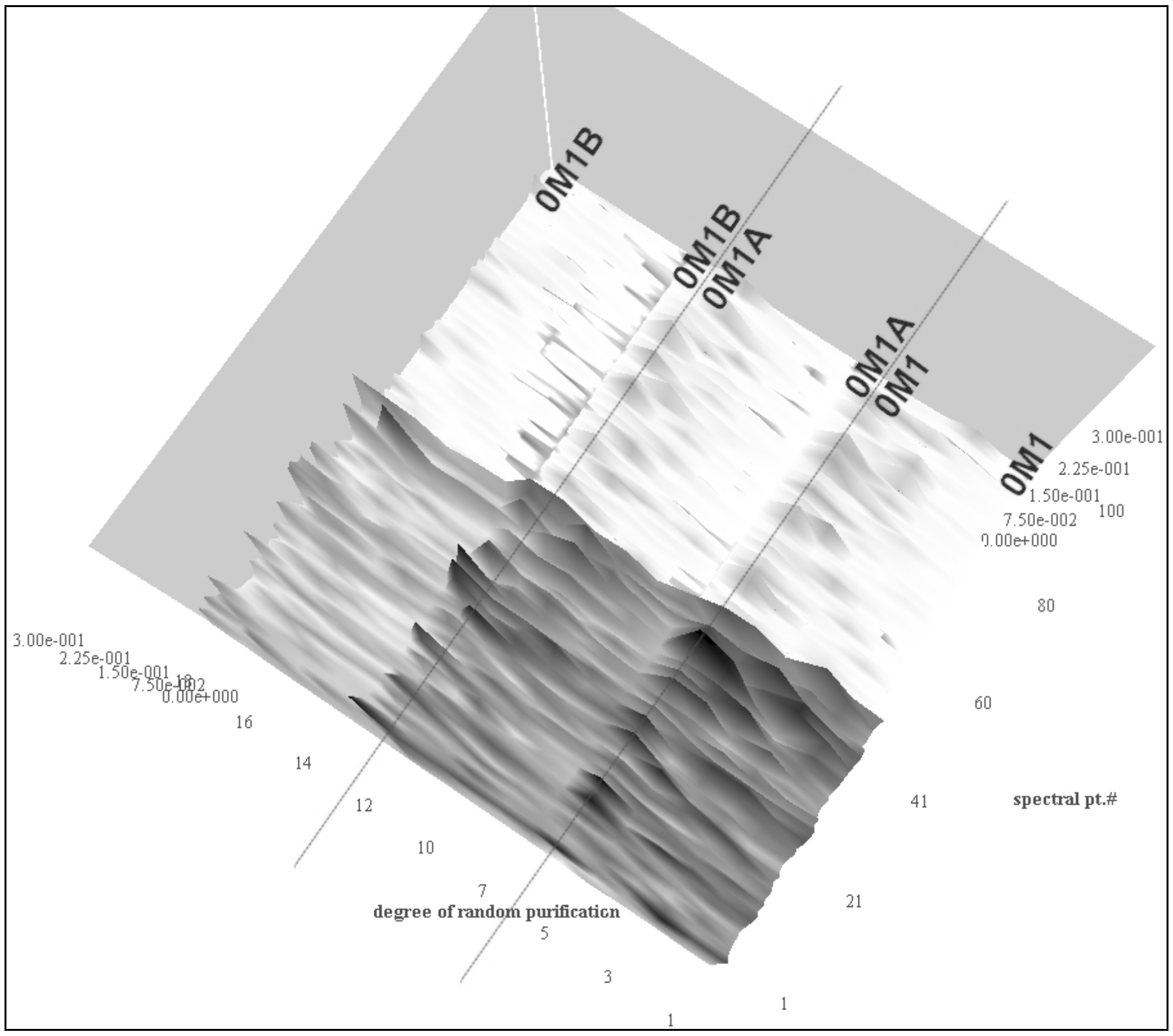

Figure 27. Plot of non-uniform drop in model-response to random purification, with increase in data density as dataset size increases. Here the degree of random purification progresses linearly from 0-50% for each model. From 0%-polluted (raw) data only. Vertical axis in var%. Below: washout-imaging effect stresses the graduation.

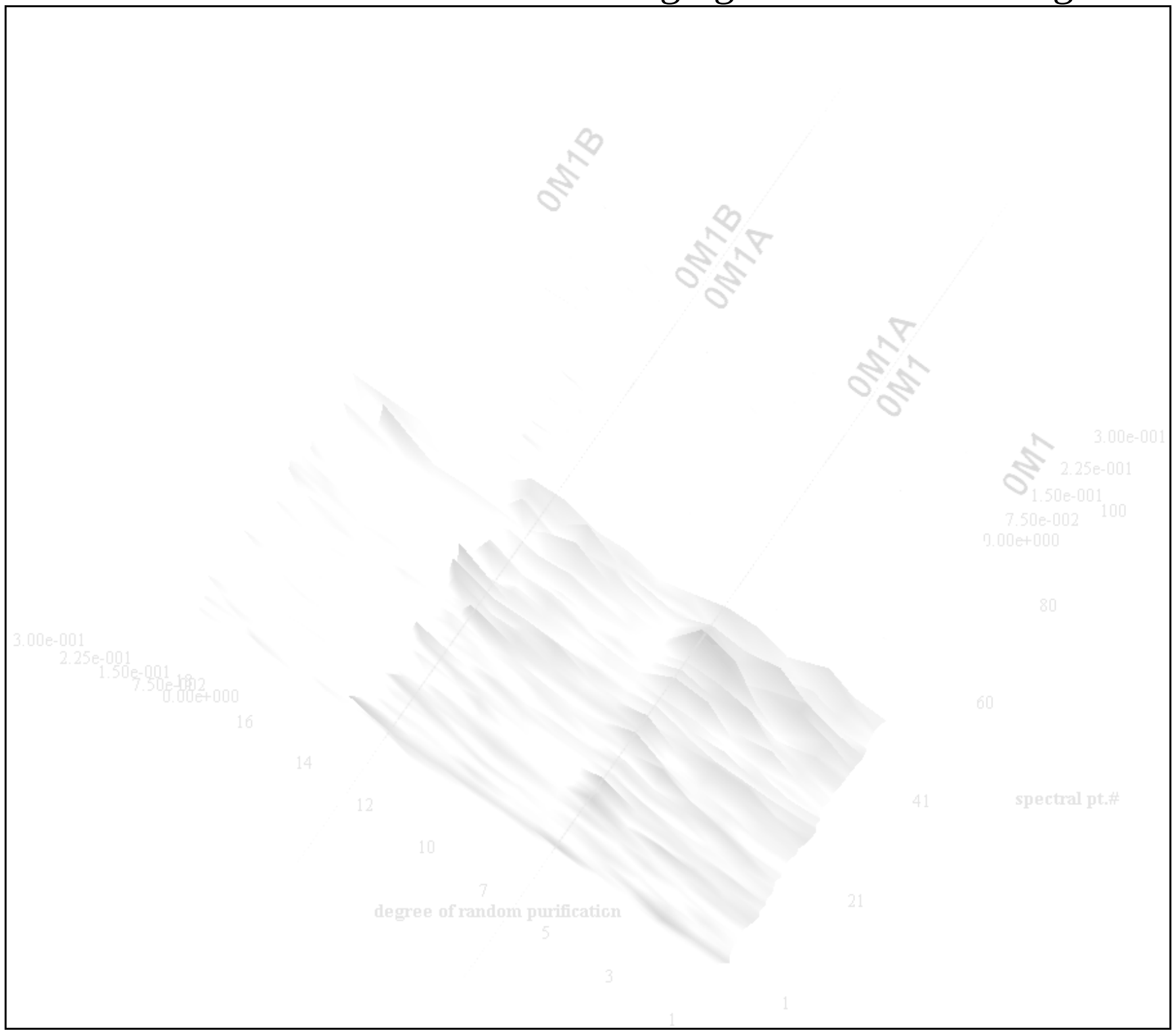

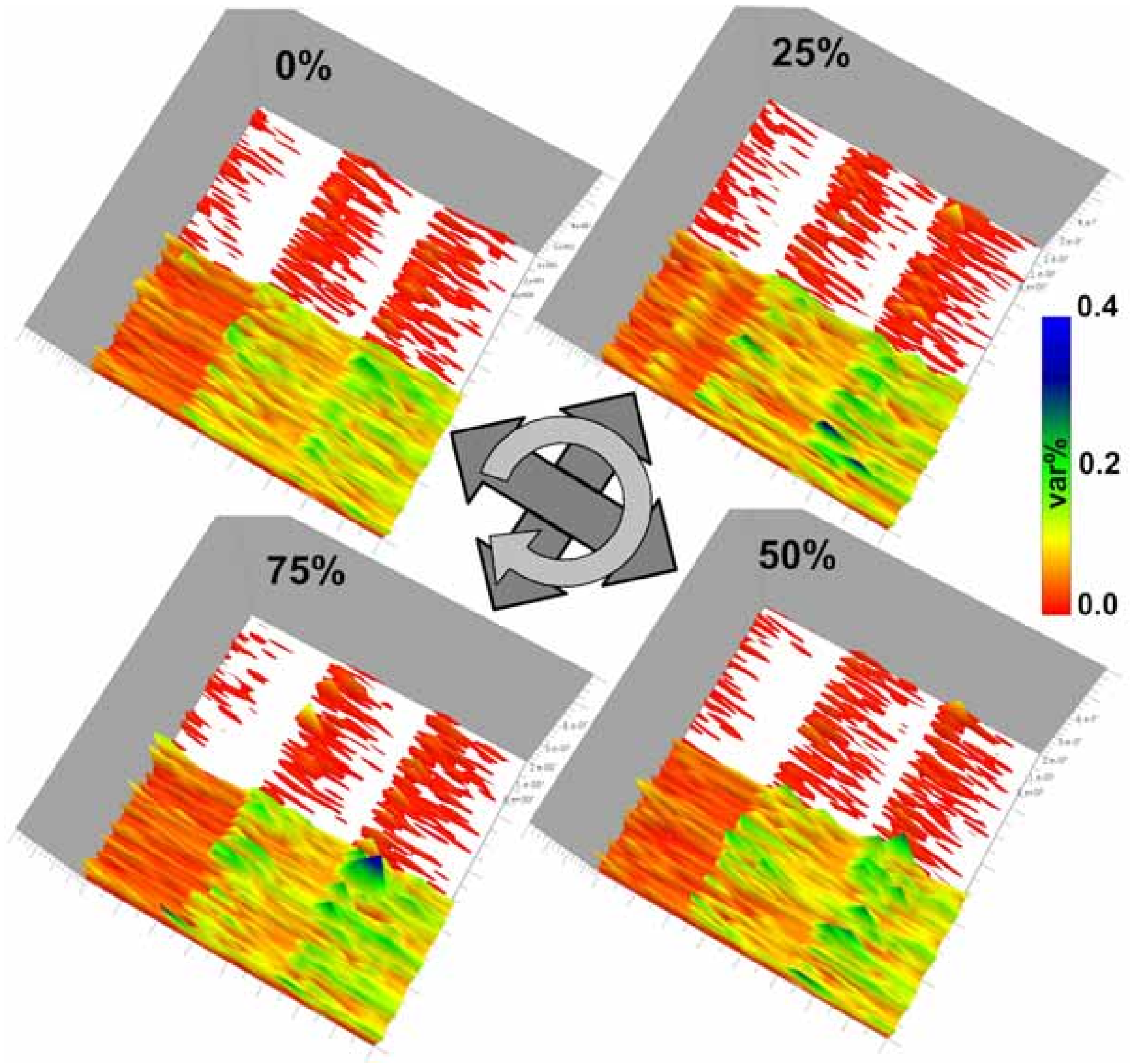

Figure 28. Effect of stochastic resonance as found from uniformly-random, linearly progressive purification (0%, 10%, 20%, 30%, 40%, 50%) of 25%- v. same of 50%- v. same of 75%-normal-randomly-polluted synthetic-trace data. Note a parity (best seen in high frequencies: the red solid-most upper halves of spectrograms) between models' behavior (as a little-to-non-gradual models-response with increase in complexity) in cases of 0%- v. 50%-polluted, and an opposite parity between models' behavior (as a gradual models-response with increase in complexity) in cases of 25%- v. 75%-polluted data. In the latter parity graduation is more obvious than in the former. Panels depicting varying levels of random pollution ordered clockwise. Axes and models order as on Fig.27.

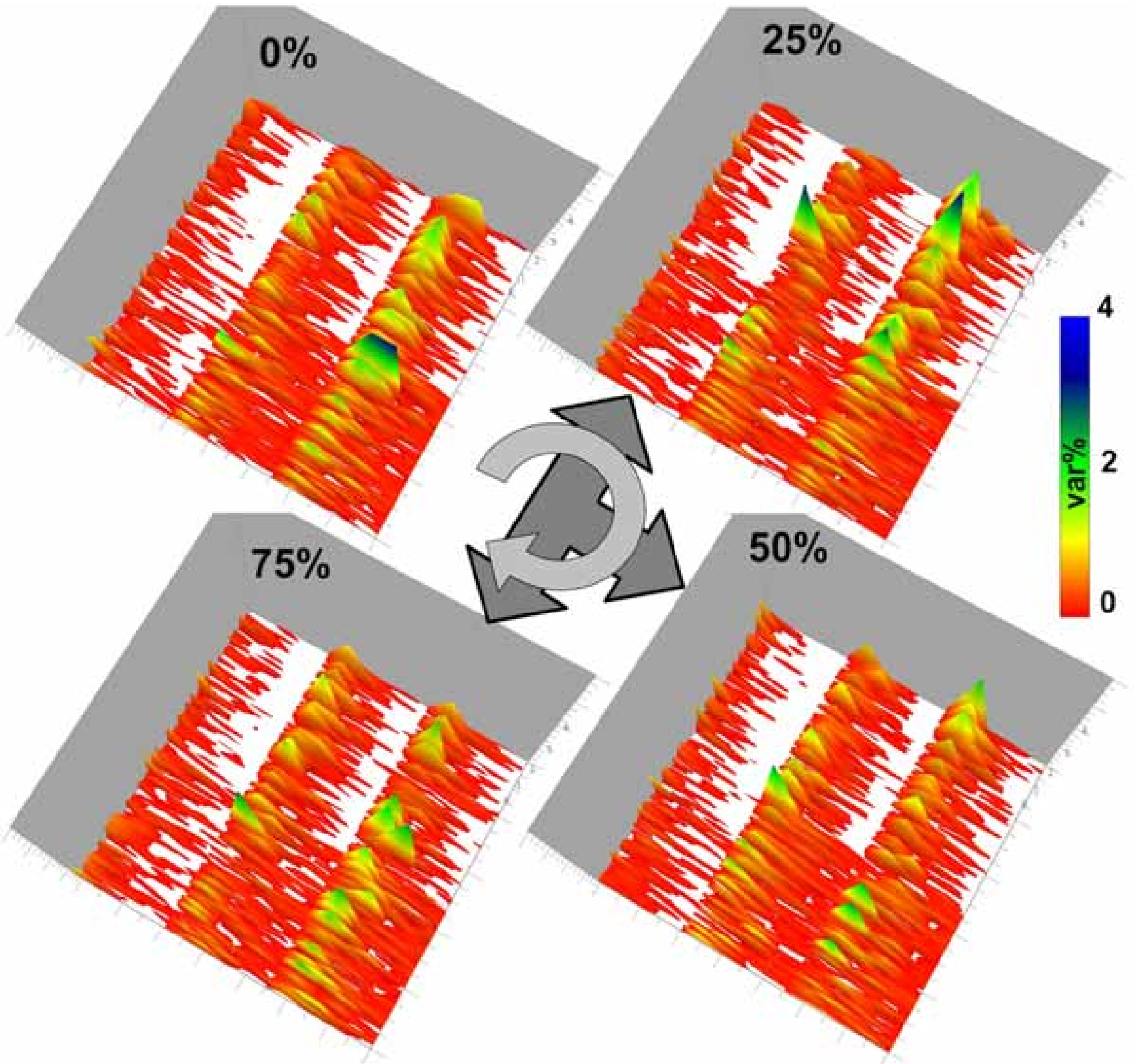

Figure 29. Effect of stochastic resonance as found from uniformly-random, exponential-progressive purification (50%, 75%, 87.5%, 93.75%, 96,875%) of 25%- v. same of 50%- v. same of 75%-normal-randomly-polluted synthetic-trace data. Note the solitary (best seen in high frequencies – upper halves of spectrograms; as on Fig.28) model behavior (as a non-gradual models-response with increase in complexity) in case of 0%-polluted, and parities between model behaviors (as a gradual models-response with increase in complexity) in cases of 25%-, 50%- and 75%-polluted data. Panels depicting varying levels of random pollution ordered clockwise. Axes and models order as on Fig.27. Hint: observe M1A model for all levels of pollution – since all cases are viewed under a same tilt angle, the larger the area under the solid redness of M1A in case of row data (0%-pollution) v. M1A in all other cases (levels of pollution) means worse relative response than in any other cases.

Note that it did not make sense herein to perform any statistical testing on the assessments of Figs.32 and 33, as the GVSA variance-spectral magnitudes are not distributed normally but they follow the $\beta$–distribution instead (Steeves, 1981).

## Discussion

There is no single reason why the Fourier and the Wavelet methods should be the last words of wisdom when it comes to subduing the S/N as a general concept from information theory. Also, harmonic noise analysis is closely related to filtering in the time-domain if proper methods are used, for in the strict sense filtering is merely a functional relationship so that the simplest filter (and the simplest model) of all reads:

$$f(x) = y,$$

where the real argument could be, say, a mass-redistribution, i.e., a Euclidean-spatial differential.

Based on the preceding, the GVSA could then be a ready-made successor to the aforementioned approximate classes of methods, in at least three important aspects in applied geophysics: (i) time-domain data filtering, (ii) highly-accurate spectra/spectrograms generation, and (iii) spectral peak's epoch determination (Omerbashich, 2008a). Here I used synthetic data to validate the first two aspects.

The FSA-based algorithms are normally considered exact. However, this is true only in the strict computational-arithmetic sense, as floating-point errors get neglected. Obviously, this reasoning is biased in the sense that it disregards the importance of any input data-related errors on the output. Namely, errors can arise from issues such as the completeness of data, data quality and reliability, as well as data overall or even segment-wise reliability. The higher the frequencies, the graver the effect of mixture and of totality of such errors on the result. Nonetheless, the Fast Fourier Transform (FFT) and its variations had been worked out, which in the end got all too easily regarded exact, in a 1990-ies euphoria built entirely on computer-only warranted aesthetics, e.g., the Cooley-Tukey algorithm. Even worse yet (for true i.e. input-independent exactness): the so-called approximate FSA-based methods have been developed which not only disregard significantly important information on the input information, but also trade computational correctness for shorter computer execution times.

To rely on such "time engineering" can seem acceptable in situations where the ultimate judge of the output's quality is the human being, with his/her utterly imperfect five senses. Meaning, that sort of manipulation could look appealing to, say, audio and video/imaging industries, or to manufacturers of consumer electronic equipment in general. It appears that things have moved in that same direction in exploration geophysics, for it too has obviously followed far more closely the developments in electrical engineering than those in the physical sciences. But, electrical engineering deals with highly-well-behaved streams of energy/signals that are at the same time well-organized into well-structured and well-posed apparatuses and information strata. Geophysical data on the other hand are pretty much everything opposite to that! Obviously, the preceding discourse was made from the point of view of a suspecting researcher interested in meticulous approaches in order to achieve the highest accuracy and precision possible in the real-life albeit idealized (all-scale; band-wide) applications, using unspecified methodologies. As a result of not focusing its resources on understanding and appreciating the data themselves, only minute improvements to the solutions within the frame of FSA fundamentals, had been worked out in the past. The wavelet "miracle" of 1980-ies scores none better either, for it too was just a patch-up for one of the many shortcomings of the FSA methods.

Examples from geophysics, of things going astray or circles at best can be read occasionally in reviews, such as: "*time-frequency analysis has unique features, such as the uncertainty principle, which add to the richness and challenge of the field*" (Cohen, 1989; p. 942). From the point of view of fundamental theoretical discussion however, this is merely a cover-up. In fact, such outlandish and thereby worrisome statements are a sign of dodging of the issue of external ("against the real world") limitations of the FSA-based methods in exploration geophysics altogether. Unfortunately, that course seems to have become typical of other engineering disciplines as well, which had simply and blindly copied methodologies from each other, without giving much or any consideration to the type and the scope of the problem *they* are faced with. Instead of being approached in a really painstaking fashion, meaning on a problem-to-problem basis, individual problems are being "solved" *en masse*, i.e., according to different groups of problems they belong to, rather than in accordance with the physics that each problem individually necessitates invoking. Consequently, resources get wasted on numerous non-physical approaches that are concerned with ways how to massage data considered as a whole.

Thus for instance, most workers nowadays see the FSA methods as the ultimate way to obtaining the true epoch of a spectral peak (i.e., when exactly a significant increase in intensity had occurred). This is normally done by bandpass filtering with windowing, i.e., sliding of the input data for a prescribed interval along the timescale, followed by a spectral analysis of the data each time the data are slid. Also, the SFT is used for detecting significant intensity changes that had occurred over intervals shorter than the prescribed sliding interval. But other, fundamentally different approaches are possible also; say *via* using the GVSA as a time-domain filter. Consequently, such an application would make the GVSA a ready-made tool for creating highly-accurate spectra as well, where "highly" apprises relative to the FSA and Wavelet methods at least. Namely, the same which can be attempted by using the SFT (that unavoidably deforms data and spectra) can be fully accomplished by using the GVSA – as a data-non-invasive and spectrum-non-editing method of spectral analysis. Thus a peak's true epoch can be determined by sliding the time interval albeit not in a uniform but random fashion instead (Omerbashich, 2008a). This is a spectral-method-independent operation, which therefore enables external, e.g. physical verifications of the whole spectrum. In other words, random sliding is spectrum-friendly in the sense that it contains no mathematical trickery which could preclude the spectrum or its part(s) from attaining a physical meaning needed for interpretation. Then in order to estimate the epoch on the most recent extreme magnitude of a statistically significant peak in a dataset, thus extending the classical approach of simple fit, one makes use of a unique feature of the GVSA: background noise levels tend to take a virtually linear form in GVS, enabling relative measurement of field dynamics (Omerbashich, 2007a).

At the same time, this makes a very important physical property of the GVSA too, as it makes possible now to apply a spectral analysis method directly within the time-domain in order to determine the effect of dataset size (the missing or/and underperforming values considered absent) on a period estimate. This in turn makes a GVS a least ambiguous (since noise-fed!) measure of the signal's relative strength in a noise-burdened information. All this while the S/N is made to vary purposely albeit not only with intention to test-process synthetic data (like in classical scenarios), but for actually measuring the signal's relative strength on the real data instead – thereby obtaining the now enhanced "signal" (as records are polluted thereby fully enhanced in the sense of the GVSA). For details on the GVSA as an epoch-estimation utility, see Omerbashich (2008a).

Based on the preceding discussion, it should be straightforward to apply the GVSA with the purpose of avoiding the fixed-resolution windowing too. Various so-called multiscale solutions have been developed in the past to handle the fixed resolution issues, however those solutions too are approximations only; the Fast Wavelet Transform being the most known one amongst such approximations. Namely, most real-life problems in physical sciences require for longer periods to reduce to a good frequency-resolution already in the first approximation, followed by ever-improved refinements with good time-resolutions for shorter periods (the *from-larger-to-smaller* modeling principle). Unlike with the wavelets however, the spectral resolution itself is not approximated in the GVSA but is extremely precise already at the outset, and depends only on the computing power available as the processing workload increases with the increase in spectral resolution. Here the spectral resolution is achieved *via* the *variance-spectral zooming*; a process which can be shown to work in long periods even for relatively very large and largely polluted natural datasets. As mentioned earlier, in one such case a simple spectral zooming from 2000 to 50.000 computed points picked up a lunar-synodic periodicity of the total-Earth's (atmosphere-to-inner core) gravity field's decadal oscillation down to the original instrumental abilities (Omerbashich, 2007a).

Even if not immediately applicable to the shortest-periods analyses, this still demonstrates that in high frequencies the GVSA is probably less prone to known issues with S/N than the FSA methods are. The importance of this comparative advantage is highlighted by recent advances in applied geophysics, aimed at using also the longer-period information, which in the past was classically disregarded altogether. Broadening the band of interest, as well as utilizing more and more information previously thought of as noise ("junk"), have in recent years become hopeful trends in many natural sciences such as theoretical geophysics and, more famously, genetics. In a synthesis then, data contained within a window can themselves be purified. This renders the classical FSA-based fixed-resolution problem obsolete when the GVSA is used, as windows can now attain practically varying resolution; conceptually this is akin of vector weighting. Then in order to alleviate the fixed-resolution-window problem (windows still can be of a fixed width but non-uniform density as less reliable data get thrown out or purposely under-weighted; Omerbashich, 2007b), one purifies one or more windows where spectra were (to be) computed in the GVSA.

Obviously in the sense of a physical criterion according to which data are to be manipulated under no circumstances whatsoever, a spectrogram compiled from so computed GVS will also be of the highest accuracy possible, in theory at least. Of course it would not be possible to apply the GVSA in the described manner on real-world problems without first tuning the procedure empirically so as to decouple thermal processes and velocity/field as normally seen in high frequencies. Furthermore, so generated spectra would not be free from all the drawbacks of the classical methods, such as the imperfections in the windowing function (e.g., the Hamming tapering, applied often in natural sciences) when used in combination rather than instead of the windowing. However, the impact of such classical drawbacks will likely be lessened in a GVS. How can this be expected?

Recall that a successful power-spectral analysis supposedly is one which resolves the adjacent signals of comparable strengths at least as well as the non-adjacent signals of significantly different strengths. But this definition is a setback in itself as it is method (FSA) dependent. So not only that this definition says nothing of information quality and throughput, but it also relies on an overall-useless magnitude background in the FSA; this is at least a waste of resources, but likely of the accuracy as well. On the other hand the success of a variance-spectral analysis is not measured by overall uniformity of spectral clarity in the frequency-domain, but locally – by the totality of single peaks' significances in the time-domain. Namely, given the full physical i.e. time-domain meaning of the overall magnitude background in the GVSA (Omerbashich, 2007b), a conventionally successful GVSA is one which detects statistically significant peaks regardless if they are well-resolved or not, as relative to each other (besides, who is to declare a *well-enough separation* in any method, since this too falls under the "art" part of the analysis?). This is not so with the FSA, which uses the above described false "internal check" of "comparing" how well (sic) it allegedly resolves frequency-adjacent but similar, v. frequency-distanced but dissimilar signals, still all under the same set of method-resulting limitations! By pretending this to be a universal check, the FSA methods have somehow ended up being regarded as a universal tool as well; the case now in most sciences.

## Conclusion

I successfully identified the stochastic resonance effects in synthetic data typical of exploration geophysics. The result suggests that SR by itself is insufficient for achieving the goal of fully recovering the across-band systematic contents in the information which is normally and on whole regarded useful to exploration geophysics. But the fact that the stochastic resonance, as a working method albeit of unknown theoretical background can be detected using the GVSA as a method of also (fundamentally) unknown theoretical background, should be telling in the sense that the two methods perhaps share or are susceptible to the same as of yet unknown properties of nature. Further research into SR enhancements to S/N in applied geophysics is hereby justified. This is important given that ever-escalating energy demands along with supplies lessening make SR potentially vital in the time to come.

# SUPPLEMENT

<u>Internal checks for usefulness of the GVSA in geophysical applications</u>

### 1.1.1 Testing on a single synthetic seismogram with simplistic modeling (one harmonic)

The synthetic trace-seismogram was created using Matlab, Fig.1. The seismogram is a complete dataset with harmonic noise of amplitude 1, in the $f \in 10\text{-}200$ Hz band. I designate such an unrealistic (with just one harmonic) scenario as the no-modeling benchmark.

In order to verify the GVSA as a valid spectral analysis tool for applied geophysics, in the following I first perform several internal checks by comparing the GV variance- v. GV power-spectra for the stationary, no-modeling synthetic sweep seismograms (with and without data manipulations – here the zero-padding v. no-padding cases), Figs.1-2.

Note here that zero-padding is commonly introduced in applied geophysics in order to enable cross-correlation of spectra, but note also its classical purpose – to artificially boost power in (Fourier-derived) power-spectra.

For the benchmark scenario, the GV VS and GV PS of zero-padded v. non-padded synthetic seismogram, Figs.1 v. 2, are compared on Figs.3-6. The respective spectral differences and spectral ratios are shown on Figs.7-10.

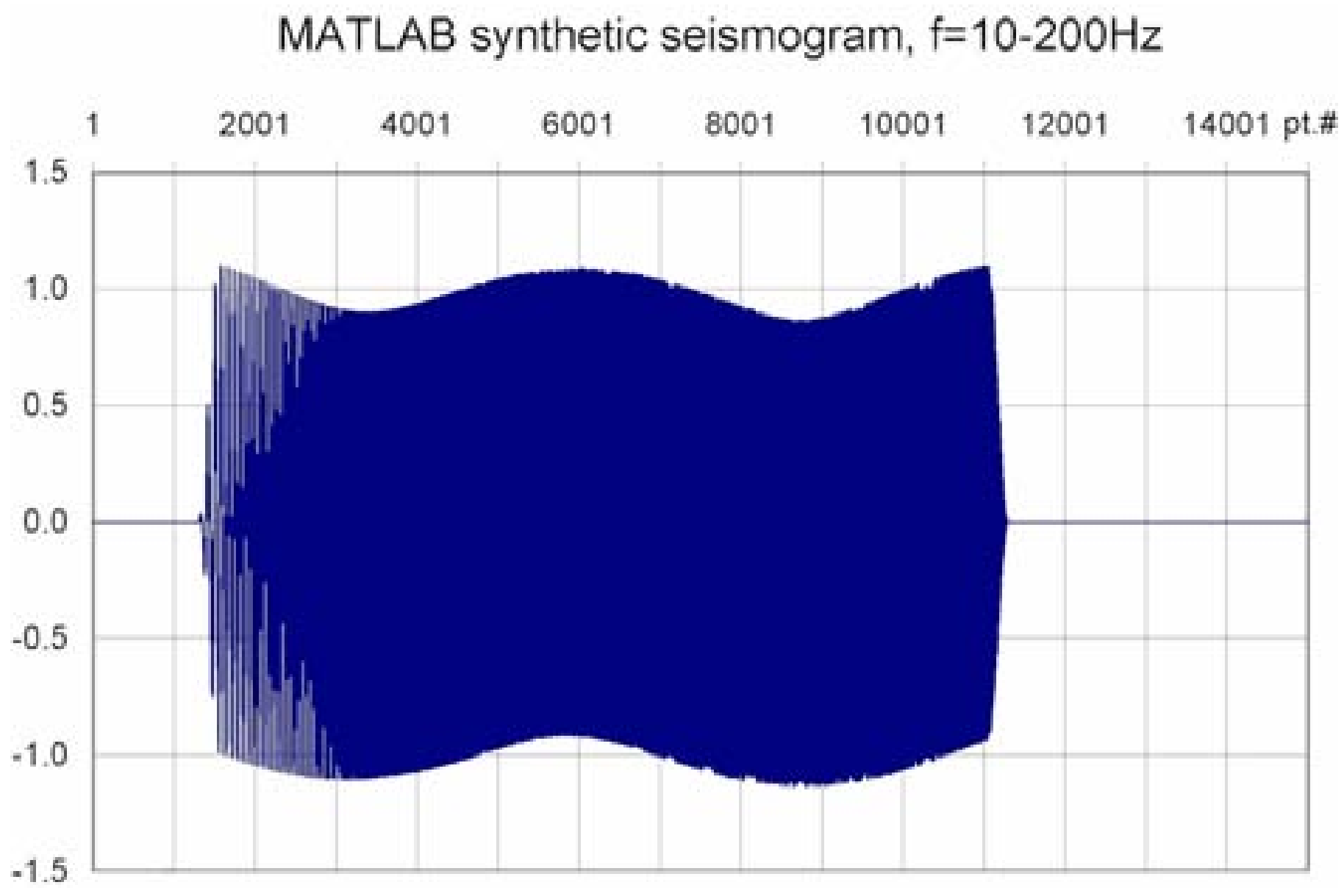

Figure 1. Matlab-created complete synthetic seismogram, f=10-200Hz, zero-padded, no modeling.

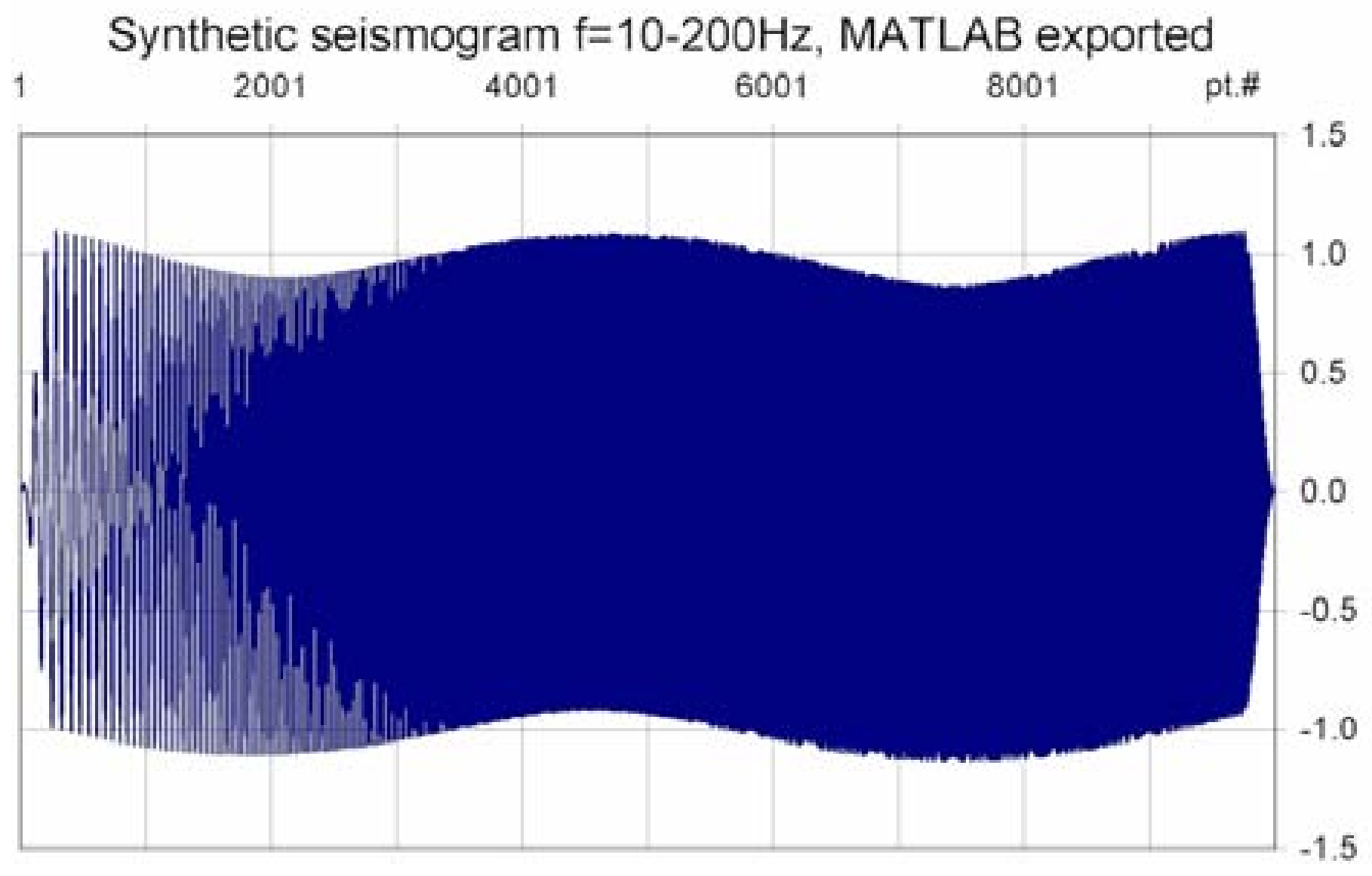

Figure 2. Matlab-created complete synthetic seismogram, Fig.1, no padding, no modeling.

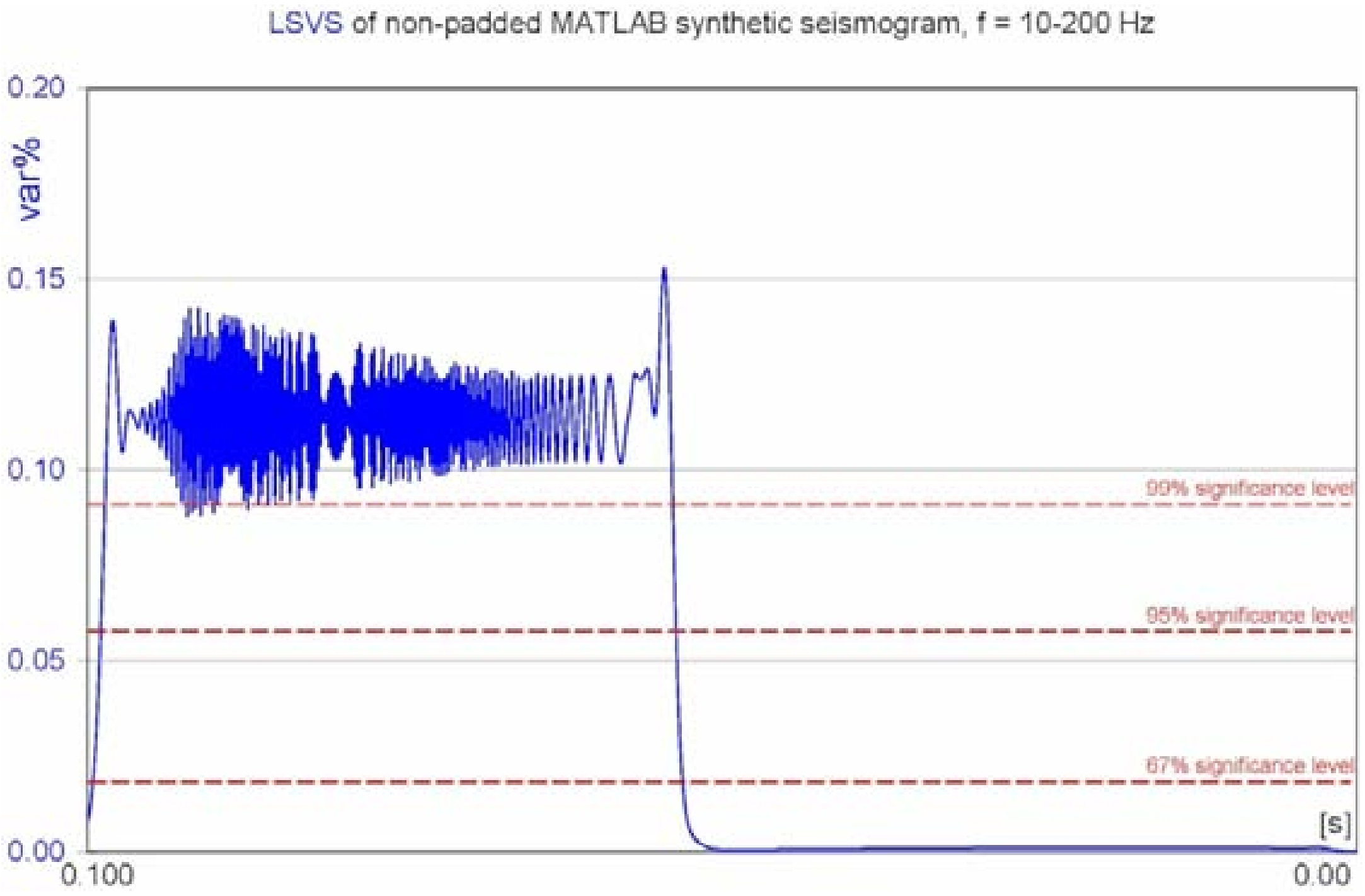

Figure 3.

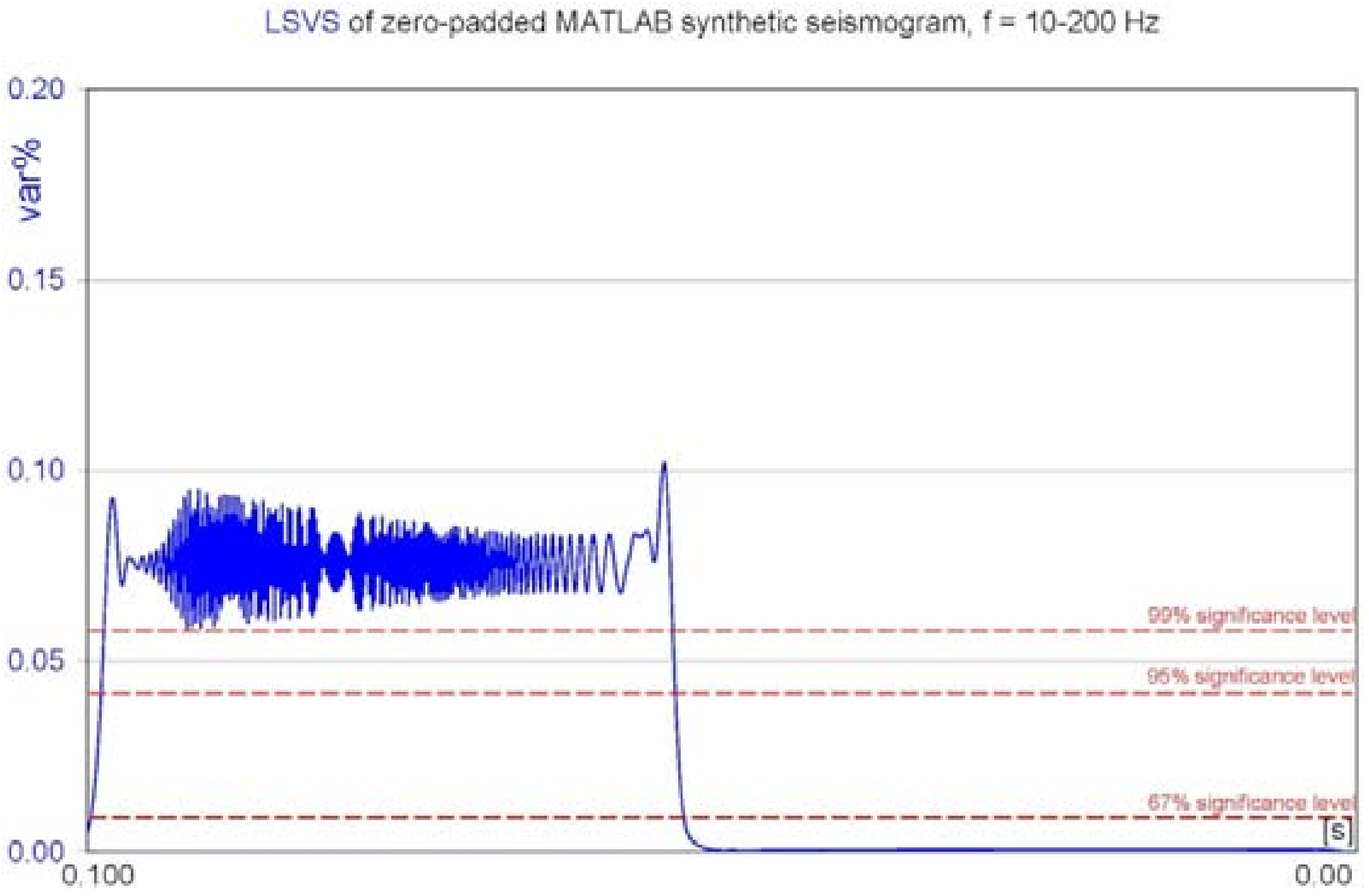

Figure 4.

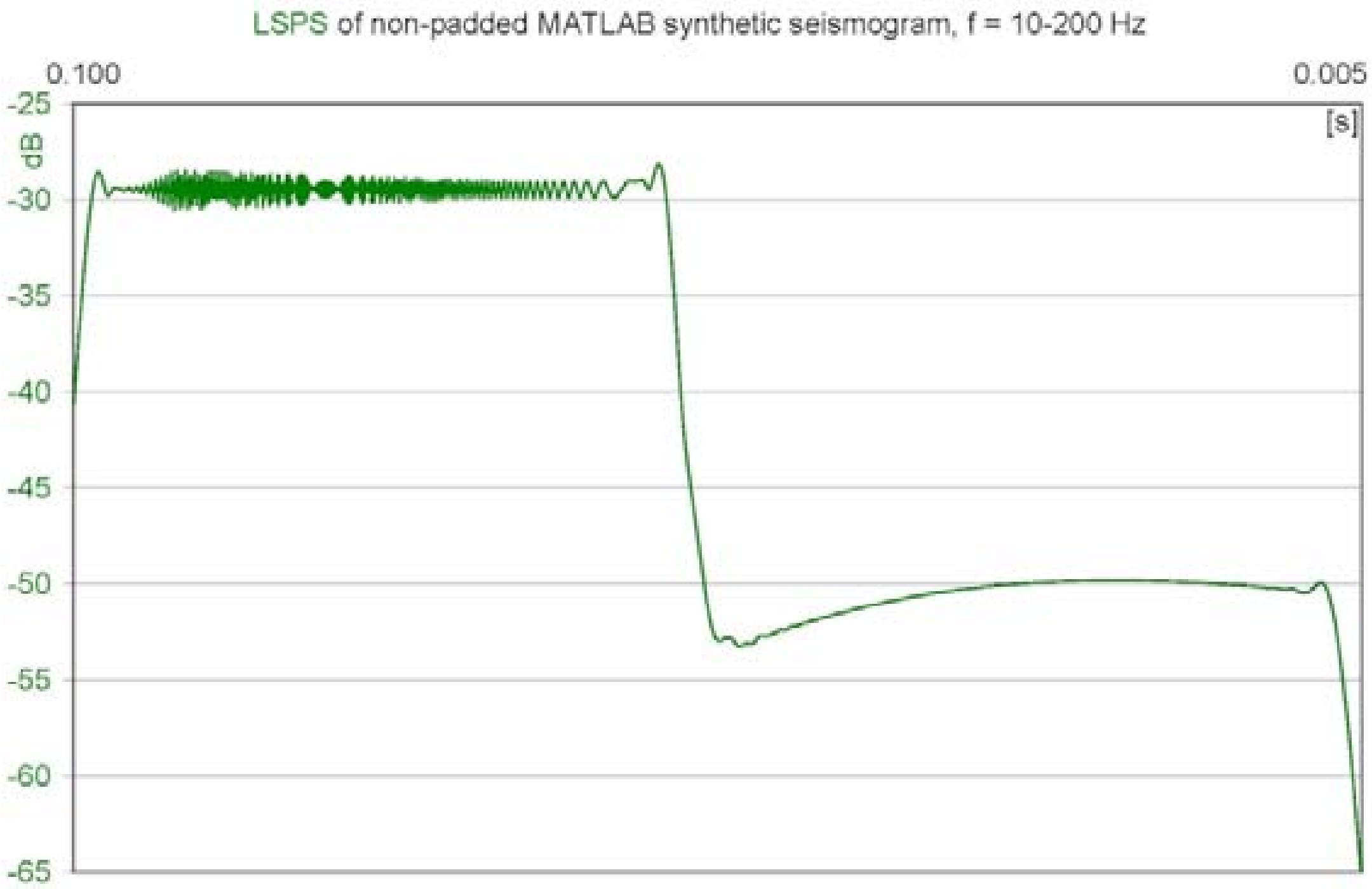

Figure 5.

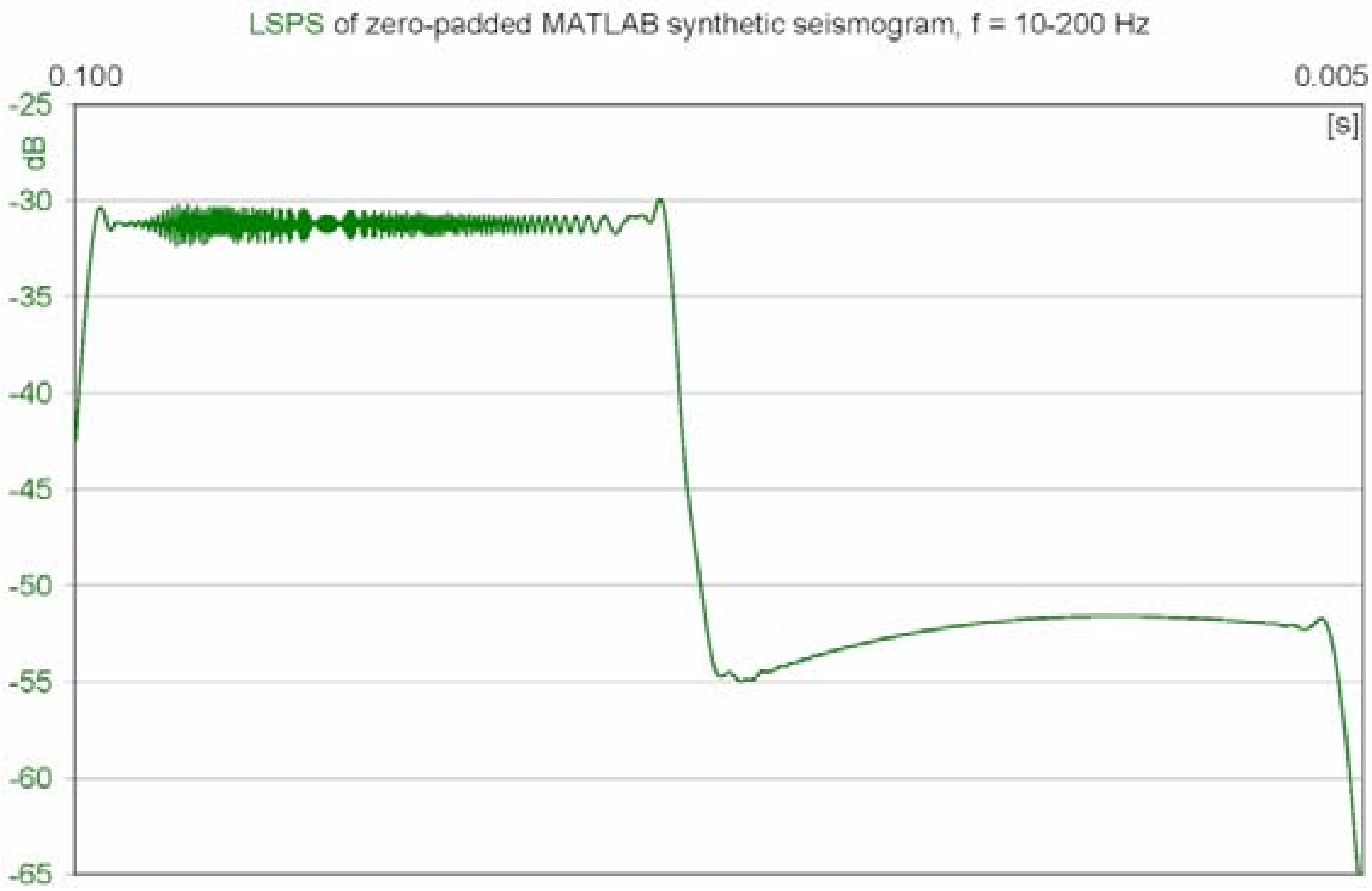

Figure 6.

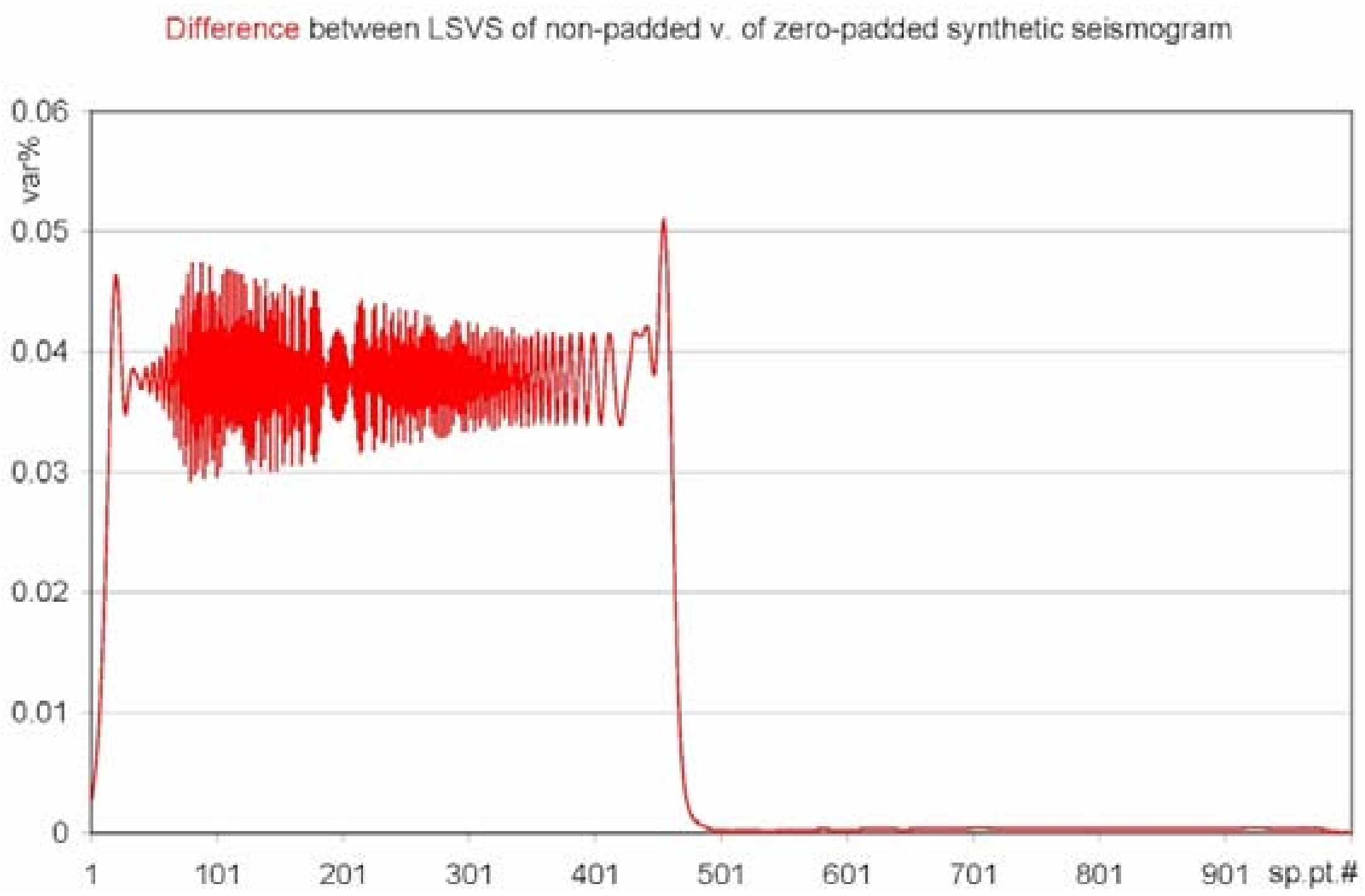

Figure 7.

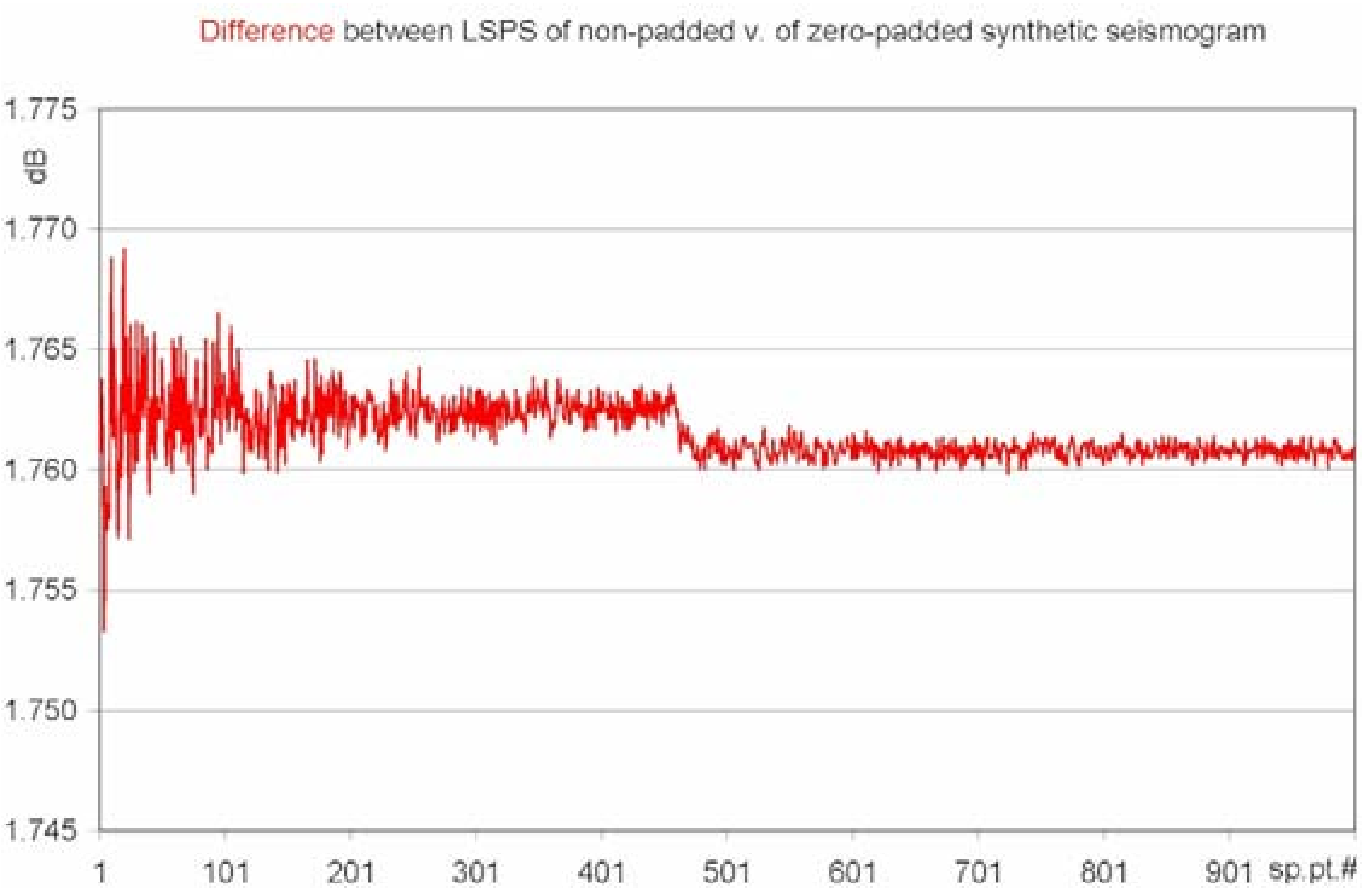

Figure 8.

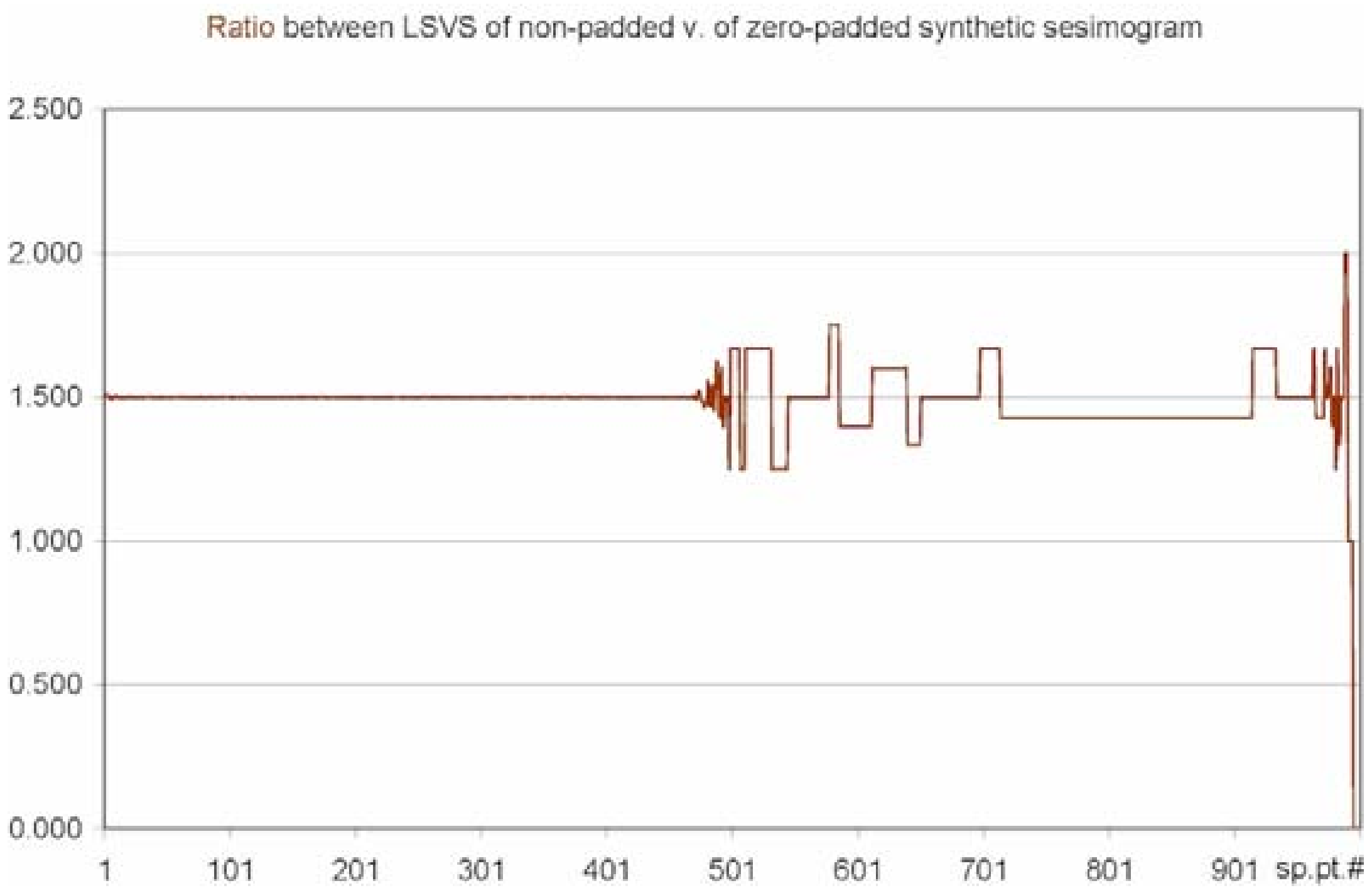

Figure 9.

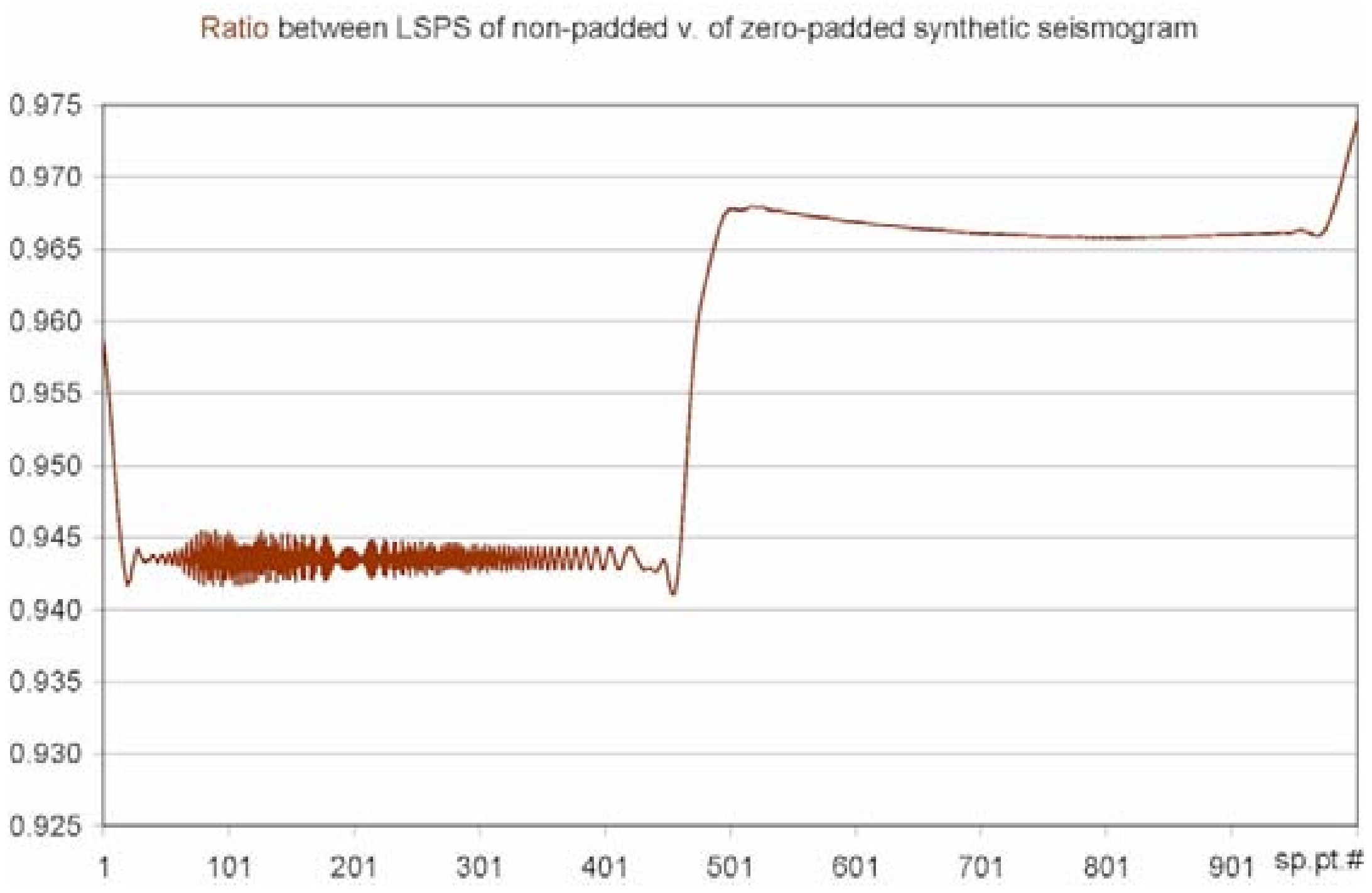

Figure 10.

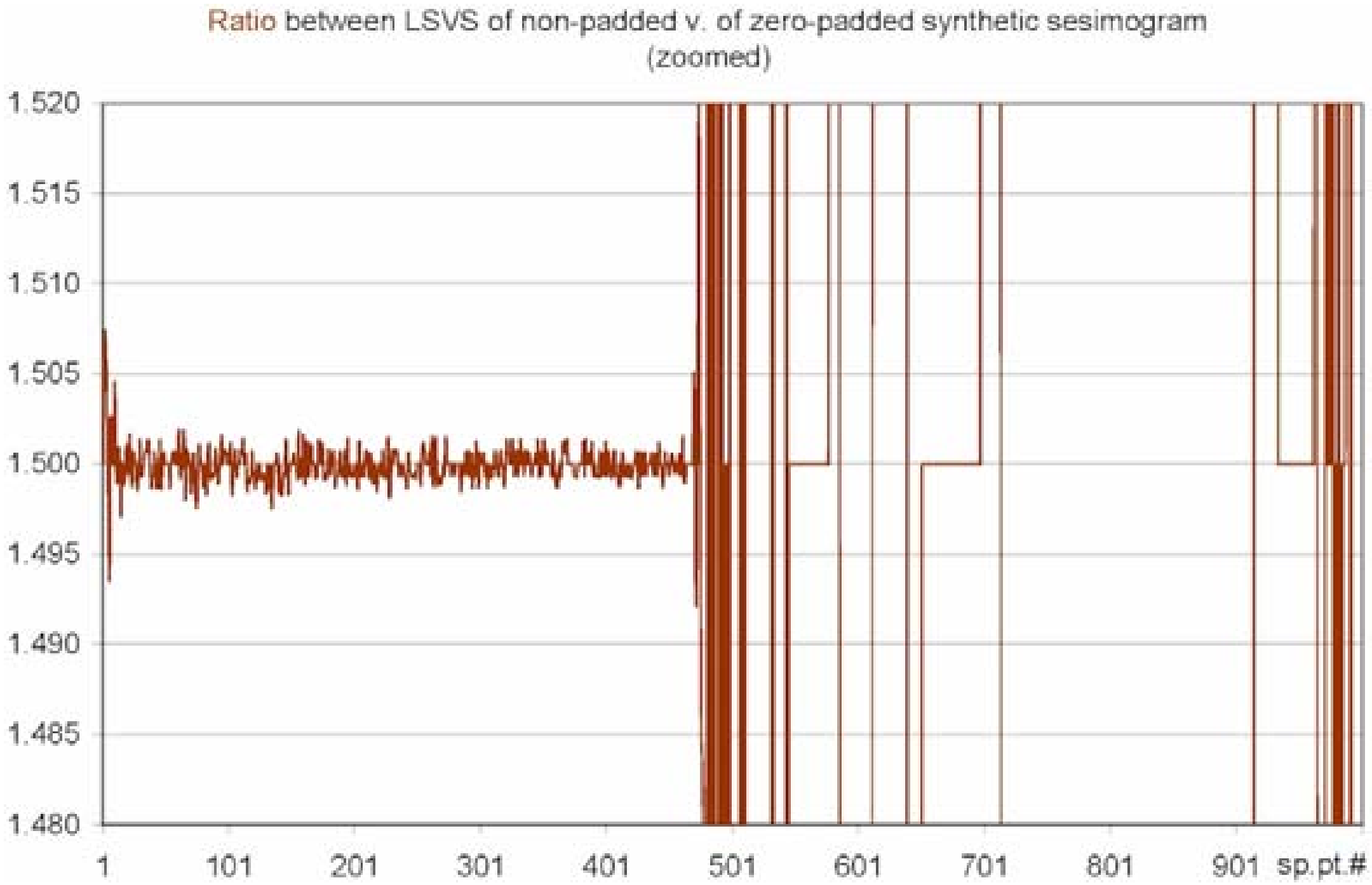

Figure 9a. Zoomed-in detail from Fig.9.

It can be seen from Figs.3-6 that the GV VS responded correctly to about 30% of data-size increase attained by zero-padding, which extinguished classical noise (here: the signal, as for testing purposes using the GV VS; see Omerbashich, 2006a). At the same time it shrank the signal envelope from about 0.5 var% to about 0.3 var%. On the other hand, the GV PS seemed insensitive to data scaling, albeit responsive to an artificial power boost as due to the zero-padding, gaining about 4dB in power; see Fig.5 v. Fig.6.

Figs.7-10 depict different residual behaviors of the GVSA, accurately all the way to the method's ability to pick up a purely theoretical signal imbedded solely within the information of interest. It is worth mentioning that, when one is faced with systematic noise, one would normally proceed on to spectrally analyze in the GVSA also the residuals from the original GVSA.

### 1.1.2 Tests on a single synthetic seismogram with a simple realistic model M1

A somewhat more realistic model M1 of a synthetic seismogram, which includes a second spike with 0.1 amplitude and 2 s shift, was created using Matlab, Figs.11-12. The seismogram without padding, Fig.12, was then used for the subsequent analyses presented herein.

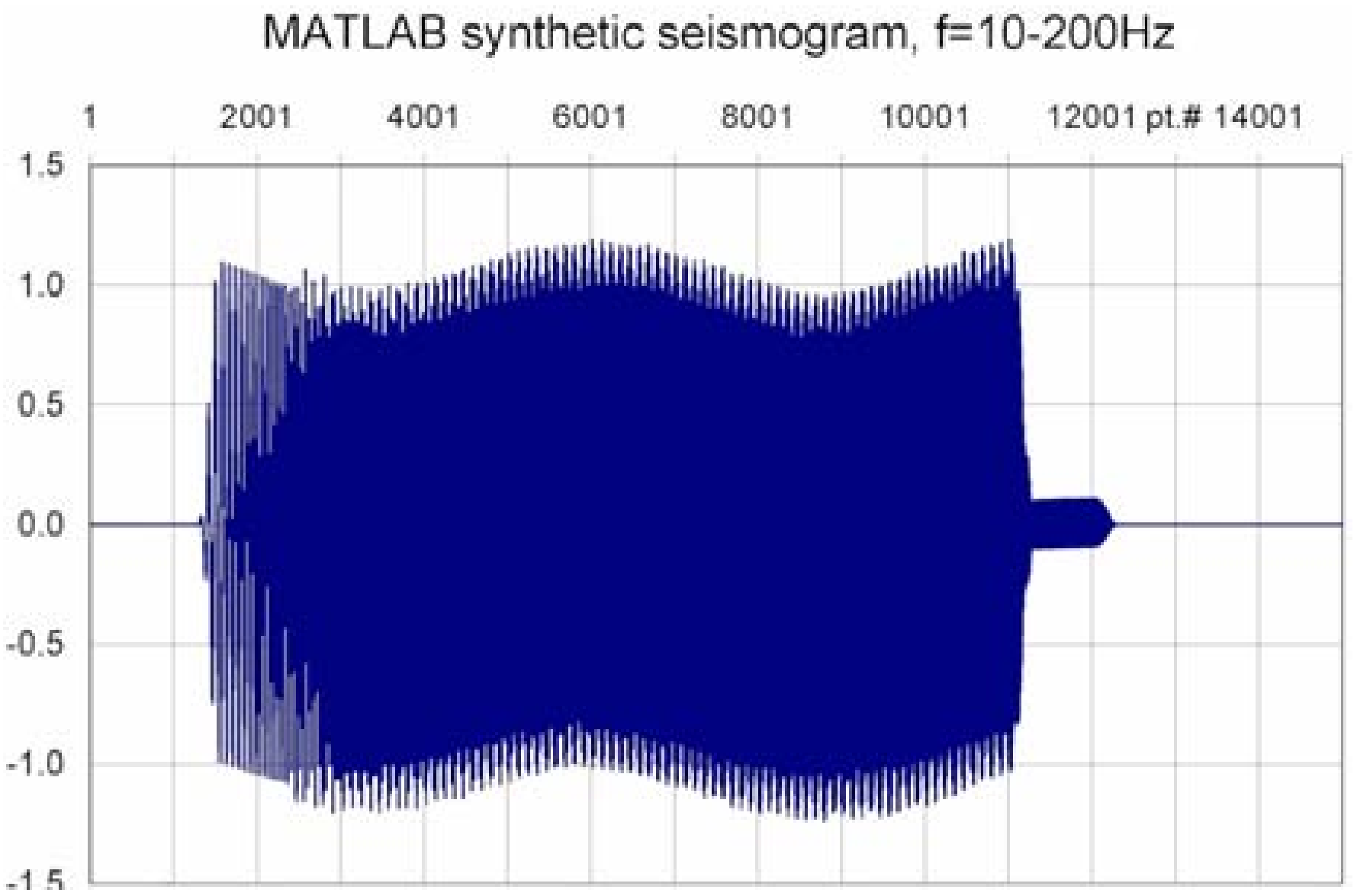

Figure 11. Matlab-created complete synthetic seismogram, zero-padded, model M1.

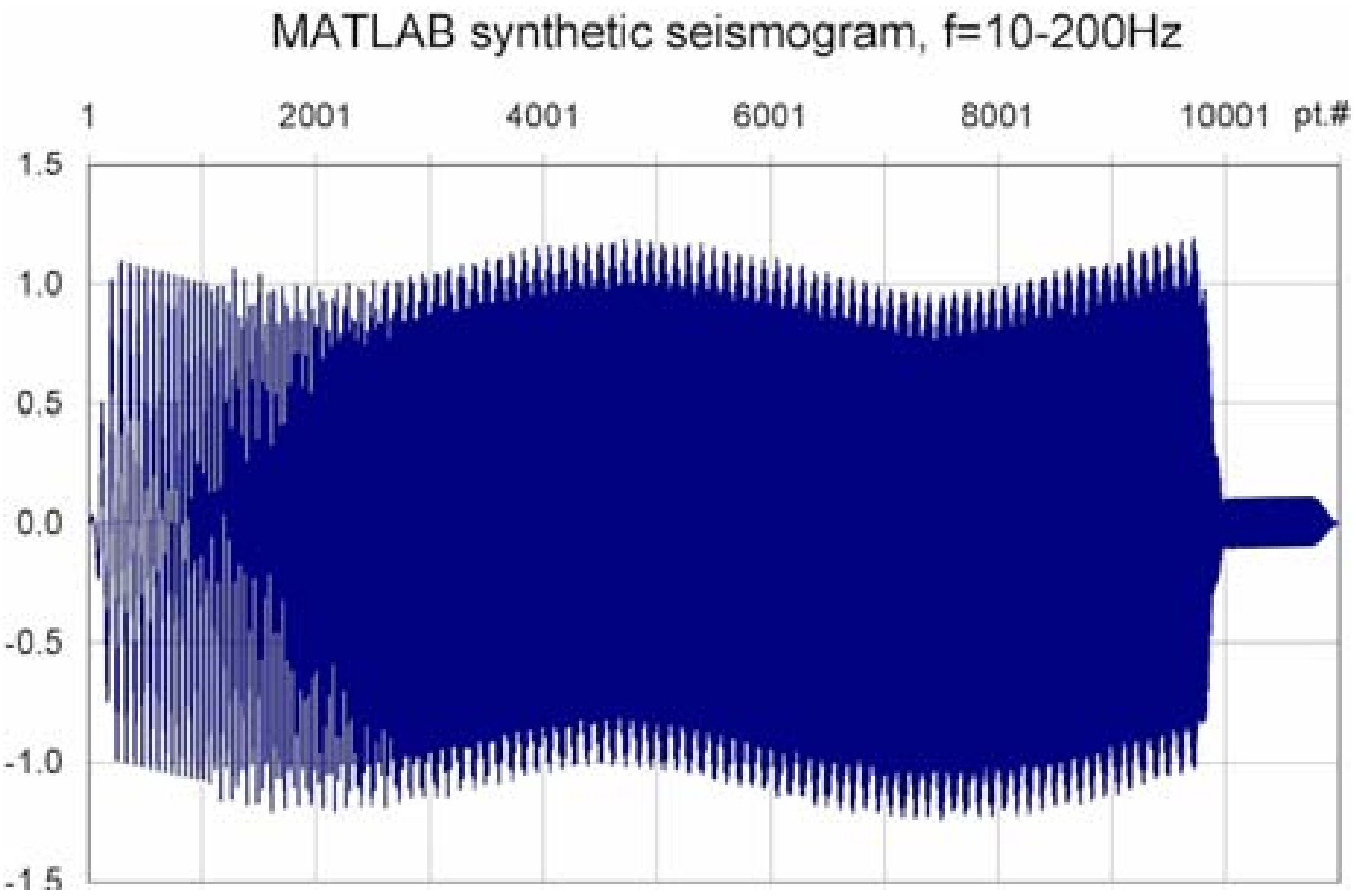

Figure 12. Matlab-created complete synthetic seismogram Fig.11, no padding, model M1.

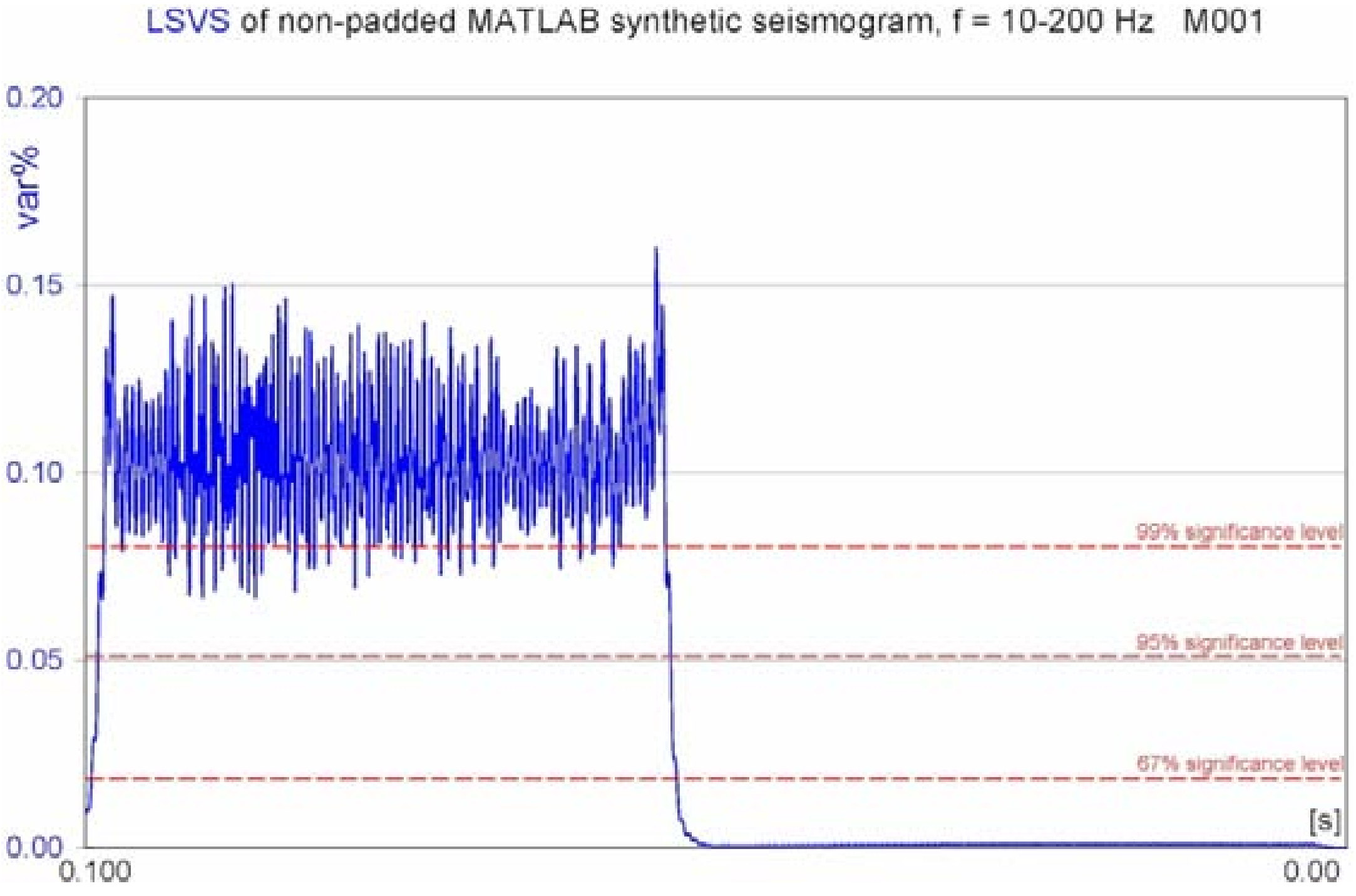

Figure 13.

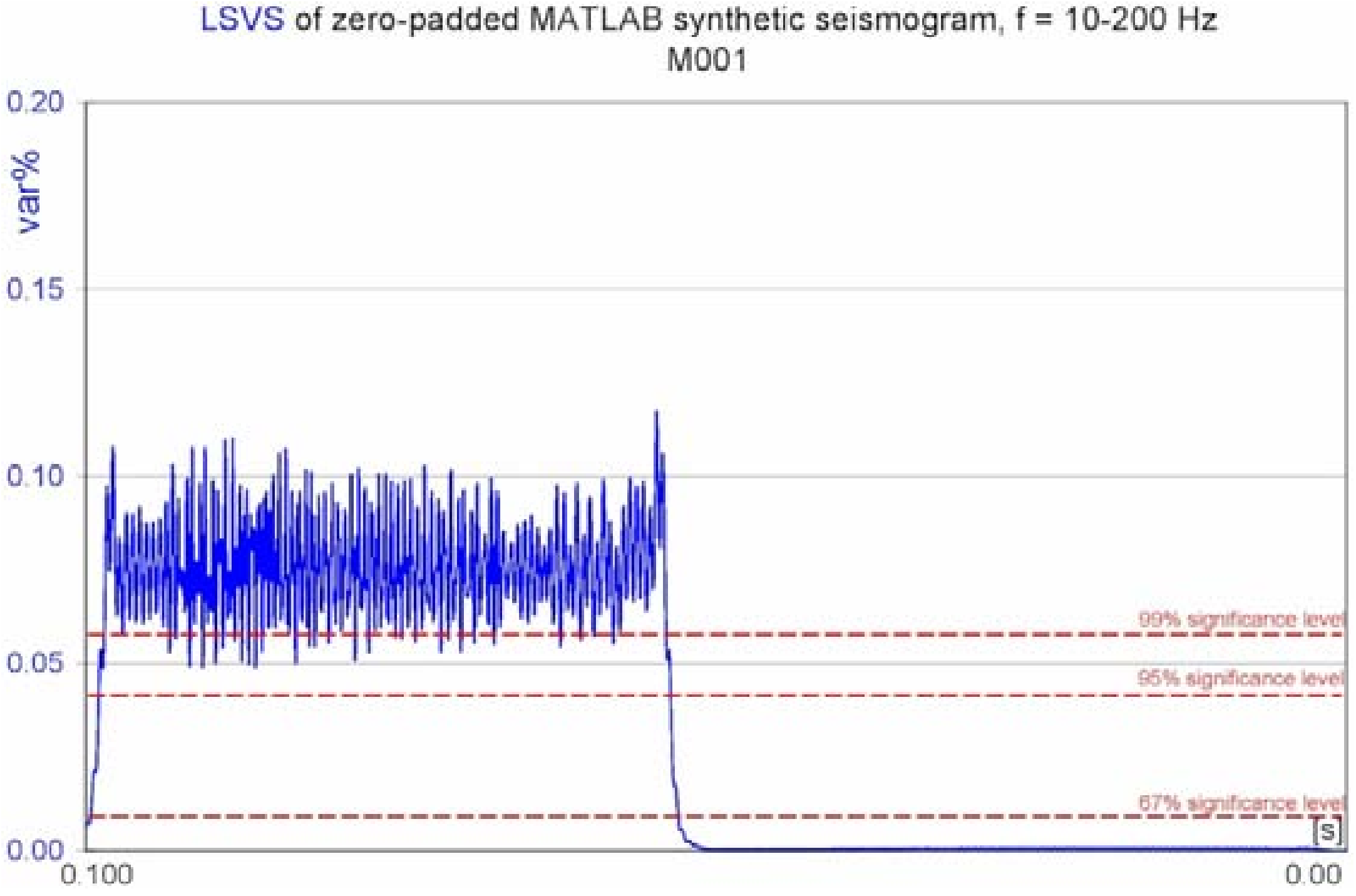

Figure 14.

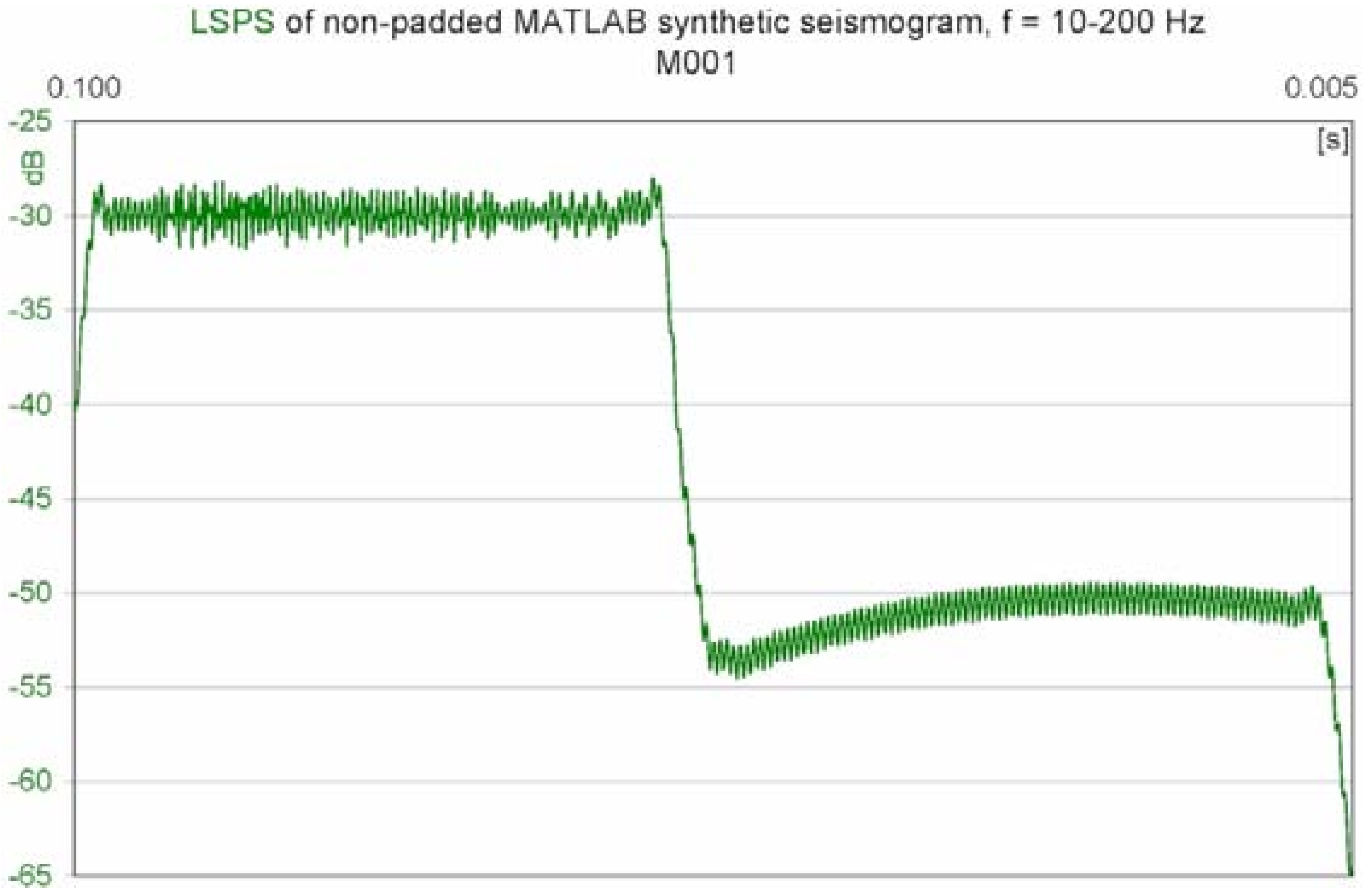

Figure 15.

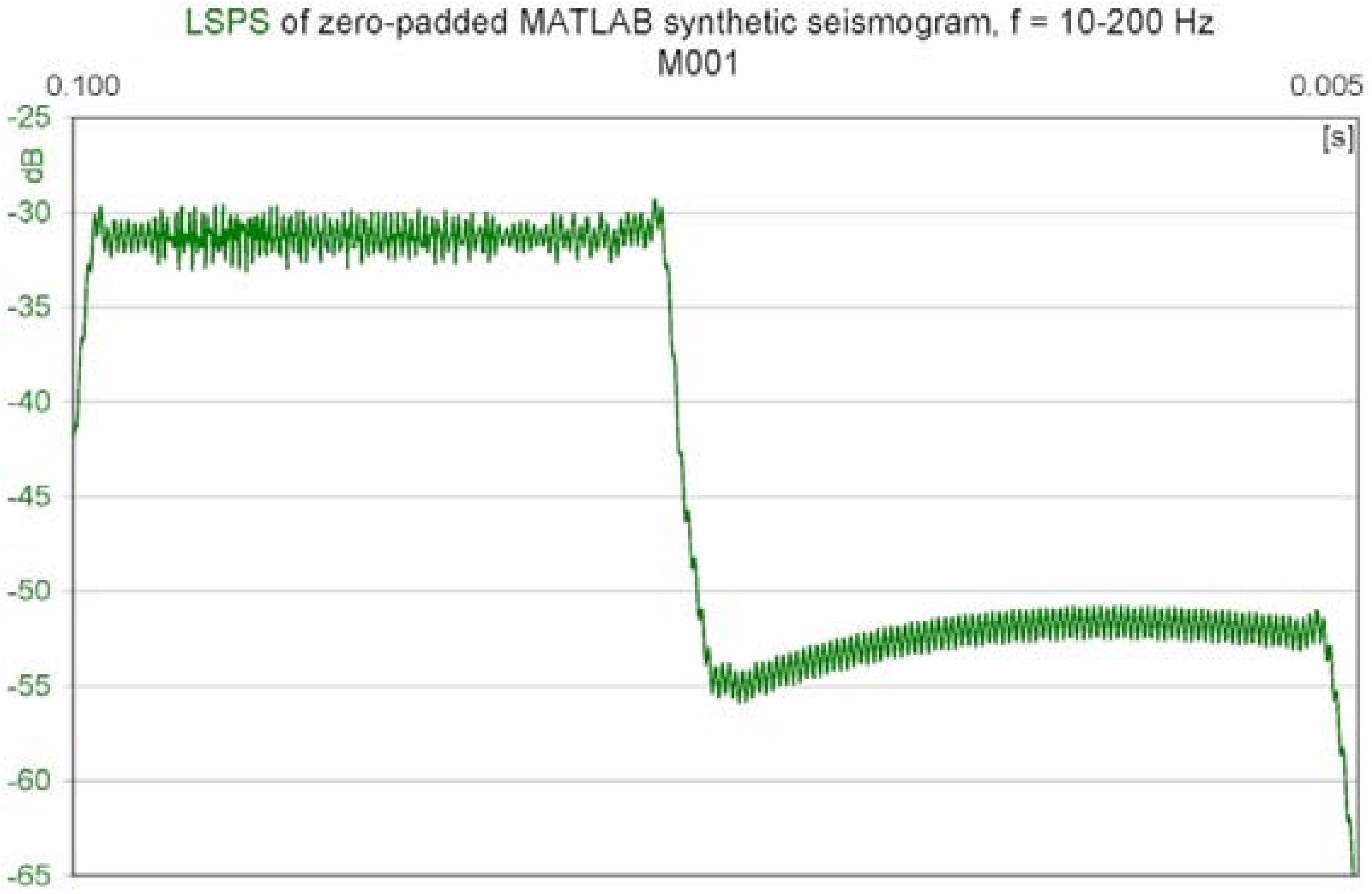

Figure 16.

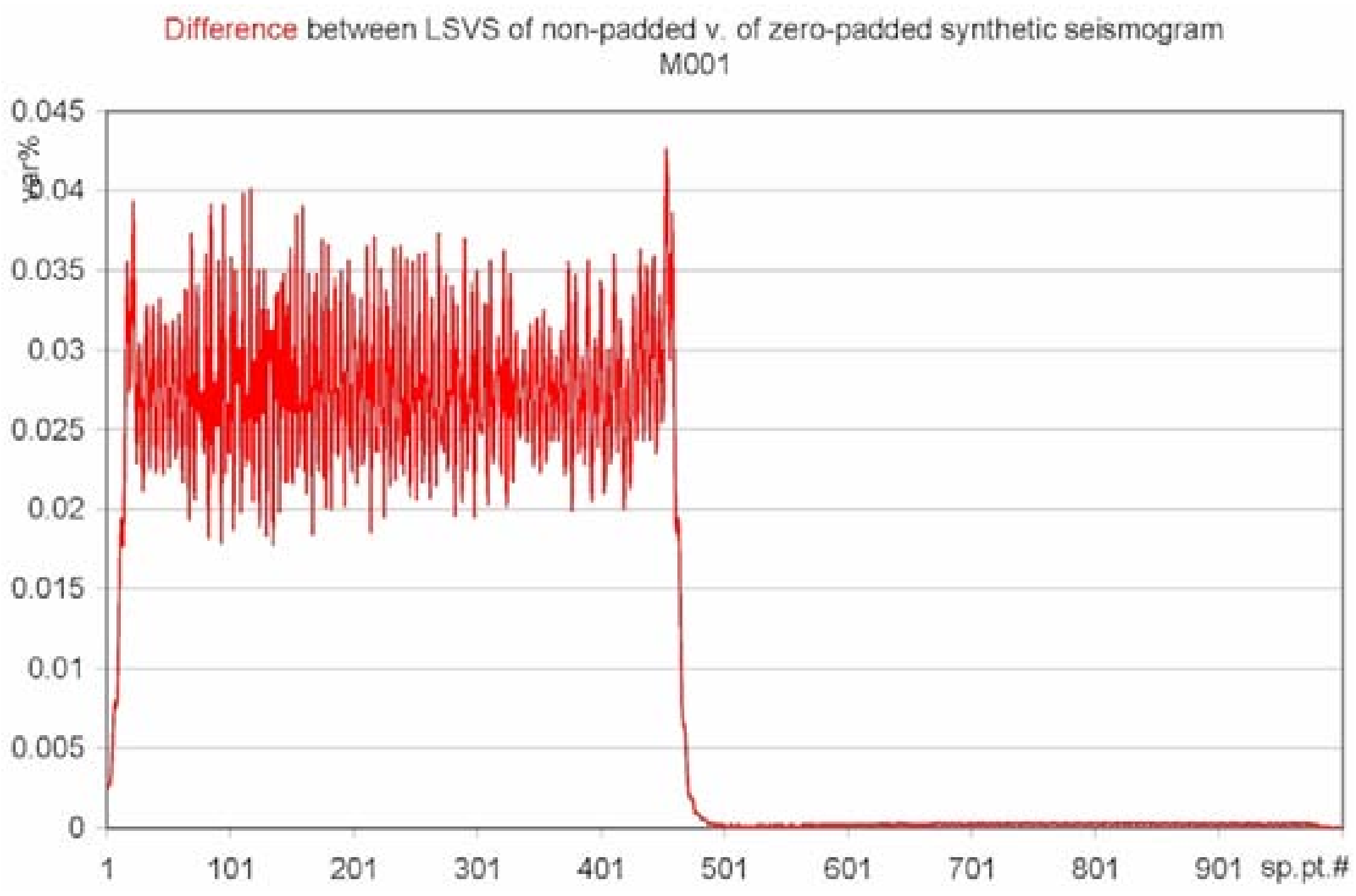

Figure 17.

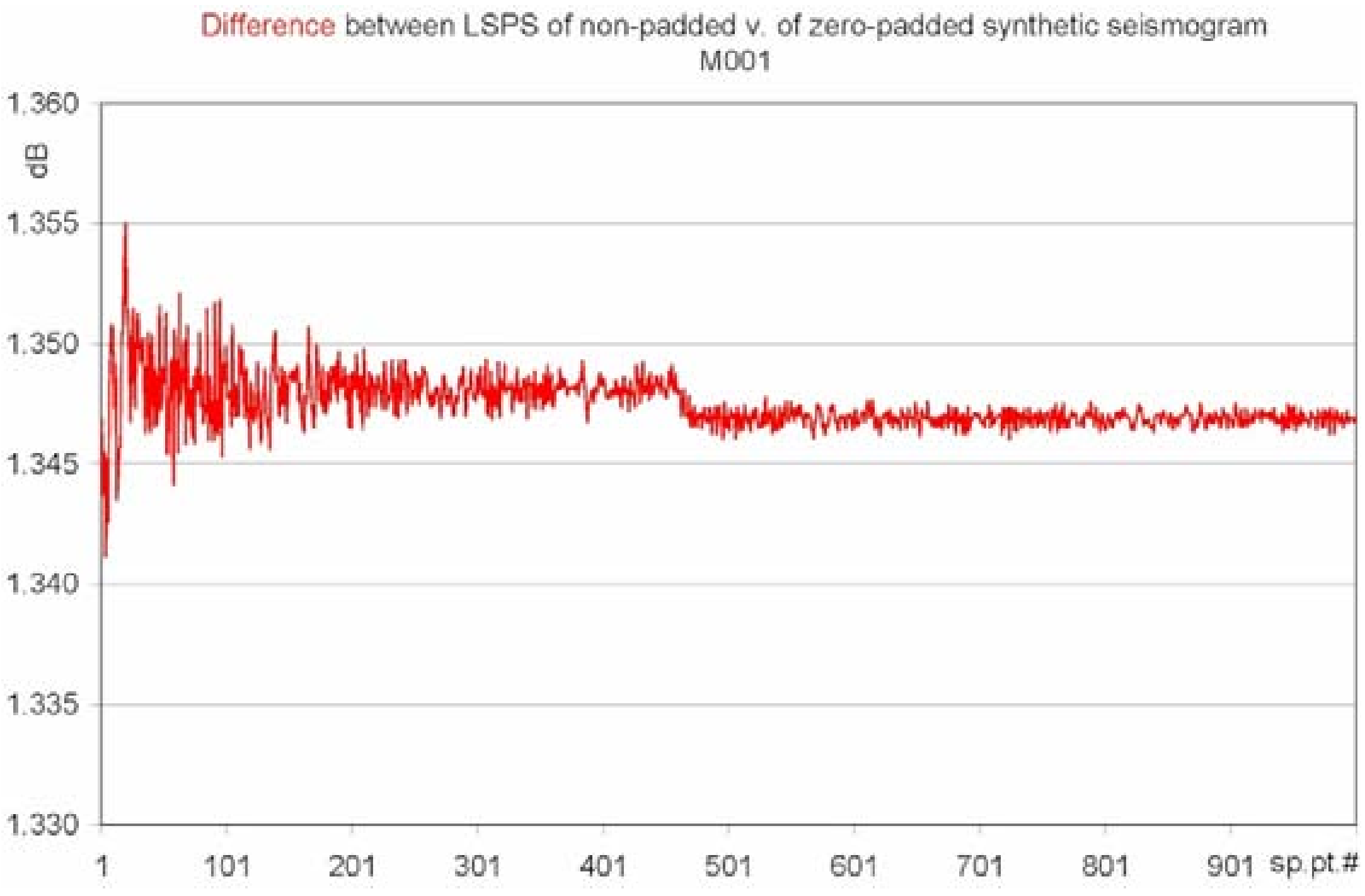

Figure 18.

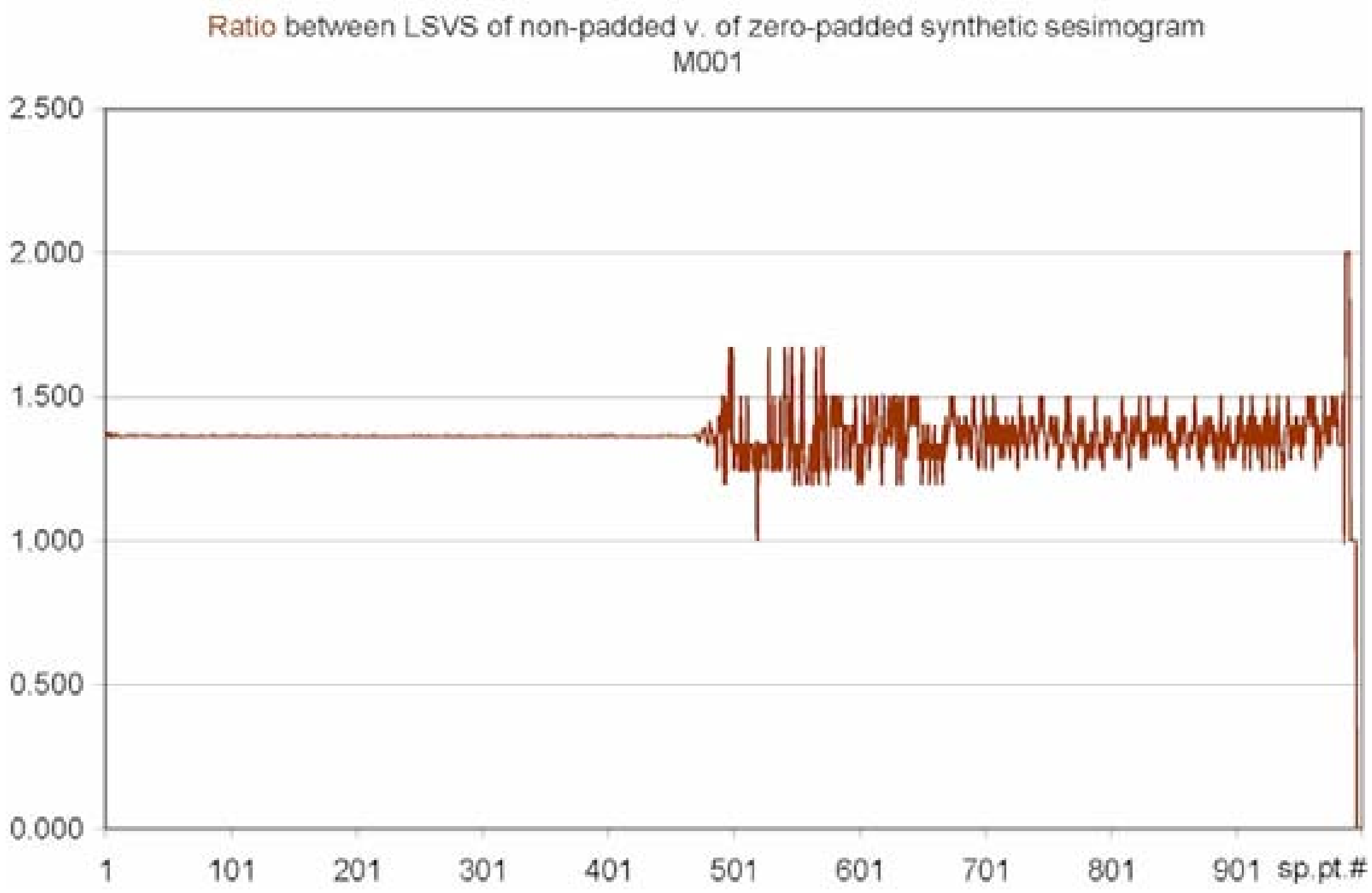

Figure 19.

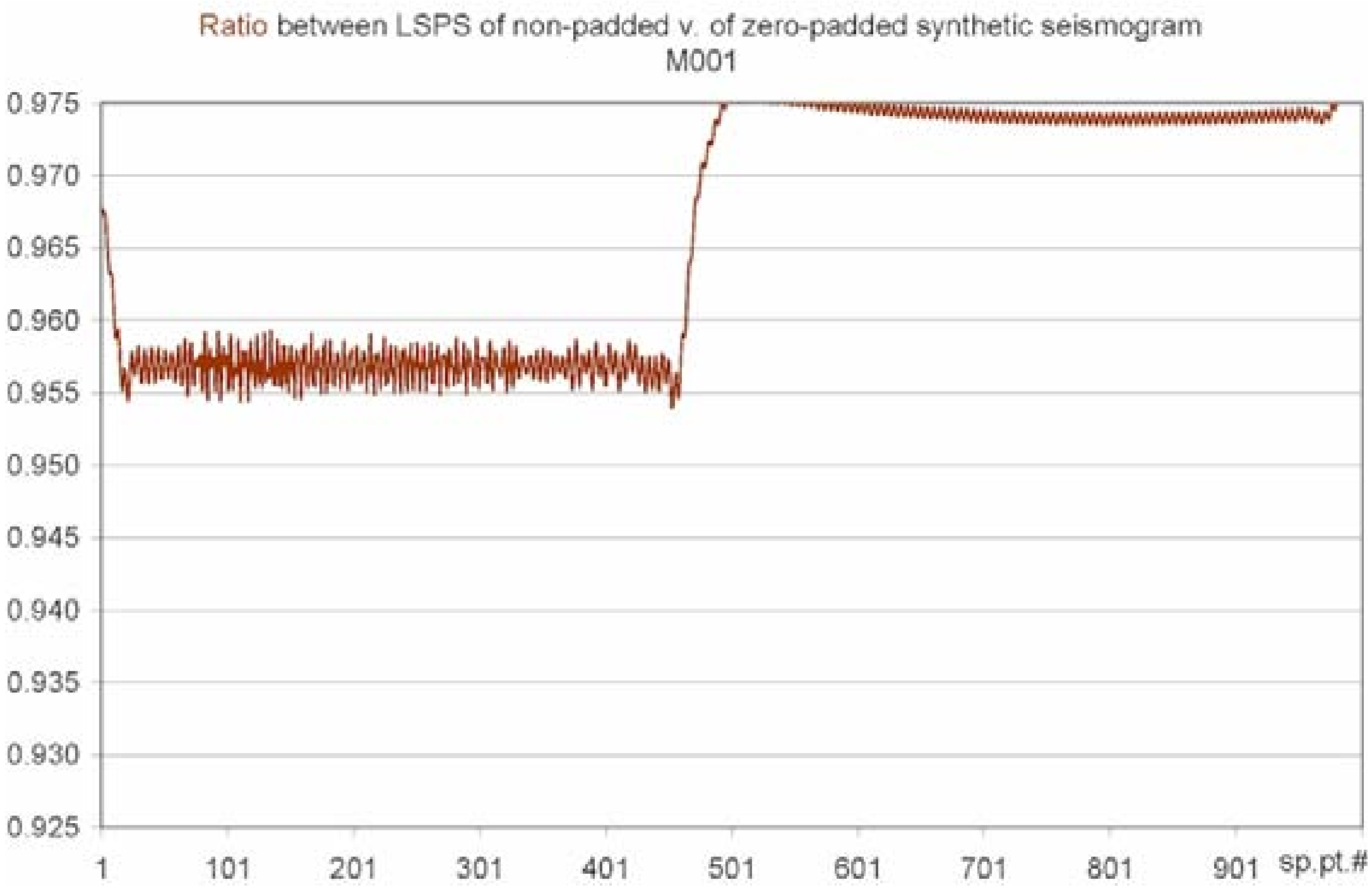

Figure 20.

It can be seen from comparing Figs.13-16 with Figs.3-6, that the GV VS behave in the expected manner for a simple realistic model M1 as well. The GV VS depict the signal as being more realistic than in case of the corresponding no-modeling signal; see Fig.13 v. Fig.3. At the same time, the GV VS responded correctly to an about 30% of data-size increase (as accomplished by zero-padding), which extinguished the classical noise (here: the signal for testing purposes using variance spectrum; see Omerbashich, 2006a). At the same time it shrank the signal envelope from about 0.8 var% to about 0.4 var% – virtually twice the extinguishment compared to the test's benchmark (the no-modeling scenario). On the other hand, the GV PS appeared insensitive to data scaling, albeit only half as responsive to the artificial power boost from the zero-padding as in the no-modeling case (2dB under M1 model v. 4dB for no-modeling; see Figs.15-16 v. Figs.5-6).

Thus, for a small number of harmonic-noise constituents, the variance-spectra' information extinguishing "ability" (or disability, depending on one's preference for signal v. noise) as well as the power-boost "benefit" (or detriment, as accordingly), fade away inversely proportionately with the increase in number of harmonic noise constituents.

Figs.17-20, same as benchmarked, again depict varying residual behavior of the GVSA, accurately all the way down to the method's ability to pick up a purely theoretical signal solely imbedded within the information of interest.

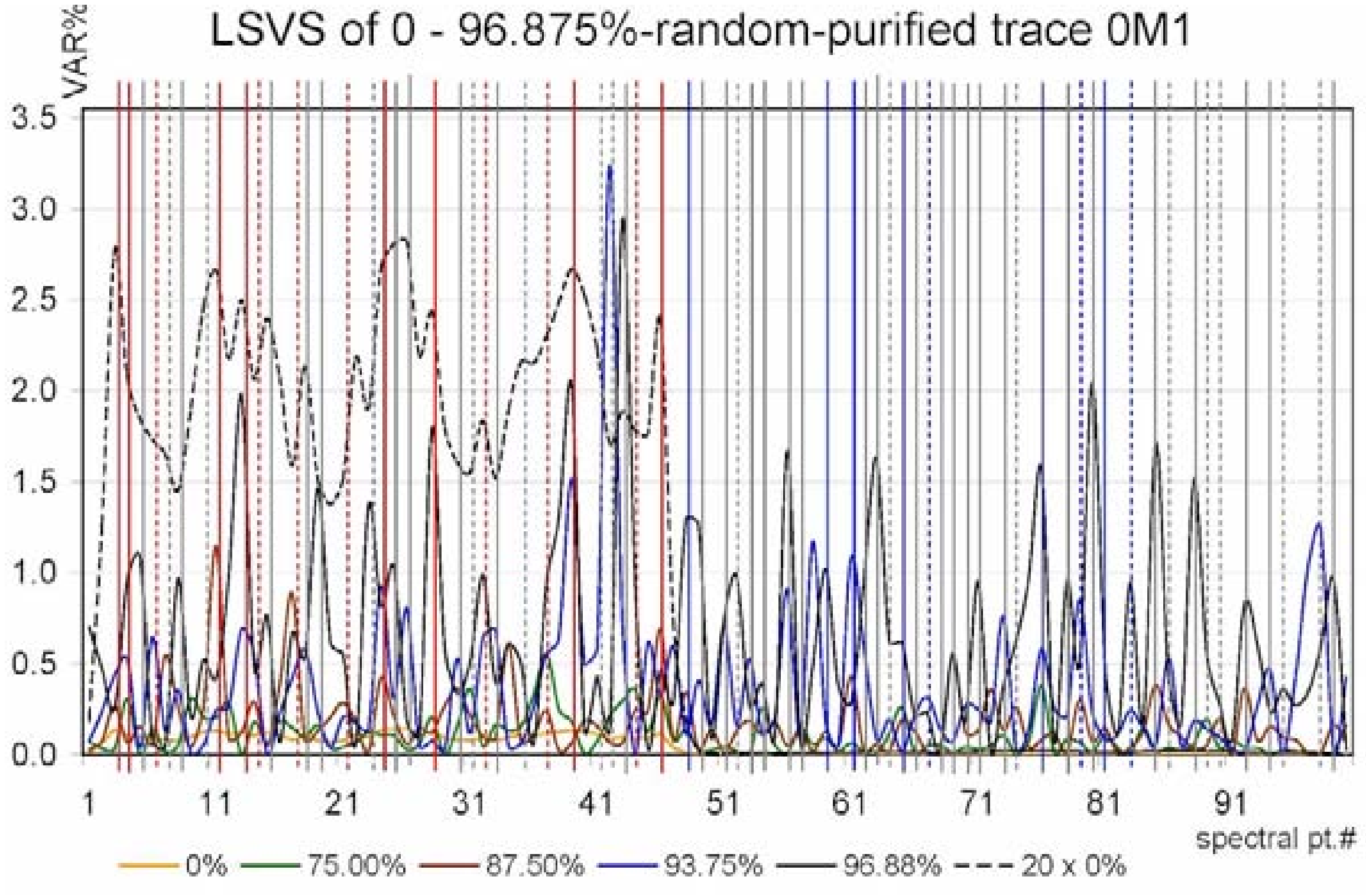

Figure 21. <u>*Common caption to Figures 21-32:*</u> Visual matching. Dashed are matches from any peaks resolved imperfectly but such that they still could be discerned from the shape or/and the slope of the respective peak. Low frequencies red, high frequencies blue.

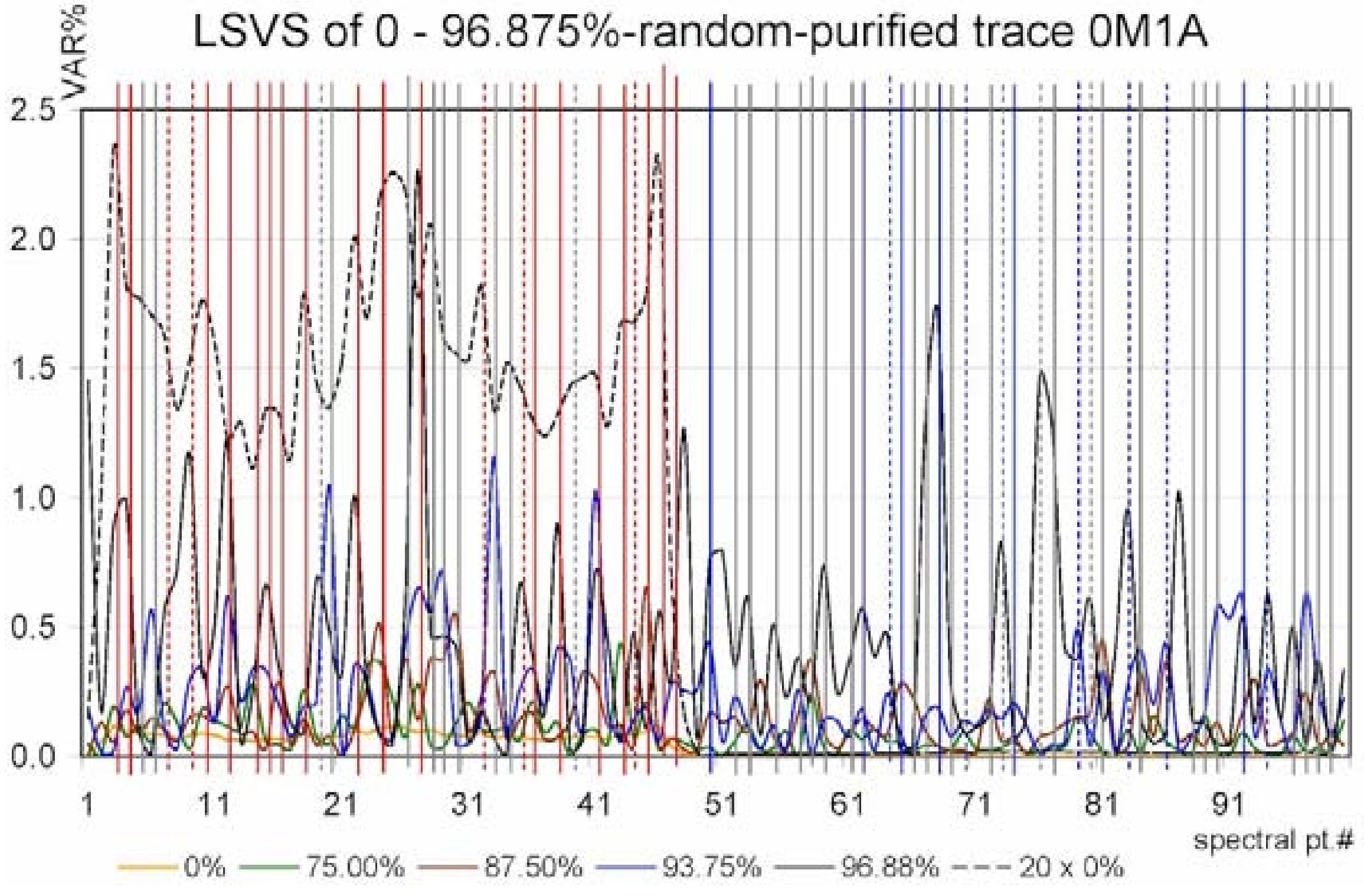

Figure 22.

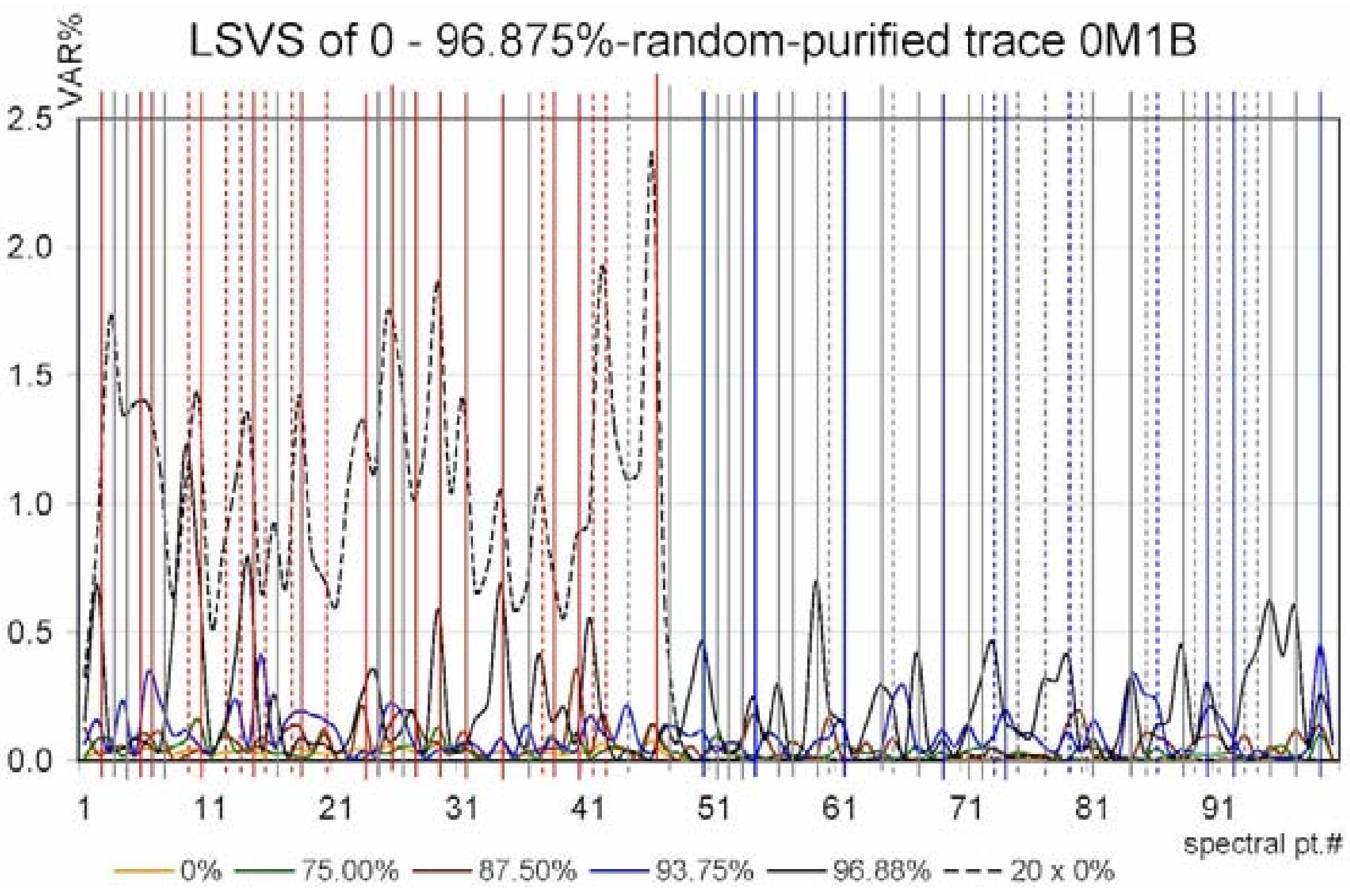

Figure 23.

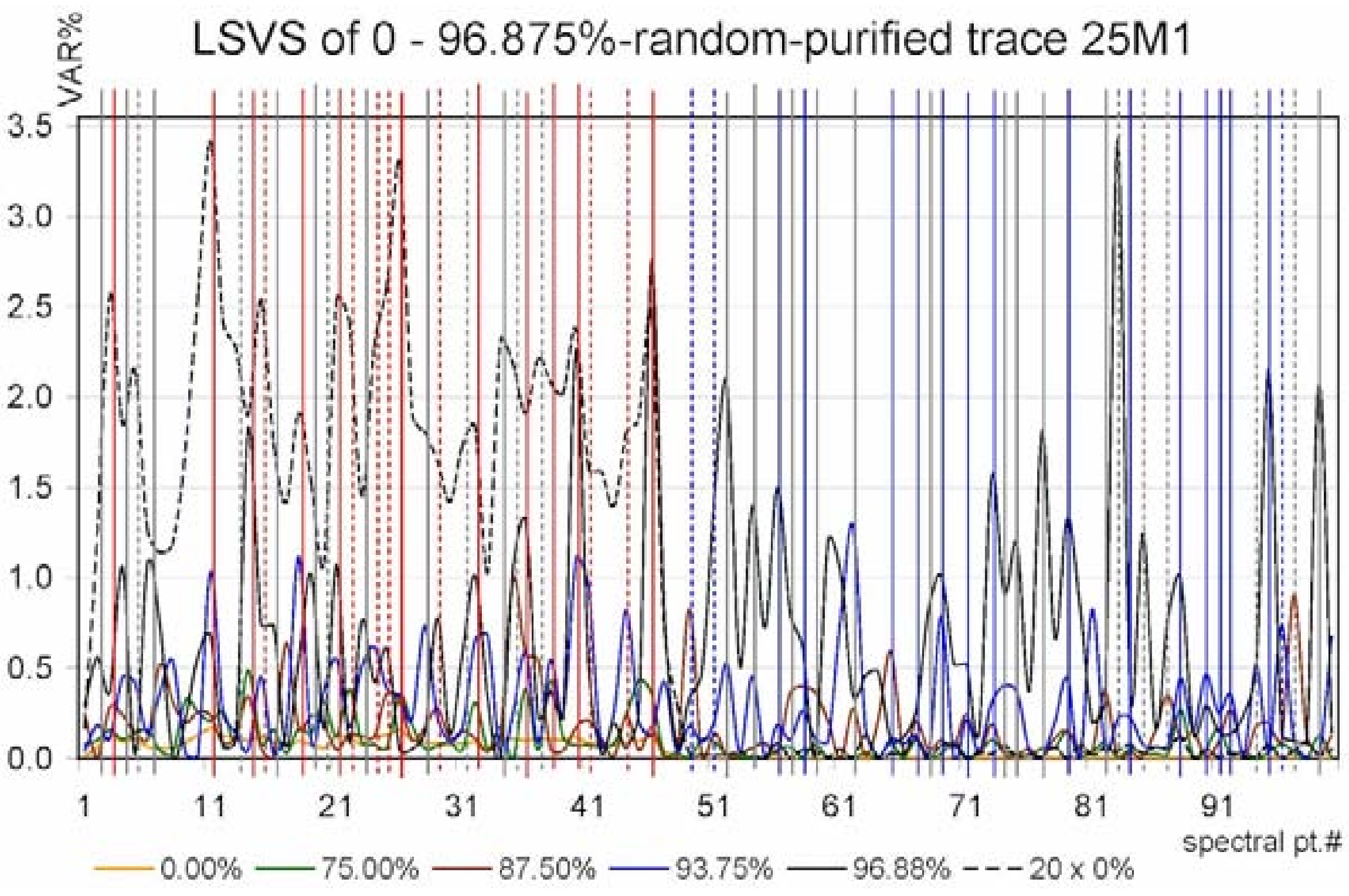

Figure 24.

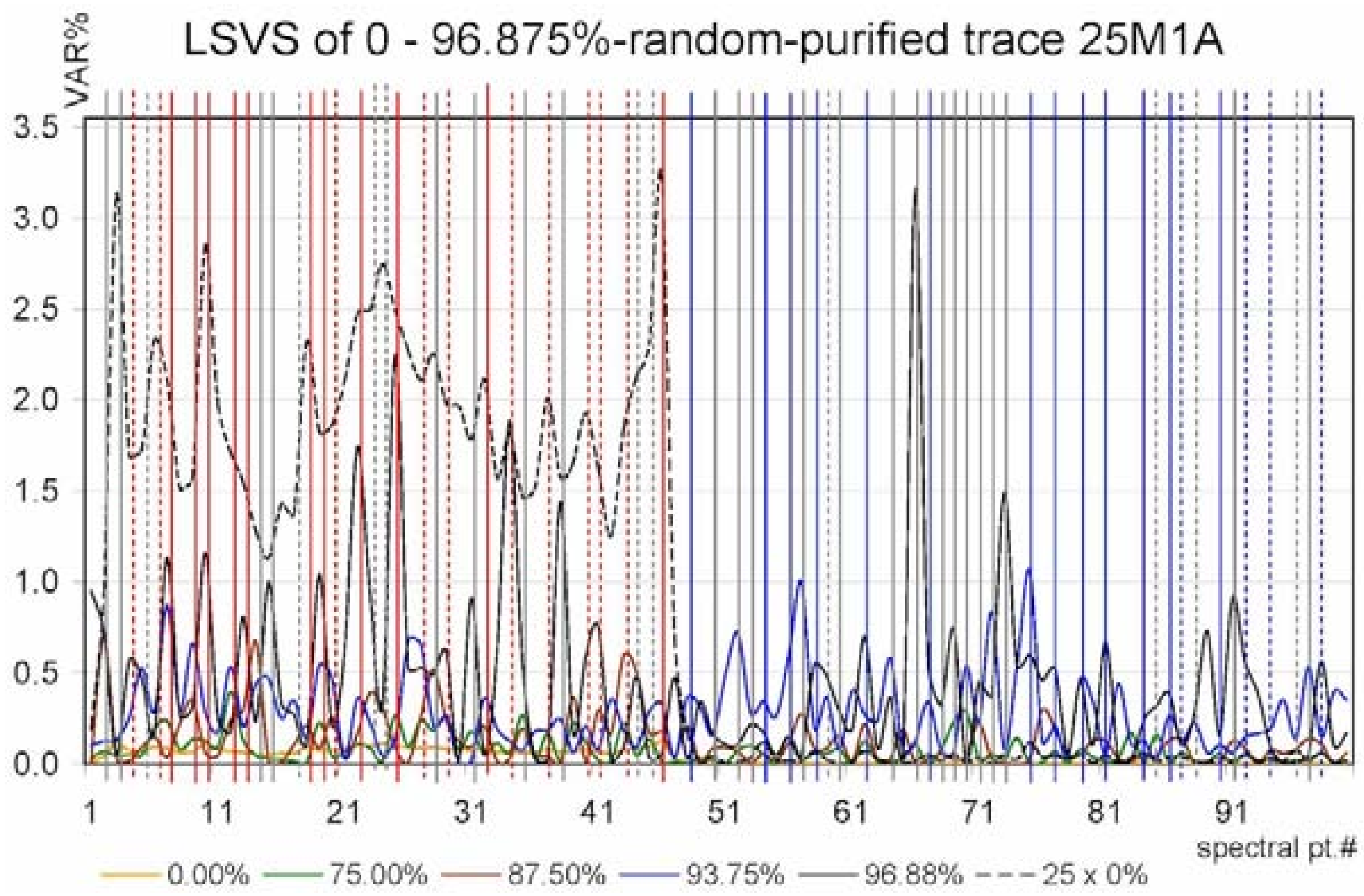

Figure 25.

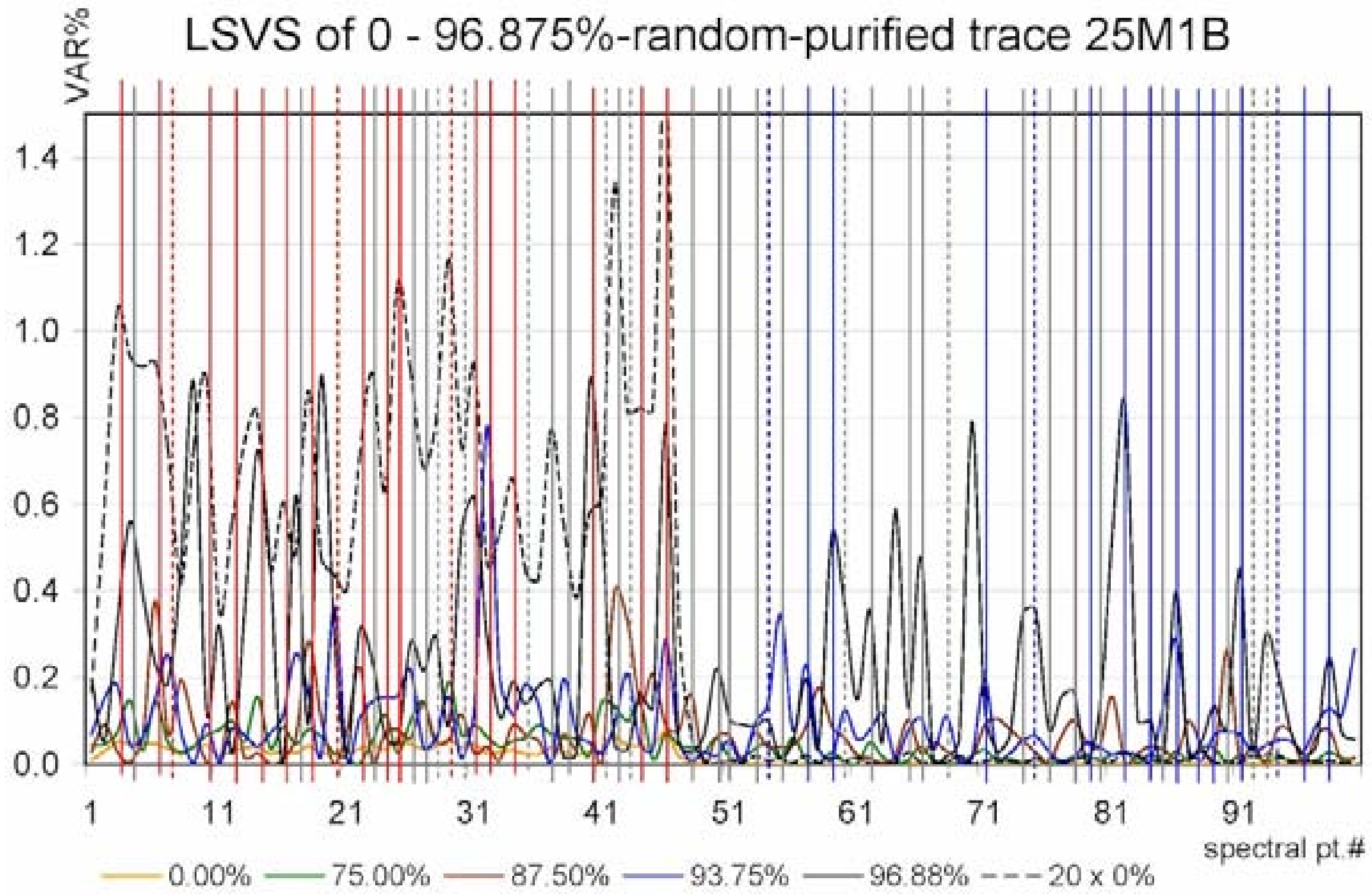

Figure 26.

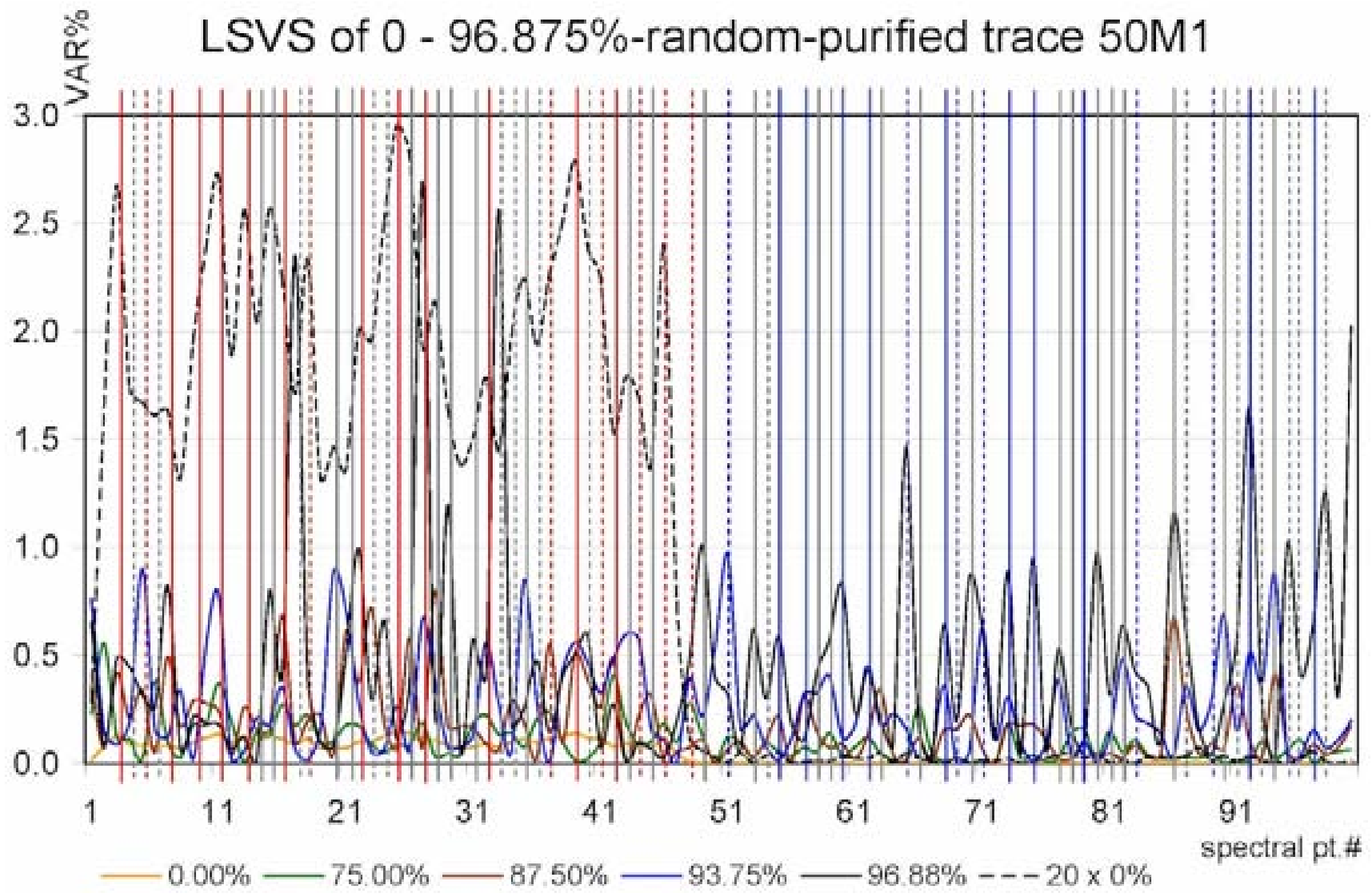

Figure 27.

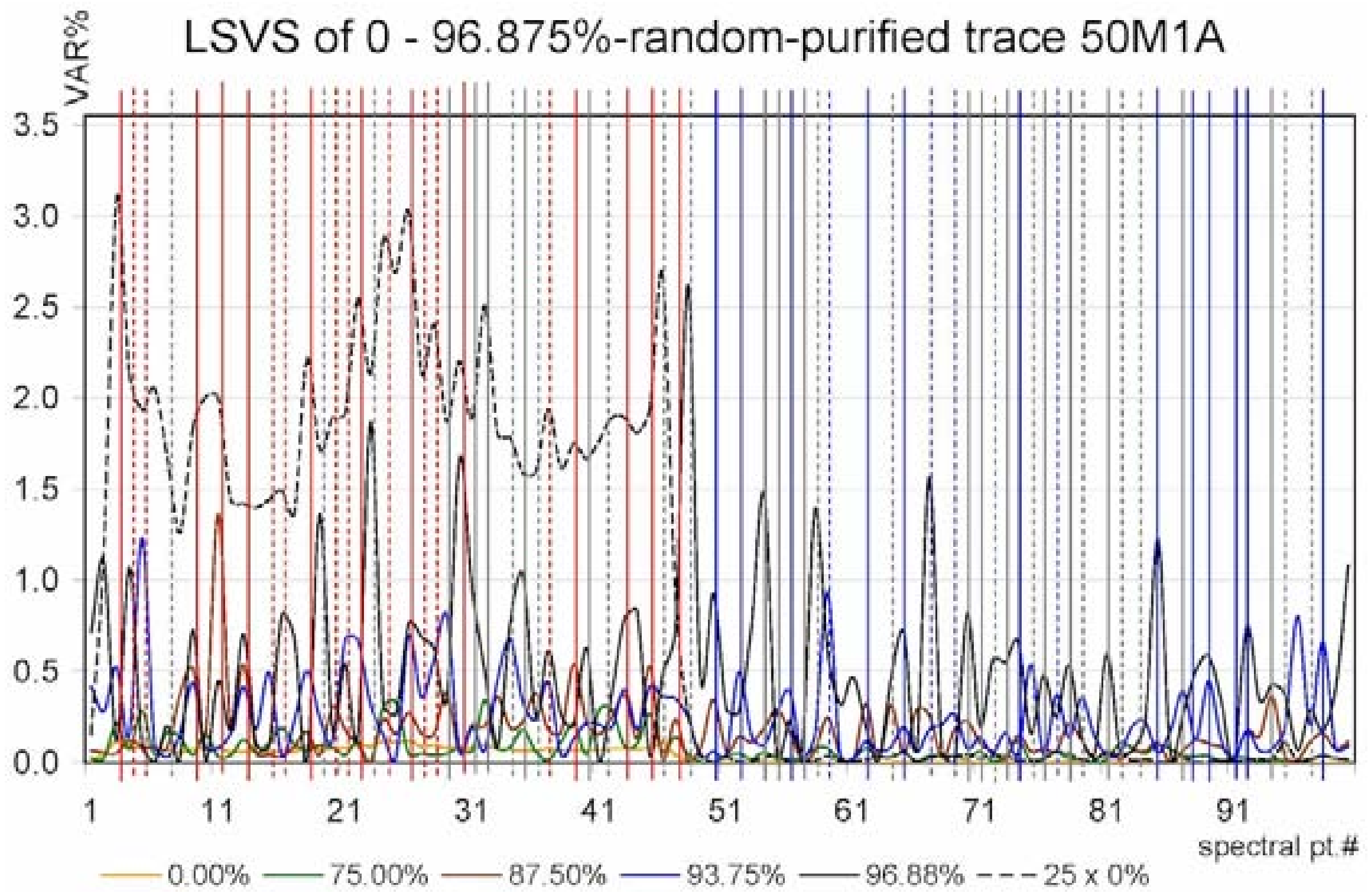

Figure 28.

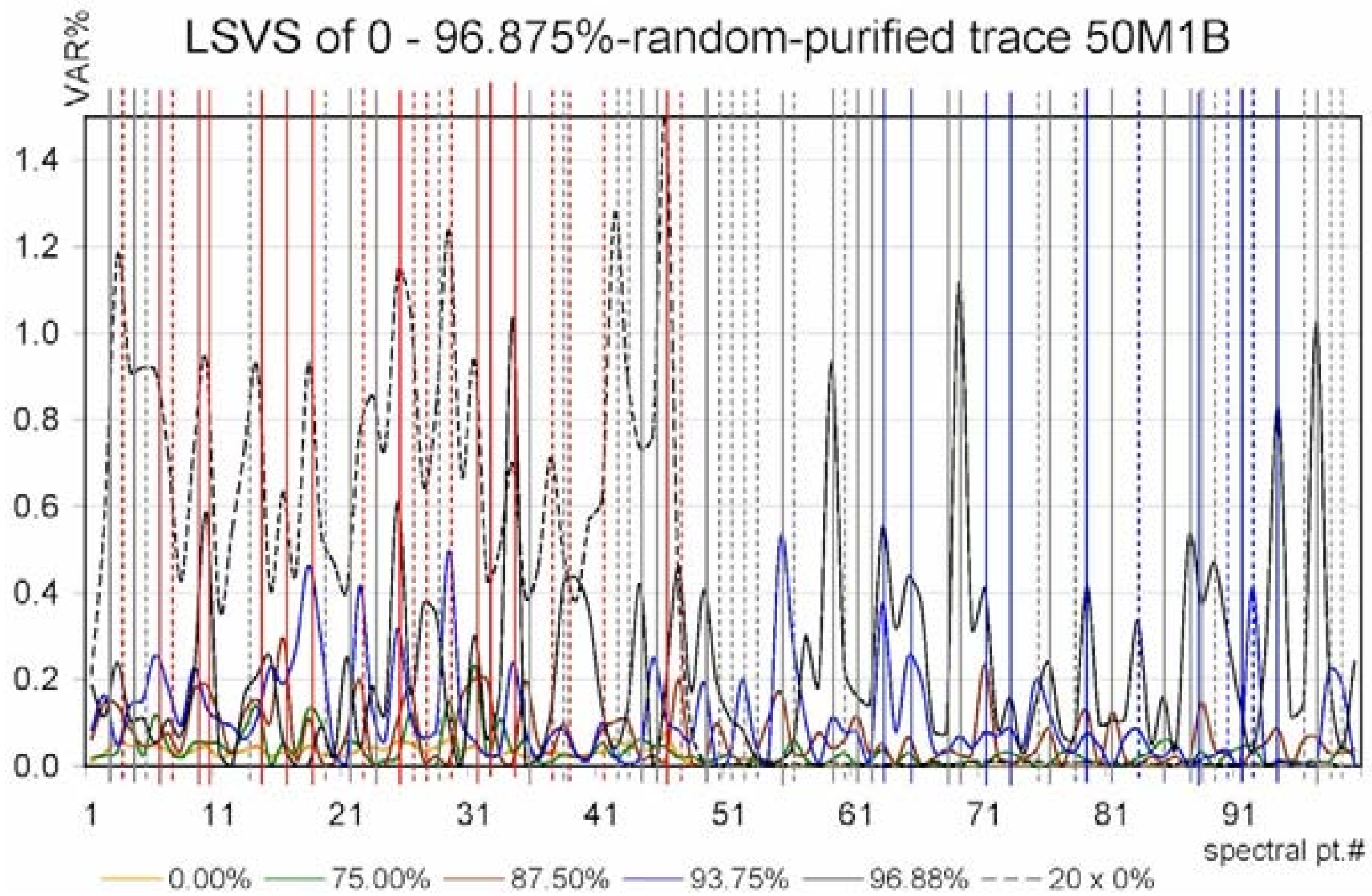

Figure 29.

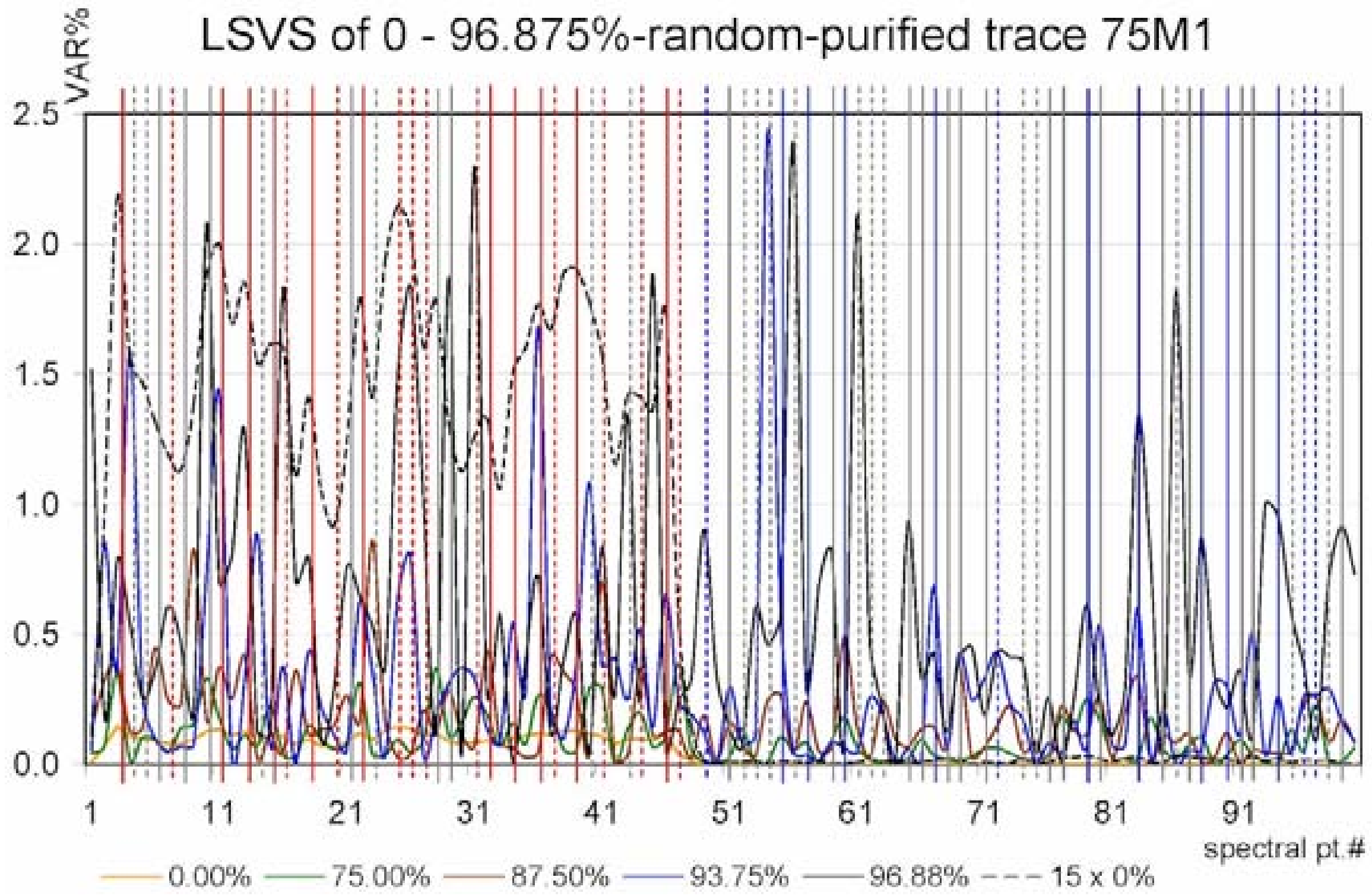

Figure 30.

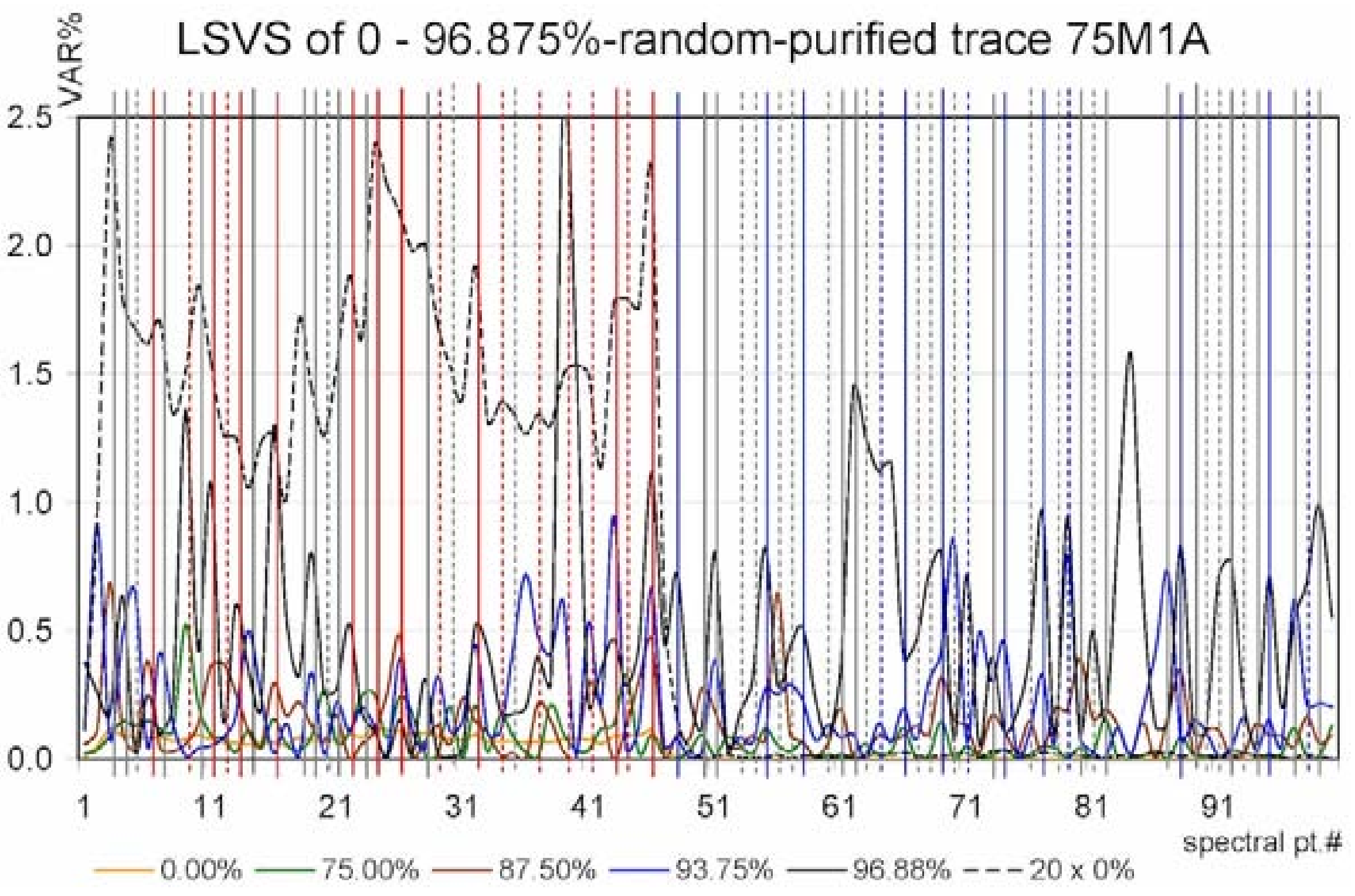

Figure 31.

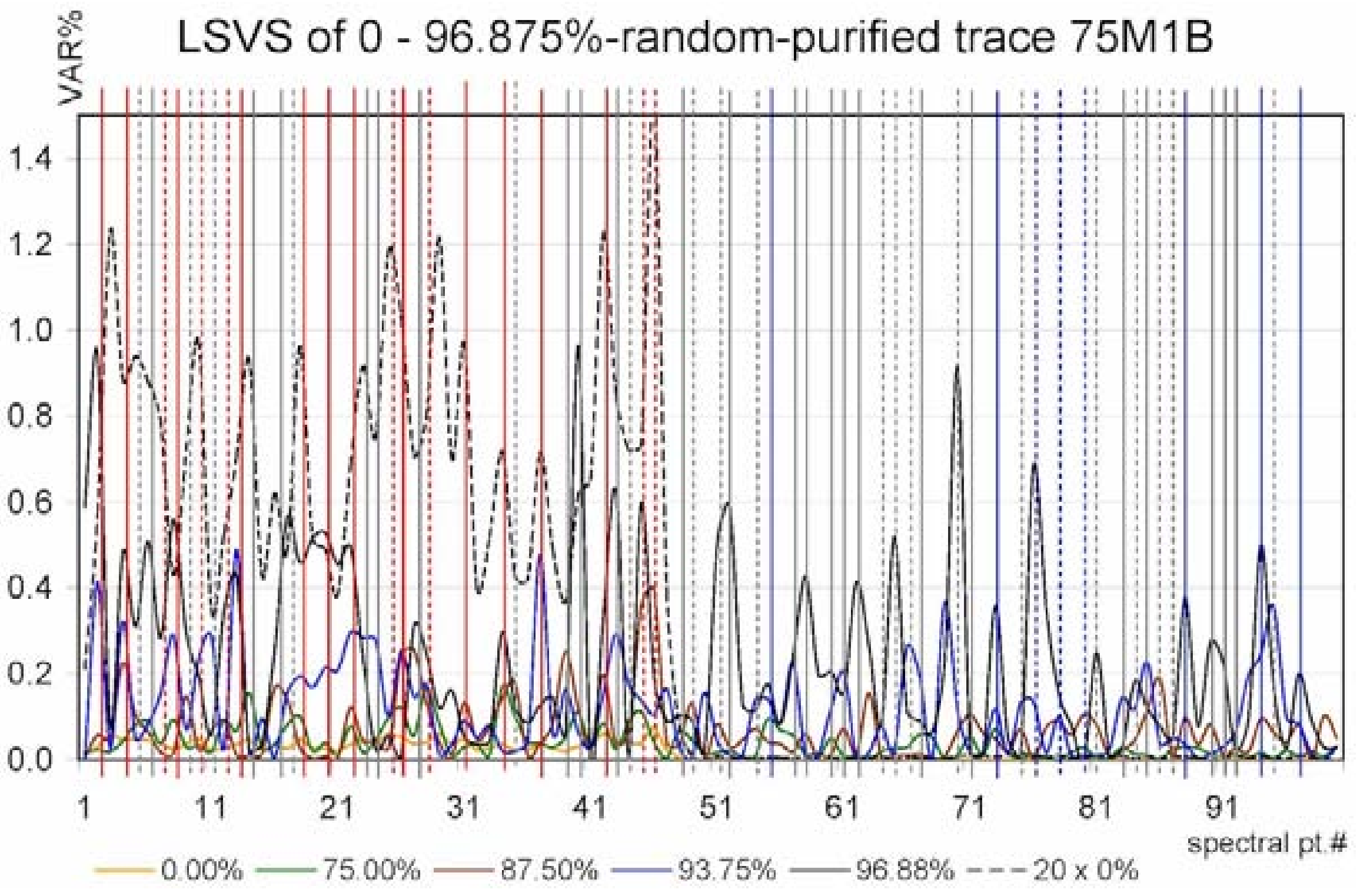

Figure 32.

<u>Additional views of Fig.5</u>

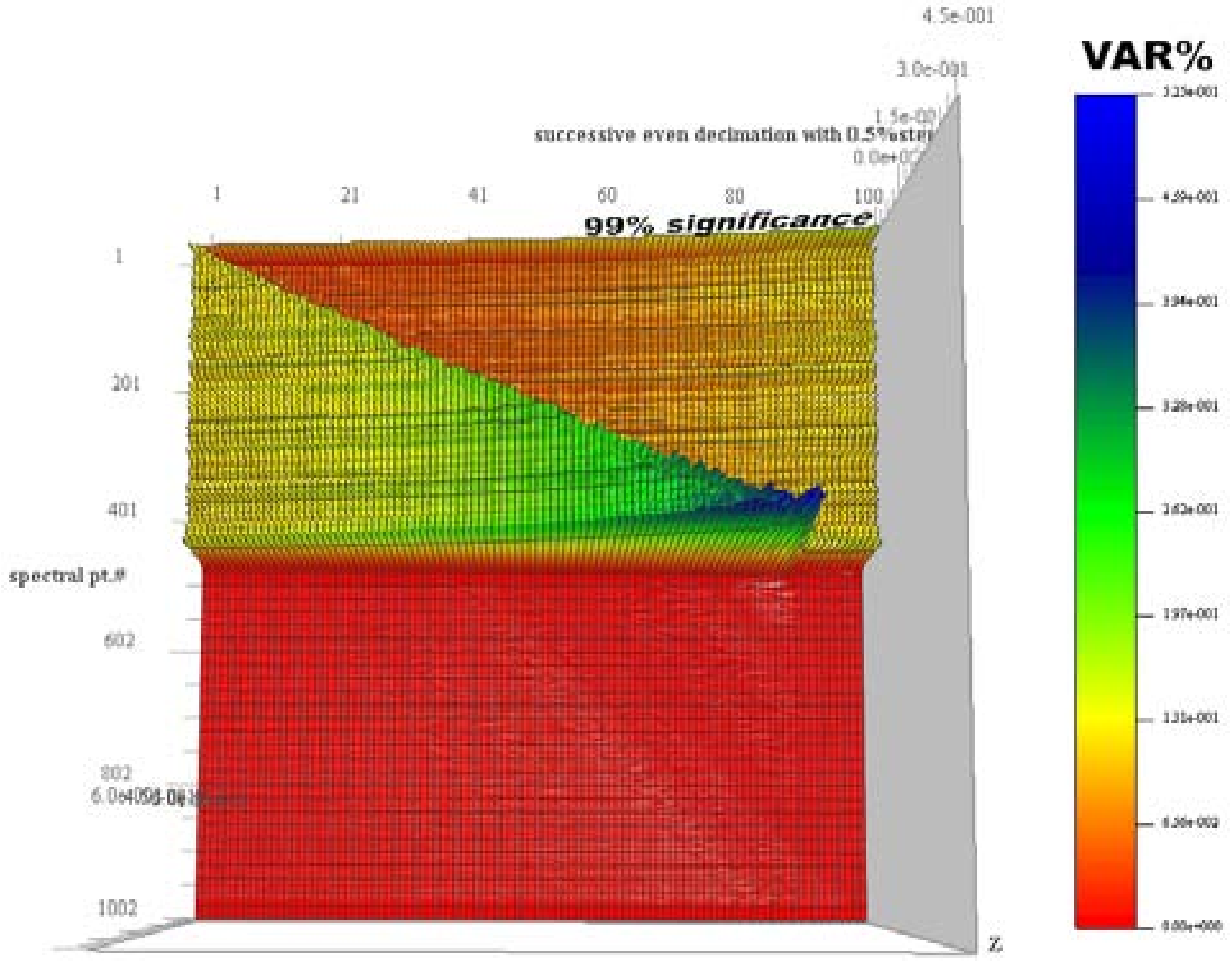

Perspective view.

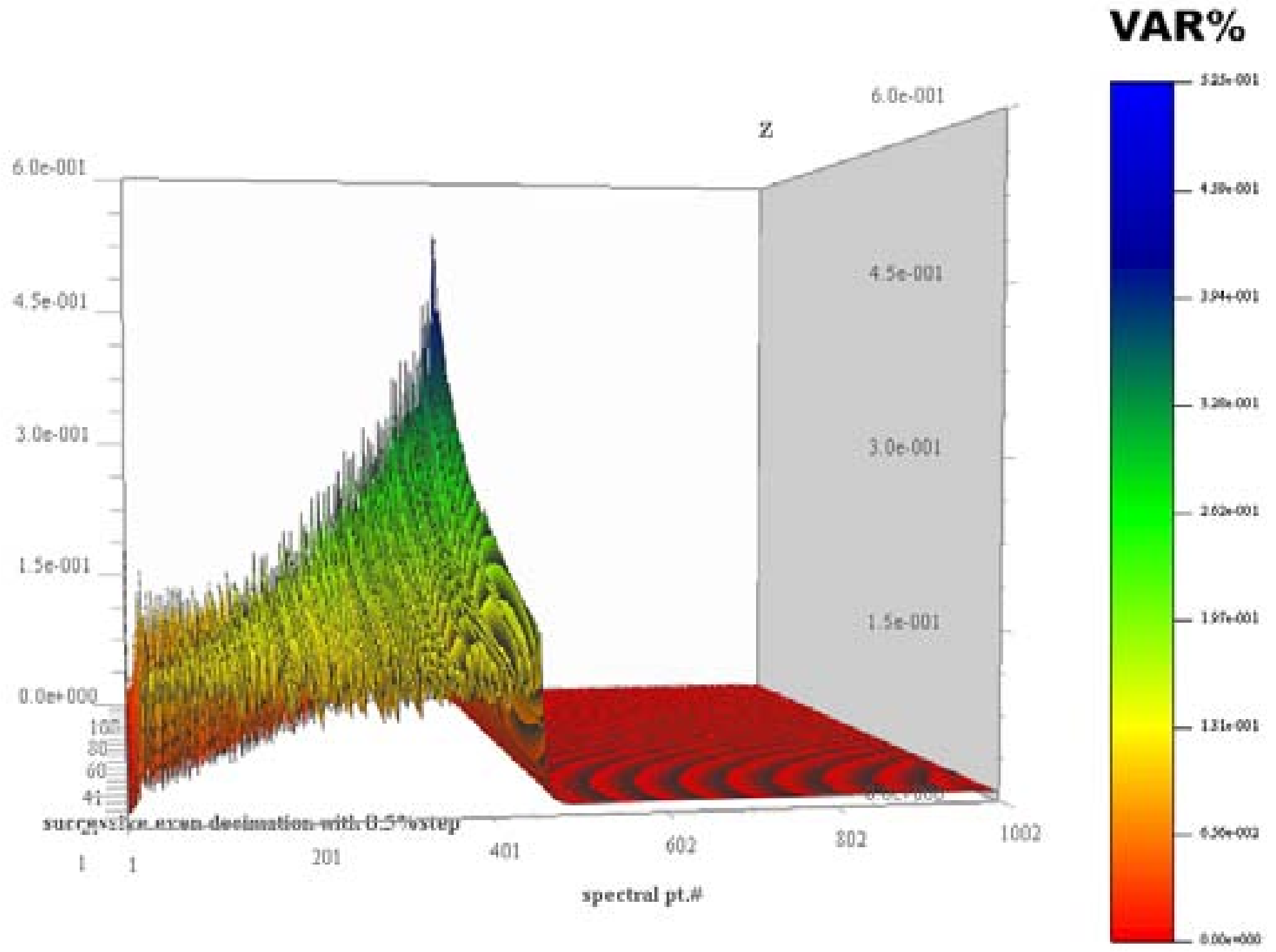

Front view.

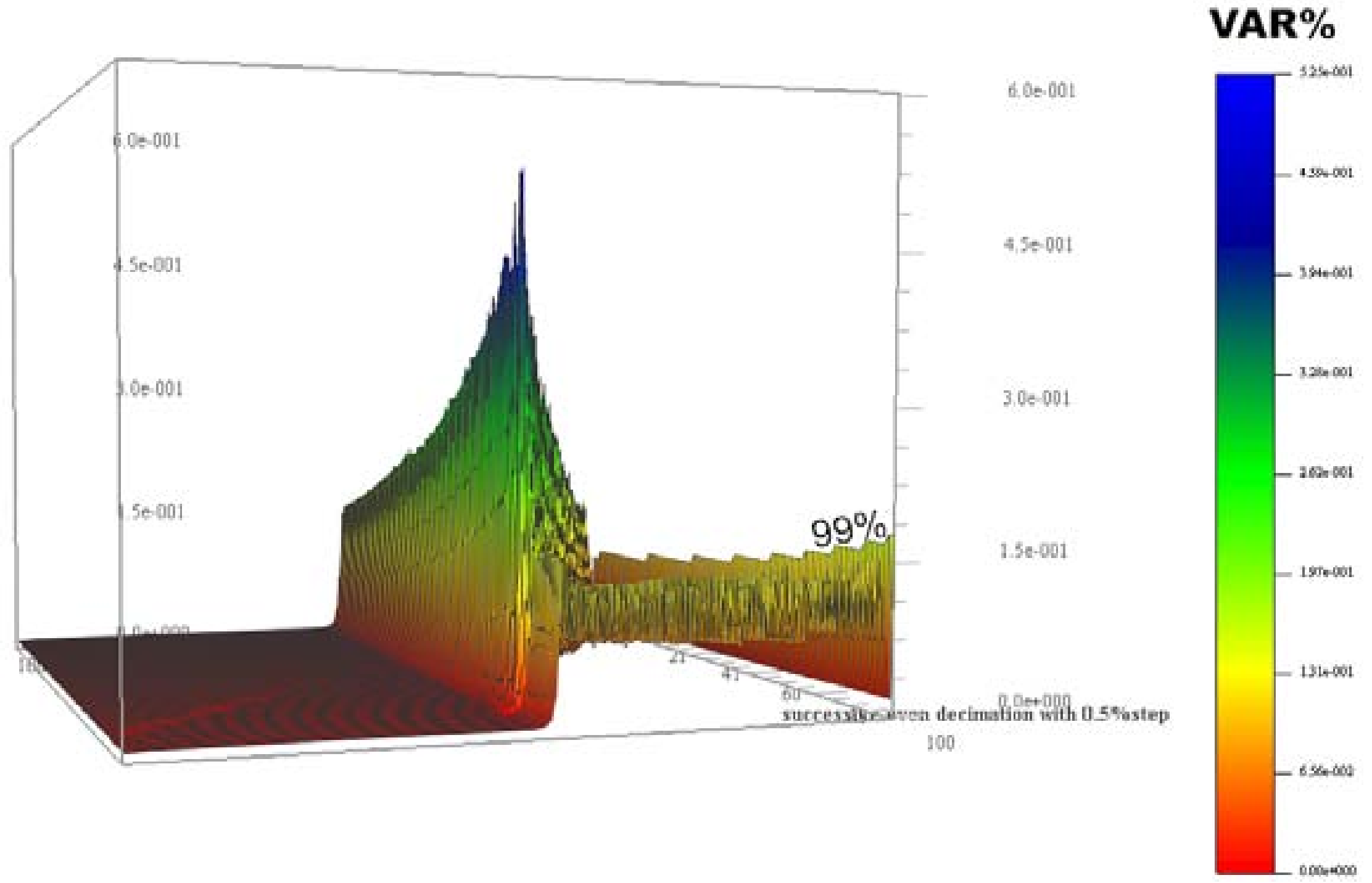

Rear view.

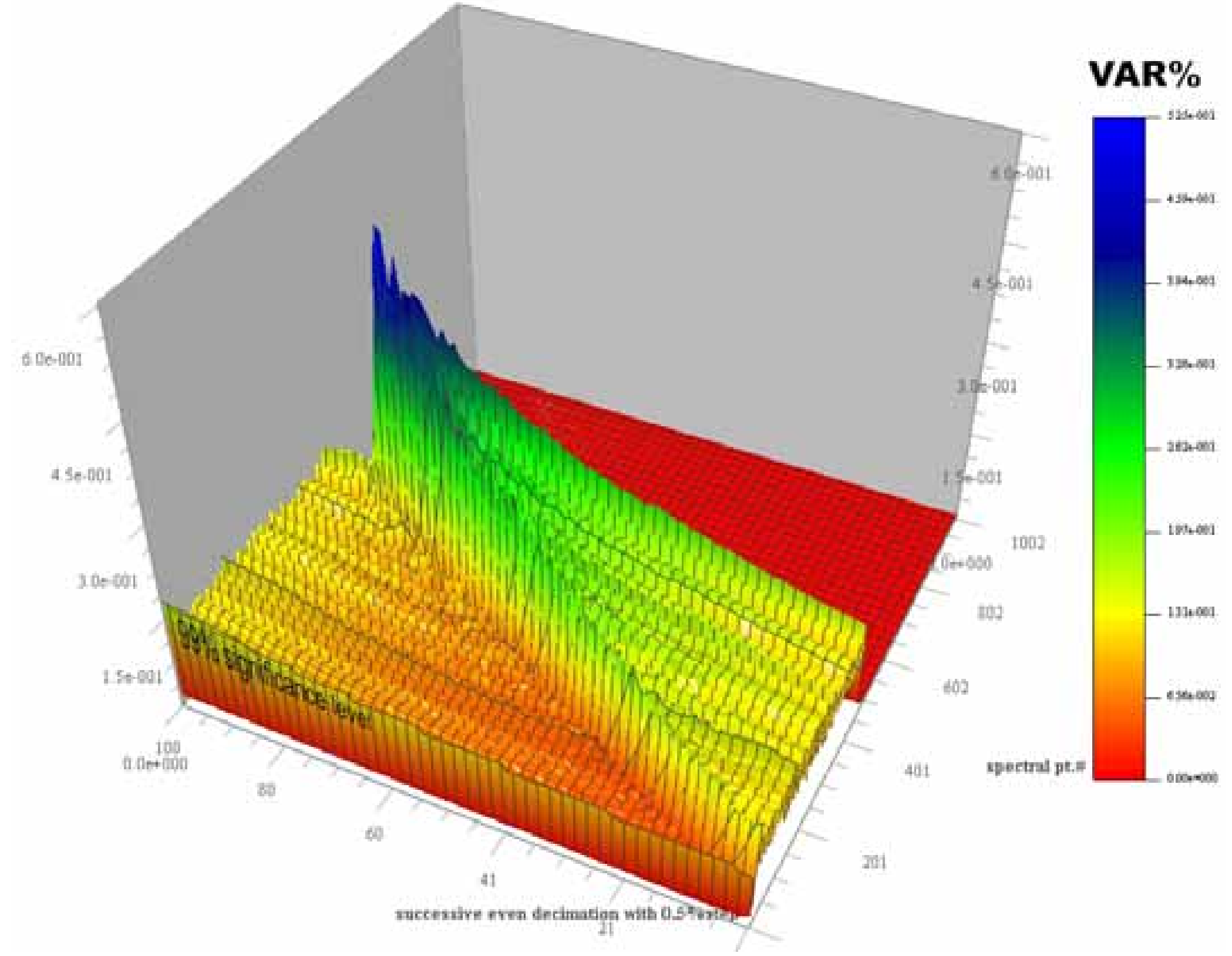

Perspective-side view.

## Computer coding

For examining the herein discussed concepts under various windowing setups, I wrote Matlab codes, Listings 1&2, which automate also the process of results extraction into separate files that are ready-made for use in the complex scientific software GVSA v.5.0.

```
% Arbitrarily windowing a series for a preset window-size
ws=550;C=1;SS=load('c:\rsrch\tests\A00.txt');n=int16(numel(SS));p=idivide(n,ws,'floor');
disp(n);disp(p);
for k=1:p;
   for i=1:ws;SW(i,1)=SS((k-1)*ws+C+i,1);end;
   F(k)=getframe;plot(SW);xlabel('window pt.#');
   sp=sprintf('c:\\rsrch\\tests\\test01_10-200Hz\\03A_windowing\\syntseis_w%d.txt',k);
   fid=fopen(sp,'wt');fprintf(fid,'%14.9f\n',SW);fclose all;clear SW;
end;
movie(F,1,5);
```

Listing 1. Matlab code for automated creation and exporting of data subsets created by windowing for a preset size of the window. Includes movie visualization for real-time overseeing of the windowing process. (By M. Omerbashich)

```
% Arbitrarily windowing a series for a preset number of windows
p=10;C=0;
for k=1:p;
SS=load('c:\rsrch\tests\A00.txt');n=int16(numel(SS));ws=idivide(n,p,'floor');disp(n);
disp(ws);
    for i=1:ws;SW(i,1)=SS((k-1)*ws+C+i,1);end;
    F(k)=getframe;plot(SW);xlabel('window pt.#');
    sp=sprintf('c:\\rsrch\\tests\\test01_10-200Hz\\03A_windowing\\syntseis_w%d.txt', k);
    fid=fopen(sp,'wt');fprintf(fid,'%14.9f\n',SW);fclose all;clear SS;clear SW;
end;
movie(F,1,5);
```

Listing 2. Matlab code for automated creation and exporting of data subsets created by windowing for a preset number of windows. Includes movie visualization for real-time overseeing of the windowing process. (By M. Omerbashich)

```
% Randomly pollute a time-stamped time series for a preset percentage “p” of values
A=load('c:\rsrch\tests\A00.txt');n=length(A);p=75;m=floor(p*n/100);q=n-
m;k=randperm(n);B(1:m,1)=A(k(1:m),1);C(1:m,1)=A(k(1:m),2)+randn;
D=[B,C];G(1:q,1)=A(k((m+1):n),1);U(1:q,1)=A(k((m+1):n),2);V=[G,U];E=[D;V];Z=sortrows(E,1);
sp=sprintf('c:\\rsrch\\tests\\_output\\polluted_%d_percent.txt',p);
fid=fopen(sp,'wt');fprintf(fid,'%8.3f %14.9f\n',Z');fclose all;clear all;
```

Listing 3. Matlab code for polluting by normally-random noise and in a uniform-random fashion a preset percentage of data values of a time-series. (By M. Omerbashich)

```
% Random-randomly purify a time-stamped time series at a preset percentage “p” of values
A=load('c:\rsrch\tests\A00.txt');n=length(A);p=96.875;m=floor(p*n/100);q=n-m;
k=randperm(n);B(1:m,1)=A(k(1:m),1);C(1:m,1)=0;D=[B,C];G(1:q,1)=A(k((m+1):n),1);
U(1:q,1)=A(k((m+1):n),2);V=[G,U];E=[D;V];Z=sortrows(E,1);for i=1:n;Y(i,2)=Z(i,2);
if Z(i,2)==0;Y(i,1)=0;else (i,1)=Z(i,1);end;end;X=Y;X(~any(Y,2),:)=[];
sp=sprintf('c:\\rsrch\\tests\\_output\\purf_%g%%.txt',p);fid=fopen(sp,'wt');
fprintf(fid,'%8.3f %14.9f\n',X');fclose all;clear all;
```

Listing 4. Matlab code for purifying in a uniform-random fashion a preset percentage of data values of a time-series. (By M. Omerbashich)